\documentstyle[aps,prc,preprint,epsf,psfig,hangcaption]{revtex}
\draft
\tightenlines
\newcommand{\K}{\enspace,}
\newcommand{\p}{\enspace.}
\newcommand{\be}{\begin{equation}}
\newcommand{\ee}{\end{equation}}
\newcommand{\bea}{\begin{eqnarray}}
\newcommand{\eea}{\end{eqnarray}}
\newcommand{\bc}{\begin{center}}
\newcommand{\ec}{\end{center}}
\renewcommand{\d}{{\rm d}}
\newcommand{\PD}{{\partial}}
\newcommand{\btu}{\bigtriangleup}
\newcommand{\btd}{\bigtriangledown}

\renewcommand{\vec}[1]{\mbox{\boldmath${#1}$}}

\newcommand{\eps}{\varepsilon}

\newcommand{\lla}{\langle}
\newcommand{\gra}{\rangle}
\newcommand{\half}{\frac{1}{2}}

\def \a {\alpha}
\def \b {\beta}
\def \s {\sigma}

\def \ro {\rho}
\def \la {\lambda}

\def \D {\mbox{D}}

\def \be {\begin{equation}}
\def \ee{\end{equation}}
\def \bea{\begin{eqnarray}}
\def \eea{\end{eqnarray}}
\def \l {\label}
\def \m {\mbox}
\def \r {\ref}

\include{scrload}
\def\gsim{\mbox{~{\raisebox{0.4ex}{$>$}}\hspace{-1.1em}
	{\raisebox{-0.6ex}{$\sim$}}~}}
\def\lsim{\mbox{~{\raisebox{0.4ex}{$<$}}\hspace{-1.1em}
	{\raisebox{-0.6ex}{$\sim$}}~}}
\newfam\scrfam                                          % Script fonts
\global\font\twelvescr=rsfs10 scaled\magstep1%
\global\font\eightscr=rsfs7 scaled\magstep1%
\global\font\sixscr=rsfs5 scaled\magstep1%
\skewchar\twelvescr='177\skewchar\eightscr='177\skewchar\sixscr='177%
\textfont\scrfam=\twelvescr\scriptfont\scrfam=\eightscr
\scriptscriptfont\scrfam=\sixscr

\begin{document}

\title{Causal Theories of Dissipative Relativistic Fluid Dynamics
 for Nuclear Collisions.}
\author{Azwinndini Muronga\footnote{Present address: {\it Institut f\"ur Theoretische
Physik, J.W. Goethe--Universit\"at, \\ \hspace*{3.3cm}D--60325 Frankfurt am
Main, Germany}}}
\address{ School of Physics and Astronomy, University of Minnesota,
         Minneapolis, Minnesota 55455, USA}

\date{\today}
\maketitle

\begin{abstract}
Non-equilibrium fluid dynamics derived from the extended irreversible
thermodynamics of the causal M\"uller--Israel--Stewart theory of dissipative
processes in relativistic fluids based on Grad's moment method is  applied to
the study of the dynamics of hot matter produced in ultra--relativistic heavy
ion collisions. The temperature, energy density and entropy evolution are
investigated in the framework of the  Bjorken boost--invariant scaling limit.
The results of these second order theories are compared to those of first order
theories due to Eckart and to Landau and Lifshitz and those of zeroth order
(perfect fluid) due to Euler. In the presence of dissipation perfect fluid
dynamics is no longer valid in describing the evolution of the matter.  First
order theories fail in the early stages of evolution. Second order  theories
give a better description in good agreement with transport models. It is shown
in which region the Navier--Stokes--Fourier laws (first order theories) are a
reasonable limiting case of the more general extended thermodynamics (second
order theories).
\end{abstract}


\section{Introduction}

The study of space--time evolution and non--equilibrium properties of matter 
produced in high energy heavy ion collisions, such as those at the 
Relativistic Heavy Ion Collider (RHIC) at Brookhaven National Laboratory, USA 
and the  Large Hadron Collider (LHC) at CERN, Geneva  using relativistic  
dissipative fluid dynamics are of importance in understanding the observables.

High energy heavy ion collisions offer the opportunity to study the  properties
of hot and dense matter. To do so we must follow its  space--time evolution,
which is affected not only by the equation of state but also by dissipative,
non--equilibrium processes. Thus we need to know the transport coefficients
such as viscosities, conductivities, and diffusivities. We also need to know
the relaxation times for various dissipative processes under consideration. 
Knowledge of the various time and length  scales is of central importance to
help us decide  whether to apply fluid dynamics (macroscopic) or kinetic theory
(microscopic)  or a combination of the two. The use of fluid dynamics as one of
the approaches in modelling the dynamic evolution of nuclear collisions has been
successful in describing many of the  observables \cite{Stocker86,Clare86}. The
assumptions and approximations of the fluid dynamical models  are another
source for uncertainties in predicting the observables.  So far most work has
focused on the ideal or perfect fluid and/or multi-fluid dynamics. In this work
we apply the relativistic dissipative fluid dynamical approach.  It is known
even from non--relativistic studies \cite{Kapusta81} that dissipation might
affect the observables.

The first theories of relativistic dissipative fluid dynamics are due to Eckart
\cite{Eckart40} and to Landau and Lifshitz \cite{Landau59}.  The difference in
formal appearance stems from different choices for the definition of the
hydrodynamical 4--velocity. These conventional theories of dissipative fluid
dynamics are based on the  assumption that the entropy four--current contains
terms up to linear order in dissipative quantities and hence they are referred
to as {\em first order theories}  of dissipative fluids. The resulting
equations for the dissipative  fluxes are  linearly related to the
thermodynamic forces, and the resulting equations of  motion are parabolic in
structure, from which we get the Fourier--Navier--Stokes  equations.  They have
the undesirable feature that causality may not be satisfied. That is, they may
propagate  viscous and thermal signals with speeds exceeding that of light.

Extended theories of dissipative fluids due to Grad \cite{Grad49}, M\"uller
\cite{Muller67}, and Israel and Stewart \cite{Israel-Stewart} were introduced
to remedy some of these undesirable features.  These causal theories are based
on the assumption that the entropy four--current  should include terms
quadratic in the dissipative fluxes  and hence they are referred to as {\em
second order theories} of dissipative fluids. The resulting equations for the
dissipative fluxes are  hyperbolic and they lead to causal propagation of
signals  \cite{Israel-Stewart,Hiscock}. In second order theories the space of
thermodynamic quantities is expanded to include the dissipative quantities for
the particular system under consideration. These dissipative quantities are
treated as thermodynamic variables in their own right. 

A qualitative study of relativistic dissipative fluids for applications to 
relativistic heavy ions collisions has been done using these first order
theories \cite{First-order,Kouno,Danielewicz85,Mornas95,Chaudhuri,Teaney}.  The application
of second order theories to nuclear collisions has just begun 
\cite{Muronga01,Muronga02a,Muronga02b}, and the results of relativistic fluid
dynamics can also be compared to the prediction of spontaneous symmetry breaking
results \cite{Lallouet}.

The rest of the article is outlined as follows.  In section \r{sect:eqtherm} we
review the basics of equilibrium thermodynamics and perfects fluids. In 
\r{sect:noneq} the  basic formulation of relativistic dissipative fluid 
dynamics will briefly introduced. In section \r{sect:bjorken} we introduce the
special hydrodynamical model of Bjorken. In section \r{sect:ideal} we discuss
the results of perfect fluids. In section  \r{sect:dissipative} we discuss
the results of dissipative fluids. We first discuss the results of first order theories and
then for the second order theories. In section \r{sect:summary} we summarize the
results and discuss the need for hyperbolic theories for relativistic
dissipative fluids.

Throughout this article we adopt the units $\hbar = c = k_{\rm B} = 1$. The
sign convention used follows the time-like convention with the signature
$(+,-,-,-)$, and if $u^\a$ is a time-like vector, $u^\a u_\a > 0$.  The metric
tensor is always taken to be $g^{\mu \nu} = \mbox{diag} (+1,-1,-1,-1)$, the
Minkowski tensor.  Upper greek indices are contravariant, lower greek indices
covariant. The greek indices used in 4--vectors go from 0 to 3 ($t,x,y,z$)
and the roman indices  used in 3--vectors go from 1 to 3 ($x,y,z$).  The
scalar product of two 4-vectors $a^\mu\, , \,\, b^\mu$ is denoted by  $a^\mu\,
g_{\mu \nu}\, b^\nu \equiv a^\mu \, b_\mu \equiv a \cdot b$. The scalar product
of two 3-vectors is denoted by bold type, namely,  $\vec{a},\, \vec{b}$, 
$\vec{a} \cdot \vec{b}$. The notations $A^{(\a\b)} \equiv 
(A^{\a\b}+A^{\b\a})/2$ and $A^{[\a\b]} \equiv  (A^{\a\b}-A^{\b\a})/2$ denote
symmetrization and antisymmetrization respectively. The 4--derivative is
denoted by $\PD_\a \equiv \PD/\PD x^\a$. Contravariant components of a tensor
are found from covariant components by $g_{\a\b} A^\a = A_\b$, $g_{\mu\a}
g_{\nu\b} F^{\a\b} = F_{\mu\nu}$ and so on.

\section{Basics of Equilibrium Relativistic Fluid Dynamics}
\label{sect:eqtherm}

The basic ingredients of equilibrium thermodynamics are the energy density
$\eps$, the number density $n$, and the entropy density $s$.  The pressure $p$
can be related to the other state variables. In this work these basic
ingredients  will be referred to as the  primary thermodynamic variables. 
From the equation of state $s=s(\eps,n)$,
the temperature $T$ and chemical potential $\mu$ emerge as partial 
derivatives:
\bea
\eps+ p &=& T s + \mu n  \label{eq:thermdyn} \K\\
T \d s &=& \d \eps - \mu \d n \label{eq:tds} \p
\eea
The chemical potential $\mu$ is the energy required to inject one extra
particle reversibly into a fixed volume. The fundamental relation
(\ref{eq:thermdyn}) then yields the last unknown thermodynamic function $p$.
Next we  introduce the  thermal potential $\alpha$ and inverse-temperature
$\beta$ by the definitions
\be
\alpha \equiv {\mu\over T}\,\,\,\,,\,\,\,\, \beta \equiv {1\over T} \p 
\ee
Then equations (\ref{eq:thermdyn}) and (\ref{eq:tds}) take the form
\bea
s &=& \beta(\eps+p) - \alpha n  \label{eq:funda} \K\\
\d s &=&\beta \d \eps - \alpha \d n \p
\eea

In relativistic fluid dynamics it is advantageous to cast the thermodynamic
quantities in terms of covariant objects, namely, the energy--momentum tensor
$T^{\mu\nu}$,  the 4--vector number current $N^\mu$ representing the net charge
current,   and the entropy 4--current $S^\mu$.  For heavy ion collisions, the
conserved charge is usually taken to be  the net baryon number.  The covariant
quantities are expressible in terms of a unique hydrodynamical 4--velocity
$u^\mu$ (normalized as $u_\mu u^\mu=1$) and  the primary state variables 
($\eps\,,\,\,n\,,\,\,p\,,\,\,s$). 
\bea
N^\mu &=& n u^\mu \K\\
T^{\mu\nu} &=& \eps u^\mu u^\nu - p \btu^{\mu\nu} \K\\
S^\mu &=& s u^\mu \p
\eea
Here $\btu^{\mu\nu}\equiv g^{\mu\nu}-u^\mu u^\nu$ is the spatial projection
tensor orthogonal to $u^\mu$,  $u^\mu=\gamma (1,{\bf v})$ ($\gamma=1/\sqrt{1-{\bf v}^2}$) is the (local) 4-velocity of
the fluid,  and $g^{\mu \nu}$ the Minkowski metric. The pressure can be related
to the energy density and the number density by an equation of state.  The
equilibrium state will therefore be described by 5 parameters
$(\eps\,,\,\,n\,,\,\,u^\mu)$.  

The thermodynamical relations can be written in a form that involves the
covariant objects $S^\mu, N^\mu$ and $T^{\mu\nu}$ directly as basic
variables:  
\be
\d S^\mu = -\alpha \d N^\mu + \beta_\nu \d T^{\mu\nu} \K\label{eq:Gibbs}
\ee
where
\be
\beta_\nu \equiv \beta u_\nu = {u_\nu\over T} \K
\ee
defines the inverse--temperature 4--vector. From (\ref{eq:thermdyn}) we have
immediately
\be
S^{\mu} = p \beta^\mu - \alpha N^{\mu} + \beta_\nu T^{\mu\nu} \p
\label{eq:therm}
\ee
It follows from (\ref{eq:Gibbs}) and (\ref{eq:therm}) that
\be
\d(p \beta^\mu) = N^\mu \d\alpha - T^{\mu\nu} \d\beta_\nu \p\label{eq:pdfuga}
\ee
Thus the basic variables $N^\mu,\,\,T^{\mu\nu}$ and 
$S^\mu$ can all be generated from partial derivatives of the fugacity 4--vector
$ p(\alpha,\beta)\beta^\mu$, once the equation of state is known.

In the covariant formulation the hydrodynamical velocity is
co--opted in a natural way as an extra thermodynamical variable. 
The  covariant equations, expressed directly in terms of
the basic conserved variables $N^\mu$ and  $T^{\mu\nu}$, can  be
manipulated more easily and transparently. Formulations for deviation from
equilibrium can be carried out without much difficulty by making an assumption or
requirement that (\ref{eq:therm}) remains
valid at least to first order in deviations (see the next section).

The four--divergence of the energy--momentum tensor, the number 4--current,  
and the entropy 4--current vanish. That is, the energy and momentum, the
number density, and the entropy density are all conserved.
The covariant forms of particle, energy, momentum and entropy conservation are
\bea
\PD_\alpha N^\alpha=0 &\quad\Rightarrow\quad& 
u^\a \PD_\a n + n \PD_\a u^\a=0 \quad 
\K\l{eq:numbercons} \\
u_\alpha\PD_\beta T^{\alpha\beta}=0 &\quad\Rightarrow\quad& 
u^\a\PD_\a \eps+(\eps+p)\PD_\a u^\a=0 \K\l{eq:Enercons}\\
\btd^\a_\b \PD_\la T^{\b\la}=0&\quad\Rightarrow\quad& 
(\eps+p)u^\b \PD_\b u^\alpha-\btu^{\a\b}\PD_\b p=0 \K\l{eq:Momcons}\\
\PD_\a S^\a = 0&\quad\Rightarrow\quad& 
u^\a\PD_\a s + s \PD_\a u^\a  =0 \p \l{eq:entrocons}
\eea
Eq. (\r{eq:entrocons}) means that the entropy is conserved in ideal  fluid
dynamics.  Note that $s$ is constant along fluid world-lines, and not
throughout the fluid in general. If $s$ is the same constant on each world-line,
that is, if $\btd_\alpha s=0$ as well as $\dot{s}=0$, so that $\PD_\alpha
s=0$,  then the fluid is called isentropic. Thus perfect fluids in equilibrium
generate no entropy and no frictional type heating because their dynamics is
reversible and without dissipation. For many processes in nuclear collisions a
perfect fluid model is adequate. However, real fluids behave irreversibly, and
some processes in heavy ion reactions may not be understood except as
dissipative processes, requiring a relativistic theory of dissipative fluids.
An equilibrium state is characterized by the absence of viscous stresses, heat
flow and diffusion and maximum entropy principle, while a non--equilibrium state
is characterized by the principle of nondecreasing entropy which arises due to
the presence of dissipative fluxes.

Perfect fluid dynamics has been successful in describing most of the observables
\cite{Stocker86,Clare86,Rischke99}. The current status of ideal hydrodynamics in
describing observables can be found in the latest Quark Matter Proceedings 
\cite{QM} and in \cite{Huovinen02,Kolb03}. Already at the level of ideal fluid
approximation constructing numerical solution scheme to the equations 
is not an easy task. This is due to the nonlinearity of the system of 
conservation equations. Much work has been done in ideal hydrodynamics for 
heavy ion collision simulations (see e.g., \cite{Rischke95}).

\section{Non-equilibrium/dissipative Relativistic Fluid Dynamics}
\label{sect:noneq}

We give a covariant formulation of non--equilibrium
thermodynamics. The central role of entropy is highlighted.  Non--equilibrium
effects are introduced by enlarging the space of basic independent variables
through the introduction of non--equilibrium variables, such as dissipative
fluxes appearing in the conservation equations. The next step is to find
evolution equations for these extra variables. Whereas the evolution equations
for the equilibrium variables are given by the usual conservation laws, no
general criteria exist concerning the evolution equations of the dissipative
fluxes, with the exception of the restriction imposed on them by the second law
of thermodynamics.

A natural way to obtain the evolution equations for the fluxes from a
macroscopic basis is to generalize the equilibrium thermodynamic theories. 
That is, we assume the existence of a generalized entropy which depends on the
dissipative fluxes and on the equilibrium variables as well. Restrictions on
the form of the evolution equations are then imposed by the laws of
thermodynamics. From the expression for generalized entropy one then can derive
the generalized equations of state, which are of interest in the description of
system under consideration. The phenomenological formulation of the transport
equations for the first order and  second order theories is accomplished by
combining the conservation of energy--momentum and particle number with the
Gibbs equation. One then obtains an expression for the entropy 4--current, and
its divergence leads to entropy production.  Because of the enlargement of the
space of variables the expressions for the  energy-momentum tensor
$T^{\mu\nu}$, particle 4--current $N^\mu$, entropy 4-current $S^\mu$, and the
Gibbs equation contain extra terms. Transport equations for dissipative fluxes
are obtained by imposing the second law of thermodynamics, that is, the
principle of nondecreasing  entropy. The kinetic approach is based on Grad's
14--moment method \cite{Grad49}.  For a review on generalization of the
14--moment method to a mixture of several particle species see
\cite{Prakash93}. For applications and discussions of the moment method 
in kinetic and transport theory of gases see e.g. \cite{cond-mat} and for
applications in astrophysics and cosmology see e.g. \cite{astro-cosmo}.

In the early stages of relativistic nuclear collisions we want to be able to
describe phenomena at frequencies comparable to the inverse of the
relaxation times of the fluxes. At such time scales, these fluxes  must be
included in the set of basic independent variables.
In order to model dissipative processes we need non--equilibrium fluid dynamics 
or irreversible thermodynamics. A satisfactory approach to 
irreversible thermodynamics is via non--equilibrium kinetic theory. 
In this work we will, however, follow a phenomenological approach. Whenever
necessary we will point out how kinetic theory
supports many of the results and their generalization. 
A complete discussion
of irreversible thermodynamics is given in the monographs 
\cite{Muller98,Jou96,Eu92}, where most of the theory and applications are non--relativistic
but includes relativistic thermodynamics. A relativistic, but more advanced,
treatment may be found in
\cite{deGroot80,Israel89,Liu86}.
In this work we will present a simple introduction to these features,
leading up to a formulation of relativistic causal fluid dynamics that can be
used for applications in nuclear collisions.

\subsection{Basic Features of Irreversible Thermodynamics and Imperfect Fluids}

The basic formulation of relativistic hydrodynamics can be found in the 
literature 
(see for example \cite{Landau59,Misner73,Weinberg72,Csernai94}). 
We consider a simple fluid and no electromagnetic
fields. This fluid is characterized by,
\begin{eqnarray}
N_A^\mu(x) & & \makebox[0.3in][l]{}\mbox{particle 4--current}\enspace,\\
T^{\mu\nu}(x) & & \makebox[0.3in][l]{}\mbox{energy--momentum tensor} \enspace,\\
S^\mu(x) & & \makebox[0.3in][l]{}\mbox{entropy 4--current} \enspace,
\end{eqnarray}
where $A = 1,...,r$ for the $r$ conserved net charge currents, such as electric
charge, baryon number, and strangeness.
$N_A^\mu$ and $T^{\mu\nu}$ represent conserved quantities:
\begin{eqnarray}
\partial_\mu N_A^\mu &=& 0 \enspace, \\
\partial_\mu T^{\mu\nu} &=& 0\enspace.
\end{eqnarray}
The above equations are the local conservation of net charge and
energy--momentum. They are the equations of motion of a relativistic fluid. 
There are $4+r$ equations and $10+4r$ independent unknown functions. 
The second law of thermodynamics requires
\begin{equation}
\partial_\mu S^\mu \geq 0 \enspace,
\end{equation}
and it forms the basis for the development of the extended irreversible
thermodynamics.

We now perform a tensor decomposition of $N^\mu,\,T^{\mu\nu},$ and $S^\mu$ with
respect to an arbitrary, time--like,  4--vector $u^\mu$, normalized as 
$u^\mu u_\mu
=1$, and the projection onto the 3-space  $\btu^{\mu\nu}$ orthogonal to 
$u^\mu$. Note that 
\begin{equation}
\btu^{\mu \nu}  = \btu^{\nu \mu}\, ,\,\,
\btu^{\mu \nu} u_\nu = 0 \, , \,\, \btu^{\mu \alpha} 
\btu_{\alpha}^\nu = \btu^{\mu \nu}\,,\,\,\,
\btu^\nu_\nu = 3  \enspace.
\end{equation}
The tensor decomposition reads:
\begin{eqnarray}
N^\mu & = & n \, u^\mu + V^\mu \enspace , \\
T^{\mu \nu} & = & \eps \, u^\mu u^\nu - p \btu^{\mu \nu}
+ 2\,W^{(\mu }\,u^{\nu)} + t^{\mu \nu} \enspace,\\
S^\mu &=& s u^\mu + \Phi^\mu \enspace,
\end{eqnarray}
where we have defined
\begin{eqnarray}
W^\mu &=&  q^\mu + h\, V^\mu \enspace,\\
t^{\mu\nu} &=& \pi^{\mu\nu} - \Pi\, \btu^{\mu \nu} \enspace.
 \end{eqnarray} 
Here $h$ is the enthalpy per particle defined by
\begin{equation}
h = \frac{(\eps + p)}{n} \enspace.
\end{equation} 
The dissipative fluxes satisfy the following orthogonality relations:
\begin{equation}
u_\mu V^\mu=0\,,\,\,u_\mu q^\mu=0\,,\,\,u_\mu\, W^\mu =0\,,\,\, u_\mu\,
t^{\mu\nu} =0 \,, \,\, \pi^\nu_\nu =0 \enspace.
\end{equation}

In the {\em local rest frame} (LRF) defined by $u^\mu=(1,\bf{0})$ the quantities appearing in the decomposed tensors have the 
following meanings:
\begin{eqnarray}
n & \equiv & \makebox[1.in][l]{$u_\mu N^\mu$} \mbox{net density of charge}\enspace,\\
V^\mu &\equiv& \makebox[1.in][l]{$\btu^\mu_\nu N^\nu$} \mbox{net flow of charge}\enspace,\\
\eps &\equiv& \makebox[1.in][l]{$u_\mu T^{\mu\nu} u_\nu$} \mbox{energy density}\enspace,\\
p + \Pi &\equiv& \makebox[1.in][l]{$\displaystyle -\frac{1}{3} \btu_{\mu\nu} T^{\mu\nu}$ } 
\mbox{pressure}\enspace,\\
W^\mu &\equiv& \makebox[1.in][l]{$u_\nu T^{\nu\lambda} \btu^\mu_\lambda $}
\mbox{energy flow}\enspace,\\ 
q^\mu &\equiv& \makebox[1.in][l]{$W^\mu - h V^\mu$}\mbox{heat flow}
\enspace,\\
\pi^{\mu\nu} &\equiv& \makebox[1.in][l]{$T^{\langle\mu\nu\rangle}$ }\mbox{stress
tensor}\enspace,\\
s &\equiv& \makebox[1.in][l]{$u_\mu\, S^\mu $}\mbox{entropy density}
\enspace,\\
\Phi^\mu &\equiv& \makebox[1.in][l]{$\btu^\mu_\nu S^\nu $} \mbox{entropy flux} \enspace.
\end{eqnarray}
The angular bracket notation, representing the symmetrized spatial and
traceless part of the tensor, is defined by
\begin{equation}
A^{\lla\mu\nu\gra} \equiv \left[\frac{1}{2}\left(\btu^\mu_\sigma
\btu^\nu_\tau
+\btu^\mu_\tau \btu^\nu_\sigma\right)
-\frac{1}{3}\btu^{\mu\nu}\btu_{\sigma\tau}\right]
A^{\sigma\tau} \enspace.
\end{equation}
The space--time derivative decomposes into
\begin{equation}
\partial^\mu = u^\mu \D + \btd^\mu \,, ~~~~~u^\mu \btd_\mu
=0 \enspace,
\end{equation}
where 
\begin{eqnarray}
D &\equiv& \makebox[1.in][l]{$u^\mu\partial_\mu $} \mbox{convective time
derivative}\enspace,\\
\btd^\mu &\equiv& \makebox[1.in][l]{$\btu^{\mu\nu}\partial_\nu $} \mbox{gradient
operator}\enspace.
\end{eqnarray}
In the LRF the two operators attain the given meanings. 
In this rest frame the projector becomes
\begin{equation}
\btu_{\mu\nu} =\btu^{\mu\nu} =
\mbox{diag}(0,-1,-1,-1),~~~~~~~~~~\btu^\mu_\nu = \mbox{diag}(0,1,1,1) \enspace.
\end{equation}
In the LRF the heat flow has spatial components only  and the
pressure tensor defined by $P^{\mu\nu}\equiv \btu^\mu_\sigma
T^{\sigma\tau}\btu^\nu_\tau = -(p\btu^{\mu\nu}+\Pi)+\pi^{\mu\nu}$ is also 
purely spatial. 

So far, $u^\mu$ is arbitrary. It has the following properties. Differentiating
$u_\mu u^\mu=1$ with respect to space--time co-ordinates, $\PD_\mu \equiv (\PD/\PD
x^\mu) \equiv (\PD_t,\vec{\nabla})$, yields
\be
u_\mu \PD_\nu u^\mu =0 \K
\ee
which is a useful relation. 

There are two choices for $u^\mu$ \cite{deGroot80}. It can be taken parallel to
the particle flux $N^\mu$,
\bea
u^\mu &\equiv& \frac{N^\mu}{\sqrt{N^\mu N_\mu}} \enspace ,\\
N^\mu &=& n u^\mu \label{eq:NE} \p
\eea
This is known as the Eckart or particle frame and in this frame $V^\mu=0$. 
It can also be taken to be parallel to the energy flow,
\bea
u^\mu &\equiv& \frac{T^\mu_\nu u^\nu}{\sqrt{u^\alpha 
T_\alpha^\beta T_{\beta \gamma} u^\gamma}} \label{eq:TL}\enspace,\\
u_{\nu} T^{\mu\nu} &=& \eps u^\mu \p
\eea
This is known as the Landau and Lifshitz or energy frame and in this frame
$W^\mu=0$. This implies that $q^\mu=-h V^\mu$.

The two choices of velocity 4--flow have different computational advantage of
each of the formulations. The Landau--Lifshitz formalism is  convenient to
employ since it reduces the energy--momentum tensor to a simpler form. The
price for this is the implicit definition of the 4--velocity. The Eckart
formalism  has the advantage when one wants to have simple integration of
particle conservation law. This choice is also more intuitive than that of
Landau--Lifshitz.   For a system with no net charge, the 4--velocity in the
Eckart formalism is not well--defined and therefore in general under this
situation one should use the Landau--Lifshitz formalism. The Landau--Lifshitz
formalism is also advantageous in the case of mixtures.

Let us consider the Eckart definition of 4-velocity. 
With respect to this frame the particle number 4-current and energy--momentum tensor thus decomposes as
follows:
\bea
N^\mu &=& n u^\mu \K\\
T^{\mu\nu} &=& \eps u^\mu u^\nu + 2q^{(\mu}u^{\nu)} + P^{\mu\nu} \K
\label{eq:TE}\\
P^{\mu\nu} &=& P^{\nu\mu}~~~,~~~P^{\mu\nu} u_{\mu}=q^\mu u_{\mu} =0 \K\\
P^{\mu\nu} &=& -(p+\Pi)\btu^{\mu\nu} +\pi^{\mu\nu}~~~,~~~\btu^{\mu\nu}
\pi_{\mu\nu}=0 \p\label{eq:succesive}
\eea
Successive terms on the right hand side of (\ref{eq:succesive}) represent the
thermodynamical pressure, bulk stress and shear stress measured by an observer
in the Eckart frame.

\subsection{Conservation Laws}

We will now consider one type of charge, namely, the net baryon number.
We also consider the Eckart frame for the 4-velocity definition. 
We insert the expressions for the number 4--current and the energy--momentum
tensor in the conservation laws and project them onto the four velocity and the
projection tensor.
Using the orthogonality properties given in the previous section we obtain the
following conservation laws.
The equation of continuity (net charge conservation) $\partial_\mu N^\mu \equiv 0$, equation
of motion (momentum conservation) $\btu^\mu_\nu\partial_\lambda
T^{\nu\lambda} \equiv 0$, and the equation of energy (energy conservation) 
$u_\mu \partial_\nu T^{\mu\nu} \equiv 0$ are, respectively,
\bea
\D\,n &=& -n\,\btd_\mu \,u^\mu \K\l{eq:E-ncons}\\
(\eps + p +\Pi) \D\, u^\mu &=& \btd^\mu (p+\Pi) -
\btu^\mu_\nu\btd_\sigma \pi^{\nu\sigma} + \pi^{\mu\nu} \D \,u_\nu \nonumber\\
&&-[\btu^\mu_\nu \D\, q^\nu + 2 q^{(\mu} \btd_\nu u^{\nu)}]
\enspace,\l{eq:E-momcons}\\
\D\, \eps &=& -(\eps + p + \Pi) \btd_\mu u^\mu +
\pi^{\mu\nu} \btd_{\lla\nu} u_{\mu\gra} \nonumber\\
&& -\btd_\mu q^\mu +2 q^\mu \D
u_\mu \enspace. \label{eq:E-energycons}
\eea

There are 14 unknowns, 
$n,\,\varepsilon,\Pi,q^\mu,\pi^{\mu\nu}$, and
$u^\mu$ and 5 equations.
To close the system of equations we need to obtain 9 additional equations
(for dissipative fluxes) in
addition to the 5 conservation equations (for primary variables) we already 
know. The derivation of the 9 equations will be presented in the following
sections. 

\subsection{Entropy 4-current and the generalized Gibbs equation}
From the phenomenological treatment of deriving the 9 additional equations 
we need the expression for the out--of--equilibrium entropy 4-current.
The off-equilibrium entropy 4-current $S^\mu(N^\mu, T^{\mu\nu})$ takes the form
\cite{Israel-Stewart}
\be
S^\mu = p(\alpha,\beta)\beta^\mu - \alpha N^\mu +\beta_\nu
T^{\mu\nu} + Q^\mu(\delta N^\mu,\,\, \delta T^{\mu\nu},\,\,...)
\label{eq:entropy}
\ee
where $Q^\mu$ is a function of deviations $\delta N^\mu$ and $\delta
T^{\mu\nu}$ from local equilibrium 
\be
\delta T^{\mu\nu} = T^{\mu\nu}-T^{\mu\nu}_{eq}\,\,\,\,,\,\,\,\, \delta N^\mu =
N^\mu - N_{eq}^\mu \K
\ee
containing all the
information about viscous stresses and heat flux in the off--equilibrium
state.

Since the equilibrium pressure is only known as a function of the equilibrium
energy density and equilibrium net charge density we need to match/fix the
equilibrium pressure to the actual state. We do this by requiring that the
equilibrium energy density and the equilibrium net charge density be equal to
the off--equilibrium energy density and off--equilibrium net charge density. In
Eckart's formalism this is equivalent to
\be
\delta T^{\mu\nu} u_\mu u_\nu=\delta N^\mu u_\mu.
\ee

With the help of the expression for the divergence of $p\beta^\mu$
\be
\PD_\mu (p\beta^\mu) = N^\mu_{eq} \PD_\mu \alpha -T^{\mu\nu}_{eq}
\PD_\mu\beta_\nu \K
\ee
and the conservation laws for $N^\mu$ and for $T^{\mu\nu}$ the generalized
second law of thermodynamics becomes
\be
\PD_\mu S^\mu = -(\delta N^\mu)\PD_\mu \alpha +\delta T^{\mu\nu} \PD_\mu\beta_\nu
+\PD_\mu Q^\mu \p
\ee

Once a detailed form of $Q^\mu$ is specified, linear relations between
irreversible fluxes ($\delta N^\mu$, $\delta T^{\mu\nu}$) and gradients 
($\PD_{(\mu}\beta_{\nu)}$, $\PD_\mu \alpha$) follow by imposing the second law
of thermodynamics, namely, that the entropy production be positive. 
The key to a complete phenomenological theory thus lies in the specification of
$Q^\mu$.

\subsection{Standard Relativistic Dissipative Fluid Dynamics}

The standard Landau--Lifshitz and  Eckart theories make the simplest possible
assumption about $Q^\mu$. That it is linear in the dissipative quantities
$(\Pi,\,q^\mu,\pi^{\mu\nu})$. In kinetic theory this amounts to Taylor
expanding the entropy 4--current  expression up to first order in deviations
from equilibrium. This leads to an expression of entropy
4--current which is just a linear function of the heat flux.

This can be understood as follows: Take a simple fluid with particle current 
$N^\mu$. Let us 
choose $\beta^\mu=u^\mu/T$ parallel to the current $N^\mu$ of the given
off--equilibrium state, so we are in the Eckart frame. Projecting
(\ref{eq:entropy}) onto the 3-space orthogonal to $u^\mu$ gives
\be
\Phi^\mu \equiv \btu^\mu_\nu S^\nu = \beta q^\mu +Q^\nu
\btu^\mu_\nu~~~,~~~~~q^\mu =u_\nu T^{\nu\la} \btu^\mu_\la \K\label{eq:Eheat}
\ee
so that 
\be
\Phi^\mu =q^\mu /T + \mbox{second order terms}  \K
\ee
which, to linear order, is just the standard relation between entropy flux
$\Phi^\mu$ and heat flux $q^\mu$. 
From (\ref{eq:Eheat}) 
this implies that the entropy flux $\Phi^\mu$ 
is a strictly linear function of heat flux $q^\mu$, 
and depends on no other variables; also that the off--equilibrium
entropy density depends only on the densities $\eps$ and $n$ and is given
precisely by the equation of state
$s = s_{eq}(\eps,n)$.

Alternatively we may begin with the ansatz for the entropy 4--current $S^\mu$: 
In the limit of
vanishing $\Pi,\,q^\mu,$ and $\pi^{\mu\nu}$ the entropy 4--current must reduce to the
one of ideal fluid. The only non--vanishing 4--vector which can be formed from the
available tensors $u^\mu, \,q^\mu$ and $\pi^{\mu\nu}$ is $\beta q^\mu$ where
$\beta$ is arbitrary but it turns out to be nothing else but the inverse
temperature. Thus the first order expression
for the entropy 4--current in the Eckart frame is given by
\be
S^\mu=s u^\mu+{q^\mu\over T} \K
\l{eq:entroflux}\ee
and one immediately realizes that 
\be
\Phi^\mu = {q^\mu\over T} \K
\ee
is the entropy flux.   
Using the expressions for $N^\mu$, $T^{\mu\nu}$ and
$S^\mu$ in the second law of thermodynamics $\PD_\mu S^\mu \geq 0$  and using
the conservation laws $u_\nu\partial_\mu T^{\mu\nu} = 0$,  $\partial_\mu
N^\mu =0$ and the Gibbs equation 
\be
\PD_\mu (p\beta^\mu) =  N^\mu \PD_\mu \alpha -
T^{\mu\nu}\PD_\mu\beta_\nu \K\label{eq:gibbs}
\ee
the divergence of (\r{eq:entroflux}) gives
the following expression for entropy production
\begin{equation}
T\,\partial_\mu S^\mu = q^\mu(\btd_\mu\beta +
D\,u_\mu) +\pi^{\mu\nu}\btd_\mu u_\nu +\Pi
\btd_\mu u^\mu \geq 0 \enspace. \l{eq:entro-prod}
\end{equation}
Notice that the equilibrium conditions (i.e., the bulk free and shear free 
of the flow and the constancy of the thermal potential,i.e., no heat flow) 
lead to the vanishing of each factor multiplying
the dissipative terms on the
right, and therefore lead to $\PD_\a S^\a=0$.
The expression (\ref{eq:entro-prod}) splits into three independent,
irreducible pieces:
\be
\Pi X -q^\mu X_\mu + \pi^{\mu\nu} X_{\lla\mu\nu\gra} 
\geq 0 \K
\ee
where
\bea
X &=& -\btd_\mu u^\mu \K\\
X^\mu &=& {\btd^\mu T\over T} -D u^\mu \K\\
X^{\lla\mu\nu\gra} &=& \left[{1\over2}\left(\btu^\mu_\s \btu^\nu_\tau
+\btu^\nu_\s \btu^\mu_\tau\right) -{1\over3}
\btu^{\mu\nu}\btu_{\s\tau}\right]\btd^\s u^\tau \p
\eea
The angular brackets extract the purely spatial, trace-free part of the tensor. 
From the second law of thermodynamics, $\PD_\mu S^\mu \geq 0$, we see that the 
simplest way to satisfy the bilinear expression (\r{eq:entro-prod})
is to impose the following linear relationships between the
thermodynamic fluxes $\Pi, q^\mu, \pi^{\mu\nu}$ and the
corresponding thermodynamic forces $\btd_\mu u^\mu, \btd^\mu \ln T - D u^\mu,
\btd^{\lla\mu}u^{\nu\gra}$: 
\begin{eqnarray}
\Pi &\equiv& -\zeta \btd_\mu u^\mu \l{eq:Pieq}\enspace,\\
q^\mu &\equiv& \lambda T \left(\frac{\btd^\mu T}{T} - D u^\mu\right)
= -{\lambda \, n \,T^2 \over \eps+p}\btd^\mu \left(\frac{\mu}{T}\right)
\l{eq:qeq}\enspace,\\
\pi^{\mu\nu} &\equiv& 2 \eta \btd^{\langle\mu}u^{\nu\rangle}
\l{eq:pieq}\enspace.
\end{eqnarray}
That is, we assume that the dissipative fluxes are linear and purely local
functions of the gradients. We then obtain uniquely, if the equilibrium state is
isotropic (Curie's principle), the above linear expressions.

These are the constitutive equations for dissipative fluxes in the standard
Eckart theory of relativistic irreversible thermodynamics. They are
relativistic generalizations of the corresponding Newtonian laws. The linear
laws allow us to identify the thermodynamic coefficients, namely the bulk
viscosity $\zeta(\eps,n)$,  the thermal conductivity $\lambda(\eps,n)$ and  the
shear viscosity $\eta(\eps,n)$.

Given the linear constitutive equations (\r{eq:Pieq}) -- (\r{eq:pieq}), the
entropy production rate (\r{eq:entro-prod}) becomes
\be
\PD_\mu S^\mu = {\Pi^2\over\zeta T}-{q_\mu q^\mu\over\lambda T^2}+
{\pi_{\mu\nu}\pi^{\mu\nu}\over2\eta T} \geq 0
\l{eq:secondlaw}\ee
which is guaranteed to be non--negative provided that
\be
\zeta\geq 0\,,\quad\lambda\geq0\,,\quad\eta\geq0 \p
\ee
Note that $q^\mu q_\mu < 0$ which can be most easily proven from $q^\mu u_\mu
=0$ in the LRF.

Using the fundamental thermodynamic equation of Gibbs 
the entropy evolution equation can be written in the following convenient form:
\begin{equation}
T \partial_\mu S^\mu = \sigma_{\mu\nu}\pi^{\mu\nu} -\Pi \theta -\partial_\mu
q^\mu +q^\mu a_\mu \enspace, \label{eq:entropyequation}
\end{equation}
which can be found with the help of the {\em fluid kinematic identity}:
\be
\PD_\nu u_\mu = \sigma_{\nu\mu} +\omega_{\nu\mu} +{1\over 3} \theta\btu_{\mu\nu}
              +a_\mu u_\nu \K
\ee
where the fluid kinematic variables are given by
\begin{eqnarray}
        a_\mu &\equiv & \makebox[2.in][l]{$u^\nu \partial_\nu u_\mu$}  
	\mbox{4--acceleration of
	the fluid}\enspace,\nonumber\\
	\omega_{\mu\nu} & \equiv & \makebox[2.in][l]{$ \displaystyle \btu_\mu^\a
	\btu_\nu^\b \partial_{[\b} u_{\a]}$} 
	\mbox{vorticity tensor}\enspace,\nonumber\\
	\theta_{\mu\nu} & \equiv & \makebox[2.in][l]{$ 
	\displaystyle \btu_\mu^\a\btu_\nu^\b
	\partial_{(\b} u_{\a)}$} \mbox{expansion tensor}\enspace,\\
	\theta & \equiv & \makebox[2.in][l]{$  \btu^{\mu\nu}
	\theta_{\mu\nu} = \partial_\mu
	u^\mu $} \mbox{volume expansion } \enspace,\nonumber\\
	\sigma_{\mu\nu} &\equiv & \makebox[2.in][l]{$\displaystyle 
	\theta_{\mu\nu}
	-\frac{1}{3}\btu_{\mu\nu}\theta $} \mbox{shear
	tensor}\enspace.\nonumber
\end{eqnarray}
The quantities defined here are the fluid kinematic variables.

The Navier--Stokes--Fourier equations comprise a  set of 9 equations. Together
with the 5 conservation laws $\PD_\mu T^{\mu\nu}= \PD_\mu N^\mu =0$, they
should suffice, on the basis of naive counting, to determine the evolution of
the 14 variables $T^{\mu\nu}$ and $N^\mu$ from suitable initial data.
Unfortunately, this system of equations is of mixed
parabolic--hyperbolic--elliptic type.  Just like the non-relativistic
Fourier--Navier--Stokes theory, they predicts infinite propagation speeds for
thermal and viscous disturbances. Already at the non--relativistic level, the
parabolic character of the equations has been a source of concern
\cite{Cattaneo}. One would expect signal velocities to be bounded by
the mean molecular speed. However in the non--relativistic
case wave--front velocities can be 
infinite such as the case in the tail of Maxwell's distribution wich has
arbitrarily large velocities. However, a relativistic theory which predicts
infinite speeds for causal propagation contradicts the foundation or the basic
principles of relativity and must be a cause of concern especially when one has to use the theory to
explain observables from relativistic phenomena or experiments such as
ultra--relativistic heavy ion experiments. The other problem is that these
first order theories possess instabilities:  equilibrium states are unstable
under small perturbations  \cite{Hiscock}.

Most of the applications of dissipative fluid dynamics in relativistic nuclear
collisions have used the Eckart/Landau-Lifshitz theory. However, the algebraic
nature of the Eckart constitutive equations leads to severe problems.
Qualitatively, it can be seen from (\r{eq:Pieq}) -- (\r{eq:pieq}) that if a
thermodynamic force is suddenly switched off/on, then the corresponding
thermodynamic flux instantaneously vanishes/appears. This indicates that a
signal propagates through the fluid at infinite speed, violating relativistic
causality.  This is known as a paradox since in special relativity the speed of
light is finite and all maximum speeds should not be greater than this speed.
This paradox was first addressed by Cattaneo \cite{Cattaneo} by introducing
{\em ad hoc}  relaxation terms in the phenomenological equations. The resulting
equations  conforms with causality. The only problem was that a sound theory
was needed. It is from these arguments that the causal extended theory of 
M\"uller, Israel and Stewart was developed. 

\subsection{Causal Relativistic Dissipative Fluid Dynamics}

Clearly the Eckart postulate (\r{eq:entroflux}) for $Q^\mu$ and hence $S^\mu$ 
is too simple.
Kinetic theory indicates that in fact $Q^\mu$ is second order in the
dissipative fluxes. The Eckart assumption, by truncating at first
order, removes the terms that are necessary to provide causality, hyperbolicity 
and stability.

The second order kinetic theory formulation of the entropy 4--current, see 
\cite{Grad49}, was the starting point for good work during the 
1950's  
on extending the domain of validity of conventional thermodynamics to shorter 
space--time scale. The turning point was M\"uller's 1967 paper \cite{Muller67} 
which, for the
first time, expressed $Q^\mu$ in terms of the off-equilibrium forces
($\Pi,\vec{q},\pi^{ij}$) and thus linked phenomenology to the Grad expansion 
\cite{Grad49}. 
This marked the birth of what is now known as extended irreversible 
thermodynamics \cite{Muller98,Jou96,Eu92}.

For small deviations, it will suffice to retain only the lowest--order, 
quadratic terms in the Taylor expansion of $Q^\mu$ , leading to linear
phenomenological laws. 
The most general algebraic form for $Q^\mu$ that is at
most second order in the dissipative fluxes gives \cite{Israel-Stewart}
\bea
S^\mu  &=& s u^\mu+{q^\mu\over T}-
\left(\beta_0\Pi^2
-\beta_1q_\nu q^\nu+\beta_2\pi_{\nu\la}
\pi^{\nu\la}\right){u^\mu\over 2T} \nonumber\\
{}&& -{\alpha_0\Pi q^\mu\over T}+{\alpha_1\pi^{\mu\nu}q_\nu\over T} 
\K\l{eq:Entro-4current}
\eea
where $\beta_A(\eps,n)\geq0$  are thermodynamic
coefficients for scalar, vector and tensor dissipative contributions
to the entropy density, and
$\alpha_A(\eps,n)$ are thermodynamic
viscous/ heat coupling coefficients. It follows from (\r{eq:Entro-4current}) 
that
the effective entropy density measured by comoving observers is
\be
u_\mu S^\mu = s-
{1\over 2 T}\left(\beta_0\Pi^2
-\beta_1q_\mu q^\mu+\beta_2\pi_{\mu\nu}\pi^{\mu\nu}\right) \K\l{eq:s-eff}
\ee
independent of $\alpha_A$. 
Note that the entropy density is a maximum in equilibrium. 
The condition $u_\mu Q^\mu \leq 0$, which guarantees that entropy is maximized in
equilibrium, requires that the $\beta_A$ be nonnegative.
The entropy flux is
\be
\Phi^\mu =\beta (q^\mu -\alpha_0 \Pi q^\mu +\alpha_1 \pi^{\mu\nu} q_\nu) 
\K\l{eq:entro-flux}
\ee
and is independent of the $\beta_A$.

The divergence of the extended current (\r{eq:Entro-4current}) together 
with the Gibbs equation (\r{eq:gibbs}) and
the conservation equations (\r{eq:E-ncons}), (\r{eq:E-momcons}) 
and (\r{eq:E-energycons}) lead to
\bea
T\PD_\mu S^\mu &=& -\Pi\left[\theta+\beta_0\dot{\Pi} + {1\over2}
T\PD_\mu \left({\beta_0\over T}u^\mu\right)\Pi -\alpha_0\btd_\mu q^\mu\right]\nonumber\\
{}&& -q^\mu\left[\btd_\mu\ln T-\dot{u}_\mu-\beta_1\dot{q}_\mu - {1\over2}
T\PD_\nu \left({\beta_1\over T}u^\nu\right)q_\mu -\alpha_0\btd_\nu\pi^\nu_\mu
-\alpha_1\btd_\mu \Pi\right]\nonumber\\
{}&&+\pi^{\mu\nu}\left[\sigma_{\mu\nu}-\beta_2\dot{\pi}_{\mu\nu}+
{1\over2}T\PD_\la \left({\beta_2\over T}u^\la\right)
\pi_{\mu\nu} +\alpha_1 \btd_{\lla\nu}q_{\mu\gra}\right] \p\l{eq:Secondlaw}
\eea
The simplest way to satisfy the second law of thermodynamics,  
$\PD_\mu S^\mu \geq 0$,   
is to impose, as in the standard theory, linear relationships 
between the thermodynamical fluxes and extended thermodynamic forces, 
leading to
the following constitutive or transport equations, 
\bea
\tau_\Pi\dot{\Pi}+\Pi &=& -\zeta \theta-\left[{1\over2}\zeta T
\PD_\mu \left({\tau_0\over\zeta T}u^\mu\right)\Pi\right] +\tau_0\btd_\mu q^\mu 
\K\l{eq:bulkeq} \\
\tau_q \btu_\mu^\nu\dot{q}_\nu + q_\mu &=& \lambda\left(\btd_\mu T-T\dot{u}_\mu
\right)+\left[{1\over2}\lambda T^2\PD_\nu \left({\tau_1\over\lambda T^2}
u^\nu\right)q_\mu\right] \nonumber\\
&&-\tau_0\btd_\mu\Pi -\tau_1\btd_\nu\pi^\nu_\mu  \K\l{eq:heateq} \\
\tau_\pi \btu_\mu^\a \btu_\nu^\b\dot{\pi}_{\a\b}+\pi_{\mu\nu} &=&
2\eta\sigma_{\mu\nu}-\left[\eta T\PD_\la \left({\tau_2\over 2\eta T}u^\la
\right)\pi_{\mu\nu}\right] \nonumber\\
&&+\tau_2\btd_{\lla\mu}q_{\nu\gra} \p \l{eq:sheareq}
\eea
Here the relaxational times $\tau_A(\eps,n)$ are given by
\be
\tau_\Pi=\zeta\beta_0\,,\quad\tau_q=\lambda T\beta_1\,,\quad
\tau_\pi=2\eta\beta_2 \K\l{eq:reltimes}
\ee
and the coupling coefficients $\tau_i(\eps,n)$ are given by
\be
\tau_0=\zeta\alpha_0\,,\quad\tau_1=\lambda T\alpha_1\,,\quad
\tau_2=2\eta\alpha_1 \p\l{eq:couplings}
\ee

A key quantity in these theories is the relaxation time $\tau$ of the
corresponding  dissipative process. It is a positive--definite quantity by the
requirement of hyperbolicity. 
It is the time taken by the corresponding dissipative flux to relax to its
steady state value.  
It is 
connected to the mean collision time $t_{c}$ of the  particles responsible
for the dissipative process, but the two are not the same.
In principle they are different since $\tau$ is a macroscopic time, although 
in some instances it may correspond just to a few $t_{c}$. No general formula linking
$\tau$ and $t_{c}$ exists; their relationship depends in
each case on the system under consideration. 

Besides the  
fact that parabolic theories are necessarily non--causal,
it is obvious that whenever the time scale of the problem under
consideration becomes of the order of or smaller than the relaxation time,
the latter cannot be ignored. Neglecting the relaxation time in this 
situation amounts to disregarding the whole problem under consideration.

Even in the steady--state regime the descriptions offered by parabolic and
hyperbolic theories might not necessarily coincide. The differences
between them in such a situation arise from the presence of
$\tau$ in terms that couple the vorticity to the heat flux and
shear stresses. These may be large even in steady states where vorticity is
important. There are also other acceleration coupling terms
to bulk and shear stresses and the heat flux. The coefficients for these vanish
in parabolic theories, but they could
be large even in the steady state. The convective
part of the time derivative, which is not negligible
in the presence of large spatial gradients, and 
modifications in the equations of state due to the presence of
dissipative fluxes also differentiate hyperbolic theories from parabolic ones. 
However, it is precisely before the establishment of the steady--state 
regime that both types of theories differ more significantly.
Therefore, if one wishes to study a dissipative process for times
shorter than $\tau$, it is mandatory to resort to a hyperbolic
theory which is a more accurate macroscopic approximation
to the underlying kinetic description.

Provided that the spatial gradients are not so large that the convective part
of the time derivative becomes important,  and that the fluxes and coupling
terms remain safely small, then for times larger than $\tau$ it is sensible to
resort to a parabolic  theory.  However, even in these cases, it should be kept
in mind that the way a system approaches equilibrium may be very sensitive to
the relaxation time.  The future of the system at time scales much longer than
the relaxation time once the steady state is reached, may also critically
depend on $\tau$.

The crucial difference between the standard Eckart and the extended
Israel--Stewart transport equations is that the latter are differential
evolution equations, while the former are algebraic relations. The evolution 
terms, with the relaxational time coefficients $\tau_A$, are needed for
causality, as well as for modelling high energy heavy ion collisions relaxation
effects are important. The price paid for the improvements that the extended
causal thermodynamics brings is that new thermodynamic coefficients are
introduced. However, as is the case with the coefficients $\zeta,\lambda$ and
$\eta$ that occur also in standard theory, these new coefficients may be
evaluated or at least estimated via kinetic theory. The relaxation times
$\tau_A$ involve complicated collision integrals. They are usually estimated as
mean collision times, of the form $ \tau \approx {1\over n\sigma v}
\K\l{eq:coltimes} $  where $\sigma$ is a collision cross section and $v$ the
mean particle speed.

The form of transport equations obtained here is justified by kinetic theory, 
which leads to the same form of the transport equations, but with extra terms
and explicit expressions for transport, relaxation and coupling coefficients.
With these transport equations, the entropy production rate has the same
non--negative form (\r{eq:secondlaw}) as in the standard theory. In addition to
viscous/ heat couplings, kinetic theory shows  that in general there will also
be couplings of the heat flux and  the anisotropic pressure to the vorticity. 
These couplings give rise to the following additions to the right hand sides of
(\r{eq:heateq}) and (\r{eq:sheareq}), respectively: 
\be +\tau_q
\omega_{\mu\nu}q^\nu\quad\quad\m{and}\quad\quad +\tau_\pi
\pi^\la_{\lla\mu}\omega_{\nu\gra\la} \p 
\ee 
In the case of scaling solution assumption in nuclear collisions these 
additional terms do not contribute since they vanish.
Also the resulting expression for $\PD_\mu S^\mu$  in general contains terms
involving gradients of $\alpha_A$ and $\beta_A$ multiplying second order
quantities such as the bilinear terms  $(\PD_\mu \alpha_1) q_\la \pi^{\la\mu}$
and   $(\PD_\mu \alpha_0)q^\mu \Pi$. In the present work where we will assume
scaling solution these terms do no contribute to the overall analysis.

It is also important to remember that the derivation of the causal
transport equations is based on the assumption that the fluid is
close to equilibrium. Thus the dissipative fluxes are small:
\be
|\Pi|\ll p\,,\quad \left(\pi_{\mu\nu}\pi^{\mu\nu}\right)^{1/2}
\ll p\,,\quad \left(-q_\mu q^\mu\right)^{1/2}\ll \eps \p \l{eq:noneq-eq}
\ee
These conditions will also be useful in guiding us when we discuss the initial
conditions for the dissipative fluxes.
Considering the evolution of entropy in the Israel--Stewart theory, 
equation (\r{eq:entropyequation}) still holds.

\subsection{ Hyperbolicity, Causality and Characteristic Speeds}

Let us write the system of 14 equations in full, that is, the 5 conservation 
equations and the 9 evolution equations for dissipative quantities, in one
single system. We will write down the linearized system; that is, we assume
that the gradients of the macroscopic variables may be treated as small
quantities of the order of the deviations from equilibrium. Consequently, terms
like $V^\mu D u_\mu$ etc., can be omitted because they are then of second order
in deviations from equilibrium. The system of 14 equations, in a general frame,
is then written as:
\begin{eqnarray}
&& Dn + n\btd_{\mu}u^{\mu} +\btd_\mu V^\mu =0\label{eq:ncons} \K\\
&&(\eps + p)Du^{\mu} + \btd^{\mu}(p+\Pi)
 +D W^\mu  - \btd_\nu\pi^{\mu\nu}
 =0\label{eq:mcons} \K\\
&& D\eps +(\eps + p)\btd_\mu u^\mu 
                  +\btd_\mu W^\mu =0 \label{eq:econs} \K\\
& &\beta_0 D\Pi  -\alpha_0\btd_\mu q^\mu +\frac{1}{\zeta }\Pi
+\btd_\mu \left(u^\mu+{1\over n} V^\mu\right) =0\label{eq:bulk} \K\\
& &\beta_1 D q^\mu +\alpha_0 \btd^\mu\Pi +\alpha_1\btd_\nu\pi^{\mu\nu}
+\frac{1}{\lambda T }q^\mu 
-\left(\frac{\btd^\mu T}{T}- D u^\mu \right) =0 \label{eq:heat} \K\\
& &\beta_2 D\pi^{\mu\nu} -\alpha_1 \btd^{\langle\mu}q^{\nu\rangle}
+\frac{1}{2\eta }\pi^{\mu\nu}
 -\btd^{\langle\mu}u^{\nu \rangle} -{1\over n} \btd^{\lla\mu} V^{\nu\gra}
=0\label{eq:shear} \p
\end{eqnarray}

The system of 14 equations may be written as a
quasilinear system of $14$ equations in the form
\be
M^{\alpha A}_B(U^c) \PD_\alpha U^B = f^A(U^c)\,\,\,\,\,(A,B\,=\,1,...,14)
\enspace,\label{eq:system}
\ee
where $M^{\alpha A}_B(U^c)$ and $f^A(U^c)$ can be taken to be components of
$14\times 14$ matrices and $14$ vectors.
The right hand side contains all the collision terms, and the coefficients 
$M^{\alpha A}_B(U^c)$ are purely thermodynamical functions. 

Let $\Sigma$ be a characteristic hypersurface for the system
(\ref{eq:system}) and let $\phi(x^\alpha)$ = 0 be the local equation for
$\Sigma$.
Then $\phi$ satisfies the characteristic equation
\begin{equation}
\mbox{det}\left[M^{\alpha A}_B (\PD_\alpha\phi) \right] = 0 \p 
\label{eq:characts}
\end{equation}
$\phi (x^\alpha)$ is a 3-dimensional space across which the variables $U^B$
are continuous but their first derivatives are allowed to present discontinuities
$[\PD_\alpha U^B]$ normal to the surface $\Longrightarrow \,[\PD_\alpha
U^B]=U^B(\PD_\alpha \phi)$. The characteristic speeds are independent of
the microscopic details such as cross sections. To solve the characteristic equation
(\ref{eq:characts}) we consider a coordinate system $x^\alpha$ chosen in
such a way that at any point in the fluid the system of reference is orthogonal
and comoving. 
If $\phi$ is a function of $x^0$ and $x^1$ only, the  characteristic speeds can be
determined from
\begin{equation}
\mbox{det}(v M^{A0}_B - M^{A1}_B) =0 \l{eq:veq} \K
\end{equation}
where $v$ is the characteristic speed defined by
\begin{equation}
v = -\PD_0 \phi/\PD_1 \phi \p
\end{equation}

The 14--component vector $U$ has been split into a scalar-longitudinal
6-vector $U_L = [\eps, n,\Pi,u^x,q^x,\pi^{xx}]$; two longitudinal-transverse
3-vectors (corresponding to the two transverse directions of polarization); 
$U_{LT_1} = [u^y,q^y,\pi^{xy}]$ and $U_{LT_2} =	[u^z,q^z,\pi^{xz}]$ and purely
transverse 2-vector $U_T = [\pi^{yz},\pi^{yy}-\pi^{zz}]$. Equation (\r{eq:veq})
for $v$ accordingly splits into one 6th-degree and two 3rd-degree equations. 
The purely transverse modes do not propagate. 
Hiscock and Lindblom \cite{Hiscock} proved -- as anticipated -- that this general scheme, when
applied to first order theories, always yields wave--front speeds $v$ that are
superluminal.       

We will be studying the dynamics of a pion fluid in the hadronic regime and a
quark--gluon plasma fluid in the partonic regime. It is therefore important to
check if these systems conform with the principle of relativity under small
perturbation of the equilibrium state. For a
quark--gluon plasma we consider a gas of weakly-interacting massless quarks and
gluons. We also consider such a system to have a vanishing baryon chemical
potential $(\mu_B =0)$. 
This implies also that the net baryon charge is zero
$(n_B=0)$. The equation of state is given by $p=\eps/3$. For massless particles 
or ultra--relativistic particles the bulk viscosity vanishes.

In the absence of any conserved charge the convenient choice of the
4--velocity is the Landau--Lifshitz frame. In this case the characteristic
equations for the wave--front speeds becomes very simple. For the
longitudinal modes we get only the {\em fast} longitudinal mode (associated 
with the true acoustical wave). The absence of
heat conduction has as a consequence the disappearance of the {\em slow}
longitudinal propagation mode (associated with thermal dissipation wave). The
phase velocity of the fast longitudinal mode is given by
\be
v_L^2  = {1+2 p \beta_2 \over 6 p\beta_2} \p
\ee
Using $\beta_2 \approx 3/4\times 1/p$ (see \cite{Israel-Stewart} for the coefficients $\a_i$
and $\b_i$), we see that
\be
{v_L^2} = {5\over 9} \p
\ee
If we are considering only the shear viscosity we will get
the above result.
The wave--front speed (signal speed) for the transverse plane wave is
given by
\be
v_T^2 = {1\over 8 p \beta_2} \p
\ee
Using $\beta_2 \approx 3/4 \times 1/p$, this reduces to
\be
{v_T^2} = {1\over 6} \p
\ee

For a pion fluid with vanishing chemical potential $\mu_\pi=0$ we have,
for the fast longitudinal mode, the following expression for the wave--front
speed
\be
\begin{displaystyle}
v_L^2 = { {2\over3 }\beta_0 +\beta_2 + \beta_0\beta_2 (\eps+p){\PD p
/\PD \eps } \over \beta_0\beta_2(\eps+p)} \K
\end{displaystyle}
\ee
and the wave--front speed of the transverse plane wave is given by
\be
v_T^2 = {1\over 2 \beta_2 (\eps+p)} \p
\ee 
Note that for the pion system we are in the relativistic regime. 
Then the equation of state is
taken to be that of a non--interacting gas of pions only. The bulk
viscosity does not vanish. We show the dependence of the wave--front speeds
and the adiabatic speed of sound on temperature in Fig.
\r{fig:pionspeeds}.
\begin{figure}[hbt]
\bc
\leavevmode
\hbox{
   \epsfxsize=10cm
   \epsffile{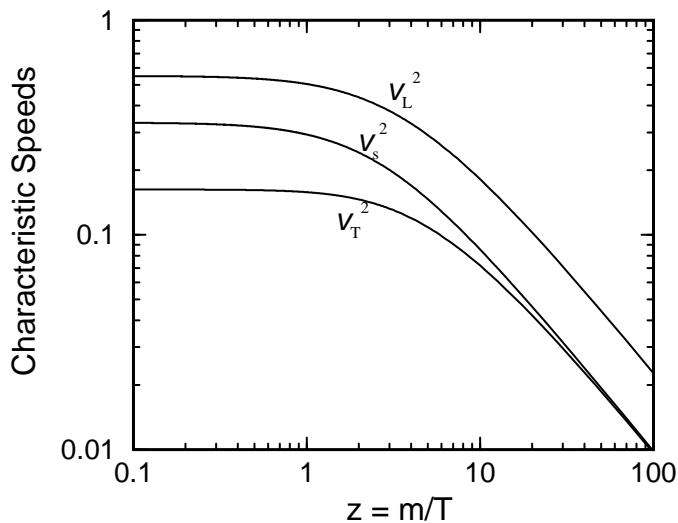}
    }
%\vspace*{-3cm}
\caption{The transverse ($v_T$), longitudinal ($v_L$) and sound ($v_s$) phase velocities in a
pion gas, as a function of $z=m/T$. $m$ is the mass of pion.}
\label{fig:pionspeeds}
\ec
\end{figure}
For completeness and as an illustration we show here the results of a
nucleon--nucleon system at a finite chemical potential. In this system all modes
are present, including the thermal mode. The results are shown in
Fig \r{fig:NNspeeds}.

\begin{figure}[hbt]
\bc
\leavevmode
\hbox{
   \epsfxsize=10cm
   \epsffile{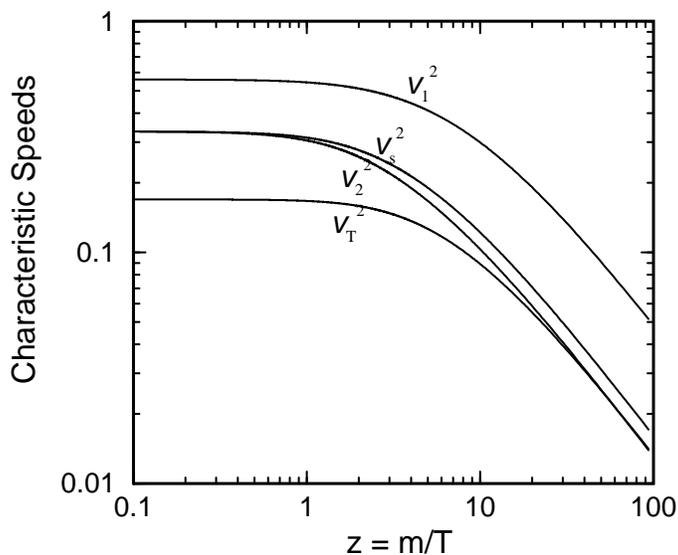}
    }
%\vspace*{-3cm}
\caption{The transverse ($v_T$), longitudinal ($v_1,\,v_2$) and sound ($v_s$)
phase velocities in a nucleon gas, as a function of $z=m/T$. $m$ is the mass of
nucleon.}
\label{fig:NNspeeds}
\ec
\end{figure}
Thus in all the systems we will consider here (see next section)
causality is obeyed.

\section{Relativistic Nucleus--Nucleus Collisions}
\label{sect:bjorken}

In heavy ion collisions there is much interest in the mid--rapidity region
because the equation of state is simple for vanishing net baryon charge. Based
on the observation that the rapidity distribution of the charged particle
multiplicity $\d N_{ch}/\d y $ is constant in the mid--rapidity region
\cite{Bjorken83}, that is, it is invariant under  Lorentz transformation in the
mid--rapidity region, it is reasonable to assume that all other quantities like
number density, energy density and dissipative fluxes also have this symmetry.
Thus these quantities depend on the proper time, $\tau$, and 
\be
\tau = t/\gamma = t\sqrt{1-v^2} = \sqrt{t^2-z^2} \p
\ee
The longitudinal component of the matter velocity is assumed to scale as
$v=z/t$ which includes a boost--invariance for the central region. This is the
Bjorken \cite{Bjorken83} scaling solution assumption for high energy nuclear collisions. In
nucleus--nucleus collision experiments a central rapidity plateau is observed
\cite{QM}. Thus one would expect that the measured quantities from the central
region will be boost--invariant. By introducing the above scaling, the
boost--invariance is automatically built into the model, simplifying the
hydrodynamical equations. The effect of the transverse hydrodynamical flow is
to remove the scale invariance at radial distances of the order of the nuclear
radius.

All particles are assumed to originate from one point in space--time, that is,
along the constant $\tau$ hypersurfaces. 
This symmetry is assumed to be achieved at the time of
thermalization after which the system evolution is governed by the
equations of relativistic fluid dynamics for given initial and boundary
conditions.

Covariantly, the four--velocity can be expressed as
\be
u^\mu={1\over \tau}(t,0,0,z) = {x^\mu \over \tau} \p
\ee
One defines the spatial rapidity $y$
as 
\be
y = \half \ln \left({t+z \over t-z}\right) \p
\ee
The spatial proper--time, $\tau$ is invariant under boosts along the $z$--axis. 
With the definitions of spatial rapidity and proper time one can find the
inverse transformations
\bea
t &=& \tau \cosh y \K\\
z &=& \tau \sinh y \p
\eea
Then the four--velocity can be written as
\be
u^\mu = (\cosh y,0,0,\sinh y) \p\label{eq:4-velo} 
\ee
The transformation of the derivatives is
\begin{equation}
\left( \begin{array}{c}
\displaystyle{\frac{\partial}{\partial t}} \\\\ 
\displaystyle{\frac{\partial}{\partial z}} \end{array} \right) =
\left( \begin{array}{cc} 
          \displaystyle{\cosh y }& \displaystyle{-\sinh y} \\ 
	  \displaystyle{-\sinh y }& \displaystyle{\cosh y} \end{array} \right)
\left( \begin{array}{c}
       \displaystyle{\frac{\partial}{\partial \tau}} \\\\ 
       \displaystyle{\frac{1}{\tau} \frac{\partial}{\partial y}}
       \end{array} \right) \enspace,
\end{equation}

We also note that using the transformation of derivatives and the definition of
the 4-velocity we can write
\bea
\theta &\equiv& \PD_\mu u^\mu = {1\over \tau} \K\\
D &\equiv& u^\mu \PD_\mu ={\PD \over \PD \tau} \p
\eea
The recipe given in this section will be used to simplify the equations of
relativistic fluid dynamics in the next sections.
In this work we consider a (1+1)--dimensional scaling solution in
which we have one nonvanishing spatial component of the 4-velocity in a (3+1)
space--time.

\section{Ideal Fluid Dynamics}
\label{sect:ideal}

Assuming local thermal and chemical equilibrium, hydrodynamical 
equations can be derived from the Boltzmann equation. 
The perfect fluid approximation is the simplest approximation. 
The relativistic hydrodynamical equations follow from the conservation 
of the baryon number, energy, and momentum. 
From the conservation of energy--momentum, 
$\partial _\mu T^{\mu \nu}=0$, multiplied by $u^\nu$, the relativistic Euler 
equation follows
\be
u^\mu\> \partial_\mu \eps + (\eps +p) \partial_\mu u^\mu =0 \p
\label{Euler}
\ee

Considering the energy--momentum conservation, $\PD_\mu T^{\mu\nu}=0$, we change the variables 
$t,\,\, z$ in the conservation laws for one--dimensional longitudinal motion in
the absence of the conserved charges,
\bea
\PD_t T^{00}+\PD_z T^{z0} &=&0 \K\\
\PD_t T^{0z}+\PD_z T^{zz} &=&0 \p
\eea
to the new variables $\tau,\,\,y$. 
We assume only a longitudinal boost--invariant
expansion, $u^\mu =x^\mu/\tau$, (Bjorken scenario \cite{Bjorken83}),
as might be the case at RHIC and LHC energies.
Then the coupled system of partial differential equations
reduces to
\bea
{\PD \eps \over \PD \tau} &=& -{\eps+p \over \tau} \K\label{eq:endens}\\
{\PD p \over \PD y} &=&0 \label{eq:forces} \p
\eea
The second equation (\r{eq:forces}) implies that there are no forces between
fluid elements with different $y$. Thus in the central rapidity region
thermodynamic quantities don't depend on $y$. From equation (\r{eq:endens}), 
together with the first law of thermodynamics, one derives the conservation of
the entropy,
\be
{\PD s \over \PD \tau} = -{s\over \tau} \K
\ee
which can be immediately integrated to give
\be
s\tau=s_0\tau_0 =\mbox{constant} \K\label{eq:entrodens}
\ee
at constant $y$. Equation (\r{eq:entrodens}) shows that the entropy density
decreases proportional to $\tau$ independent of the equation of state of the
fluid. 

For an ideal ultra--relativistic gas, such as the non-interacting
quark--gluon plasma, $\eps =3p$ and the evolution of the energy density, entropy
density and temperature, can be determined easily:
\bea
T(\tau)&=&T_0\left({\tau_0\over\tau}\right)^{1/3} \K \label{eq:0Tsol}\\
s(\tau) &=&s_0{\tau_0\over\tau} \K\\
\eps(\tau)&=&\eps_0\left({\tau_0\over\tau}\right)^{4/3} \p
\eea
Here $\tau_0$, $\epsilon_0$, $s_0$ and $T_0$ are the initial time,
energy density, entropy density and temperature, respectively. 
They are determined by the time at which 
local equilibrium has been achieved. 
\begin{figure}[htbp]
\centering
\psfig{figure=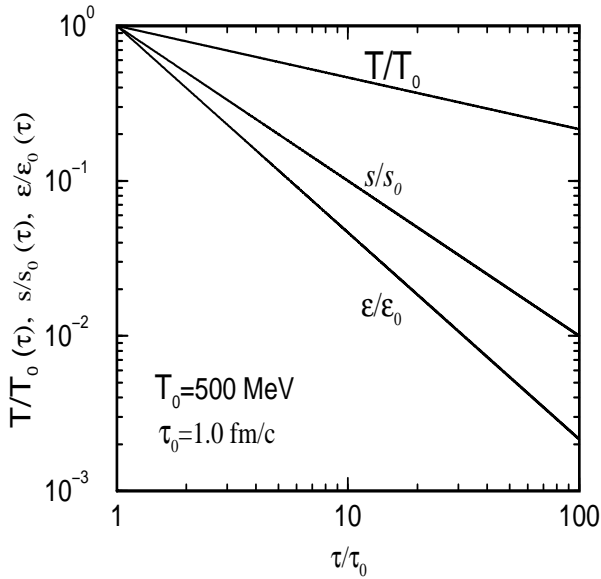,width=3.5in,height=3.5in}
\caption{Proper time evolution for temperature, energy density and 
entropy density in the Bjorken model for nuclear 
collisions (longitudinal expansion only).}
\label{fig:idealtes}
\end{figure}
The evolution of the energy density, the entropy density and the temperature
are shown as a function of proper time $\tau$ in Fig. \r{fig:idealtes}. For
(1+1)--dimensional scaling solution the scalar volume expansion decreases as
$\theta \sim \tau^{-1}$. Due to additional mechanical work performed the energy
density decreases as $\eps \sim \tau^{-4/3}$ and consequently the temperature
decreases as $T \sim \tau^{-1/3}$.

The results of a hydrodynamical model depend strongly on the choice of the
initial conditions. Therefore a reliable determination of the initial 
conditions is crucial. The initial conditions can be taken in principle from 
transport calculations describing the approach to equilibrium, such as the
Parton Cascade Model (PCM) commonly known as VNI \cite{Geiger95} which treat
the entire evolution of the parton gas from the first contact of the cold
nuclei to hadronization or Heavy Ion Jet Interaction Generator (HIJING) 
\cite{Wang97}.  For initializing a hadronic state one can use 
Ultrarelativistic Quantum Molecular Dynamics (UrQMD) \cite{Bass98}.

Another frequently used relation between the initial temperature and the 
initial time is based on the uncertainty 
principle \cite{Kapusta92}. The formation time $\tau $ of a particle with an 
average energy 
$\langle E\rangle$ is given by $\tau \langle E\rangle \simeq 1$.
The average energy of a thermal parton is about $3T$. Hence, we find 
$\tau_0\simeq 1/(3T_0)$. However, the formation time of a particle, the 
time required to reach its mass shell, is not necessarily identical
to the thermalization time \cite{Kapusta92}. Consequently, 
determination of the initial conditions is far from being
trivial. However, if data for hadron production are available, such as at SPS, 
they can be used to determine
or at least constrain the initial conditions for a hydrodynamical
calculation of observables such as the photon spectra \cite{Peitzmann02}. 

The hydrodynamical model presented above for illustration
is certainly oversimplified. 
Transverse expansion cannot be neglected, especially during the later
stages of the fireball, significantly changing the observables at 
RHIC and LHC. 
In reality the expansion of the system will not be
purely longitudinal; the system will also expand transversally.

It might not be justified 
to use an ideal fluid; dissipation might be important too. 
Hence, the Euler equation should be 
replaced by the Navier-Stokes \cite{Chaudhuri,Teaney} or even higher order dissipative
equations \cite{Muronga01,Muronga02a,Muronga02b}. 

Under the simplifying assumption of an ideal fluid, the 
hydrodynamical equations can be solved numerically
using an equation of state 
and the initial conditions, such as initial time and 
temperature, as input. The final results depend strongly on the input 
parameters as well as on other details of the model, as
in the simple one--dimensional case. Also, it is important to adopt a 
realistic equation of state, in particular for the hadron gas.

\section{Non-Ideal/Dissipative Fluid Dynamics}
\label{sect:dissipative}

In dissipative fluid dynamics entropy is generated by dissipation. The
dissipative quantities, namely, $\Pi, q^\mu$ and $\pi^{\mu\nu}$ are not set 
{\em a priori} to zero. They are specified through additional equations.
Since we will be working with a baryon--free system ($n=0$), a convenient 
choice of the reference frame is the Landau and Lifshitz frame. 
The number current, energy--momentum tensor and the entropy 4--current in this
frame are given by
\bea
N^\mu &=& n u^\mu + V^\mu \K\\
T^{\mu\nu} &=& \eps u^\mu u^\nu - (p+\Pi)\btu^{\mu\nu} +\pi^{\mu\nu}
\K\\
S^\mu &=& s u^\mu - \beta h V^\mu \K 
\eea
Taking the divergence of the entropy 4--current with the help of 
conservation of net baryon number, $\PD_\mu N^\mu=0$, conservation of energy, 
$u_\mu\PD_\nu T^{\mu\nu}=0$ 
together with the relativistic Gibbs-Duhem relation in the form
\be
{1\over n}\btd^\mu p = h{\btd\mu T \over T} +T\btd^\mu {\mu \over T} \K
\ee
gives the equation for the entropy flow:
\be
T \PD_\mu\left[s u^\mu -{h\over T} V^\mu\right] = -T V^\mu \btd_\mu\left({\mu\over
T}\right) -\Pi\theta +\pi^{\mu\nu} \btd_{\lla\mu} u_{\nu\gra} \K\label{eq:entpro}
\ee
where $s$ is the entropy density. While in the case of a perfect fluid the entropy
is conserved, here we see that the right hand side of (\r{eq:entpro}) represents
entropy production. 
In the Landau and Lifshitz frame the energy and momentum conservation laws
are, respectively,  
\bea
D \eps &=& -(\eps+p+\Pi) \theta +\pi^{\mu\nu} \btd_\nu u_\mu
\K\label{eq:enercons}\\
(\eps+p+\Pi) Du^\mu &=& \btd^\mu (p+\Pi) -\btu_\nu^\mu\btd_\s \pi^{\nu\s} 
+\pi^{\mu\nu} D u_\nu \p
\label{eq:momcons}
\eea
 
The energy--momentum tensor in the Landau--Lifshitz LRF is given by
\be
\label{Tmunulrf} T_{LRF}^{\mu \nu }=\left( 
\begin{array}{cccc}
\eps & 0 & 0 & 0 \\ 
0 & \left( p+\Pi-\pi /2\right)  & 0 &0 \\ 
0 & 0 & \left(p+\Pi -\pi /2\right) & 0 \\ 
0 & 0 & 0 & \left(p+\Pi+\pi\right) 
\end{array}
\right) \p 
\ee
This satisfy $T^\nu_\nu = \eps -3(p+\Pi)$, $\pi^\nu_\nu = \pi^{\mu\nu}u_\nu =
0$. 
To study the dynamics of the system it is necessary to obtain the
energy--momentum 
tensor as seen by an observer at rest with respect to the Minkowskian
coordinates. To this end it is necessary to apply a boost in the longitudinal
direction. Using $u^\mu = (\cosh y,0,0,\sinh y)$ we have 
\be
\label{Tmunufinal} T^{\mu \nu }=\left( 
\begin{array}{cccc}
{\mathcal{W}} \cosh^2 y -{\mathcal{P}} & 0 & 0 & {\mathcal{W}} \cosh y \sinh y \\ 
0& {\mathcal{P_\perp}} & 0 & 0 \\ 
0 & 0 &{\mathcal{P_\perp}}  & 0 \\ 
{\mathcal{W}} \cosh y \sinh y & 0 & 0 & {\mathcal{W}} \sinh^2 y +{\mathcal{P}}
\end{array}
\right) \K 
\ee
with ${\mathcal{W}} = \eps +{\mathcal{P}}$ the effective enthalpy density, 
${\mathcal{P}} = p+\Pi+\pi$ the effective pressure  and ${\mathcal{P_\perp}} =
p+\Pi-\pi/2$ the effective transverse pressure. For perfect fluids
$\pi=\Pi=0$.  It is clear that the effects of viscosity is to reduce  pressure
in the longitudinal direction and increase pressure in the transverse
direction. In the first order theory
\bea
\Pi &=& -\zeta{1\over \tau} \K\\
\pi &=& -{4\over 3}\eta {1\over \tau} \p
\eea
In the second order theory $\Pi$ and $\pi$ have to be determined from the second order
transport equations. 

Already in the first order theory regime the effects of various dissipative
processes are  clear. Bulk viscosity corresponds to the isotropic gradient and
its effect is to lower the pressure. Heat conductivity is associated with
energy flow from a temperature gradient. Shear viscosity,  which is associated
with anisotropic velocity gradients, lowers the pressure along the gradient. In
the second order theory regime there are additional terms such as relaxation
and coupling terms.  At RHIC the local longitudinal pressure scales with the
inverse proper time between collisions. Thus the pressure in the longitudinal
direction is reduced. This in turn will lead to a reduction in the longitudinal
work. This implies more transverse energy. Transverse expansion will be
accelerated.

\subsection{Standard Dissipative Fluid Dynamics}

In the first order theories the dissipative fluxes are given by 
\bea
\Pi &=& -\zeta\theta \K \label{eq:bulkeqn}\\
V^\mu &=& \lambda\left[{n T\over \eps+p}\right]^2 \btd^\mu \left(\mu\over
T\right) \K \label{eq:heateqn}\\
\pi^{\mu\nu} &=& 2\eta\left(\btd^{(\mu} u^{\nu)}  -{1\over 3}
\btu^{\mu\nu}\btd_\s u^\s\right) \K\label{eq:sheareqn}
\eea 
The current $V^\mu$ is induced by heat conduction, $V^\mu =-q^\mu/h$. 

The (1+1)--dimensional scaling solution implies that the thermodynamic 
quantities depend on $\tau$ only.
Thus the scaling solution and the relations
$\PD \tau/{\PD x^\mu} = u_\mu\,$ and $\PD f(\tau)/\PD x^\mu = u_\mu(\PD
f(\tau)/\PD \tau)$ (where $f(\tau)$ represent thermodynamic variables such as
temperature, chemical potential and dissipative fluxes) lead to
\bea
\Pi &=& -\zeta {1\over\tau} \K\\
V^\mu &=& 0 \label{eq:heat} \K\\
\pi^{00} &=& -{4\over 3}\eta{1\over \tau} \sinh^2 y \K\\
\pi^{zz} &=& -{4\over 3}\eta{1\over \tau} \cosh^2 y \p 
\eea
Note that in the energy and entropy equations the quantity that appears is
$\pi^{00}-\pi^{zz}$, which is independent of $y$.
Equation (\r{eq:heat}) implies that there is no heat conduction in the scaling
solutions. 
This is independent of the fact that $n=0$, another condition that also makes
$V^\mu$ vanish. 
The scaling solution also implies that
\be
(\eps+p)D u^\mu =0\,\,,\,\, \btd^\mu p
=0\,\,,\,\,\btu^\mu_\nu \partial_\alpha
(\pi^{\nu\alpha} - \bigtriangleup^{\nu\alpha}\Pi) = 0\enspace.
\ee

Before we continue, let us note that in (N+1) space--time 
$\delta^\nu_\nu=N+1$, $\btu^\mu_\mu = N$ and $N$ is the dimension of space. 
The symmetrized spatial and traceless part of the hydrodynamic velocity
gradient is in general
\be
X^{\lla\mu\nu\gra} = \left[\half\left(\btd^\mu u^\nu+\btd^\nu u^\mu\right)
-{1\over N}\btu^{\mu\nu} \btd_\s u^\s\right] \p
\ee
This is traceless since in $(N+1)$ space--time $\btu^\mu_\mu = N$ 

The $\tau$ dependence of thermodynamic quantities can be determined from the
entropy and energy equations. 
We will only consider a (1+1)--dimensional scaling solution in (3+1) dimensions 
since we only have shear viscosity. The energy and entropy  
conservation becomes, in first order theory,
\begin{equation}
\frac{d\varepsilon}{d\tau}+\frac{\varepsilon+p}{\tau} -
\frac{4\eta+\zeta}{3}\frac{1}{\tau^2} = 0 \enspace.
\end{equation} 
Similarly, the entropy equation becomes
\begin{equation}
\frac{d s}{d \tau} +\frac{s}{\tau} - 
\left[\frac{4}{3}\eta + \zeta\right]\frac{1}{\tau^2 T} =0 \enspace.
\end{equation}
These two equations form the basis for the first order theory results 
in this work.
In terms of the ratio of the nondissipative term to dissipative term we can
write the above equations as
\bea
{\PD \eps  \over \PD \tau}  + {\eps+p\over\tau} &=& {\eps+p\over R\tau}
\K\label{eq:de}\\
{\PD s  \over \PD \tau}  + {s\over\tau} &=& {s\over R\tau} \p\label{eq:ds}
\eea
where the ratio $R$, associated with the Reynolds number in \cite{Kouno,Gavin84}, 
is defined by 
\be
R^{-1} \equiv {{4\over 3}\eta +\zeta \over (\eps+p) \tau}
={{4\over 3}\eta +\zeta \over T s \tau} \p\label{eq:Re}
\ee
In the case of perfect fluid, $R^{-1}$ and hence the right hand side of 
Eq. (\r{eq:ds}) 
vanish. Thus the entropy density in the ideal fluid approximation is proportional to
$\tau^{-1}$, since $s \tau$ is constant.

For this exploratory study a simple equation of state is used, namely that of a weakly interacting plasma 
of massless $u\,,d\,,s\,$ quarks and gluons. The pressure is given by $p =
\varepsilon/3 = aT^4$ with zero baryon chemical potential. 
That is,  $\mu=0$,  $\eps=3 p$ or $s=4 a T^3$
, $\eta = b T^3$ and $\zeta=0$ , $a,\,b =$ constant. The solution of Eq.
(\r{eq:ds}) becomes
\bea
{T\over T_0} &=& \left[{\tau_0\over\tau}\right]^{1/3} \left\{1+{b\over 6 a
\tau_0 T_0}\left[1-\left[{\tau_0\over\tau}\right]^{2/3}\right]\right\}
\nonumber\\ 
&=& \left[\tau_0\over\tau\right]^{1/3} \left\{1+ {R^{-1}_0\over
2}\left[1-\left[\tau_0\over \tau\right]^{2/3}\right]\right\} \K\label{eq:1Tsol}
\eea
where $T_0$ and $R_0$ are the initial values of the temperature and the Reynolds
number at the initial proper time $\tau=\tau_0$.
Here 
\be
a = \left(16 + \frac{21}{2} N_f\right)\frac{\pi^2}{90} \K
\ee
is a
constant determined by the number of quark flavors and the number of gluon
colors. In the case of massless particles the bulk pressure
equation (\ref{eq:bulkeqn}) does not contribute since the bulk viscosity is
negligible or vanishes
\cite{Weinberg72}.  
The only relaxation coefficient we need is $\beta_2$ which, for massless
particles, is given by $\beta_2 = 3/(4 p)$.
The shear viscosity is given by
\cite{Baym} $\eta = b T^3$ where 
\be
b = \left(1 + 1.70 N_f\right)\frac{0.342}{(1+N_f/6)\,\alpha_s^2\ln(\alpha_s^{-1})}
\enspace,
\ee
is a constant determined by the number of quark flavors and the number of gluon
colors. Here $N_f$ is the number of quark flavors, taken to be $3$, and
$\alpha_s$ is the strong fine structure constant, taken to be $0.4$--$0.5$.

A nonvanishing $R^{-1}$ makes the cooling rate smaller, as expected. If
$R^{-1}_0=0$, Eq. (\r{eq:1Tsol}) gives
\be
{T\over T_0} = \left[{\tau_0\over\tau}\right]^{1/3} \K
\ee
as is expected for an ideal fluid.
The role of $R$ can be examined by rewriting energy and entropy equations as 
\bea
{\PD \eps\over \PD \tau} &=& (R^{-1}-1) {\eps+p\over \tau} \K\\
{\PD s\over \PD \tau} &=& (R^{-1}-1) {s\over \tau} \p
\eea
As is seen in these equations, the energy density and entropy density increase 
if $R<1$ and decrease if $R>1$. Therefore there is a peak in $s$ (and hence in
$T$ and in $\eps$) if $R_0<1$, and no peak if $R_0>1$.  $R=1$ is the critical
value for Reynolds number. One notes that at this value thermodynamics
quantities do not change with time. $R$ itself increases monotonically towards
a limiting value during the expansion stage as shown in Fig. \r{fig:TsR1}. 

\begin{figure}[hp]
\begin{minipage}[t]{7.0cm}
\centerline{\psfig{figure=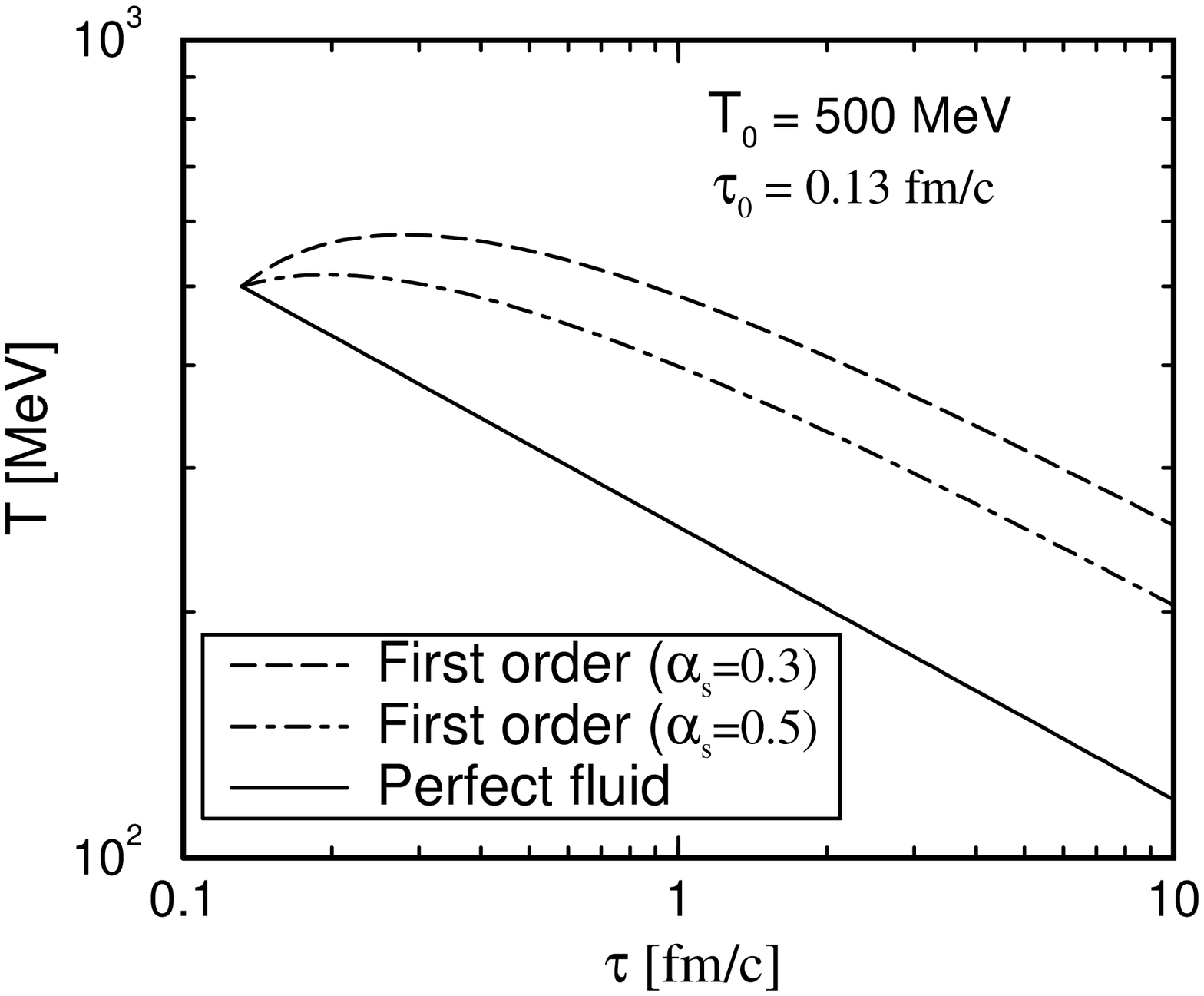,width=7.5cm}}
\end{minipage}
\hfill
\begin{minipage}[t]{7.0cm}
\centerline{\psfig{figure=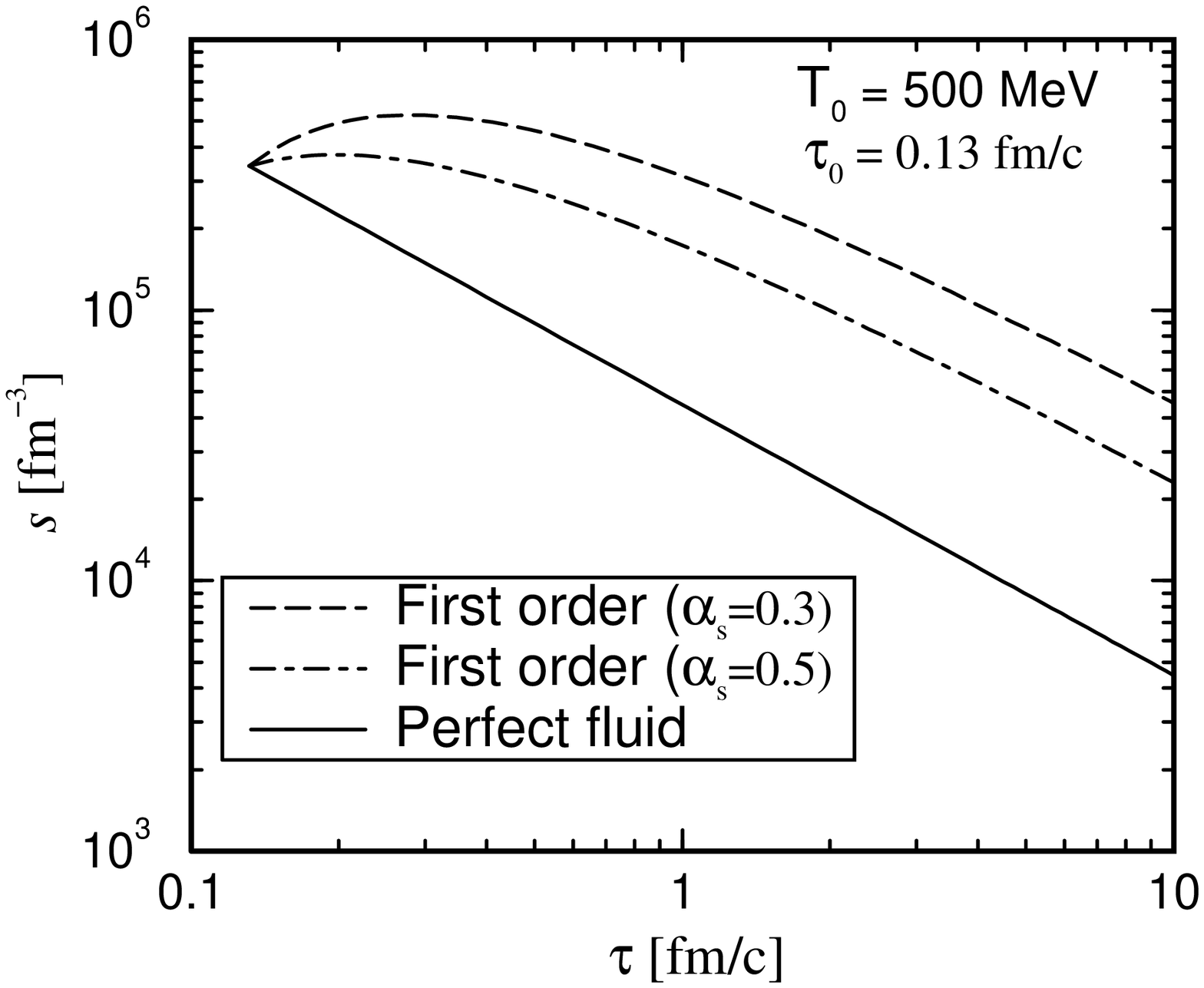,width=7.5cm}}
\end{minipage}

\vspace{0.3cm}
\centering
\begin{minipage}[b]{9.0cm}
\centerline{\psfig{figure=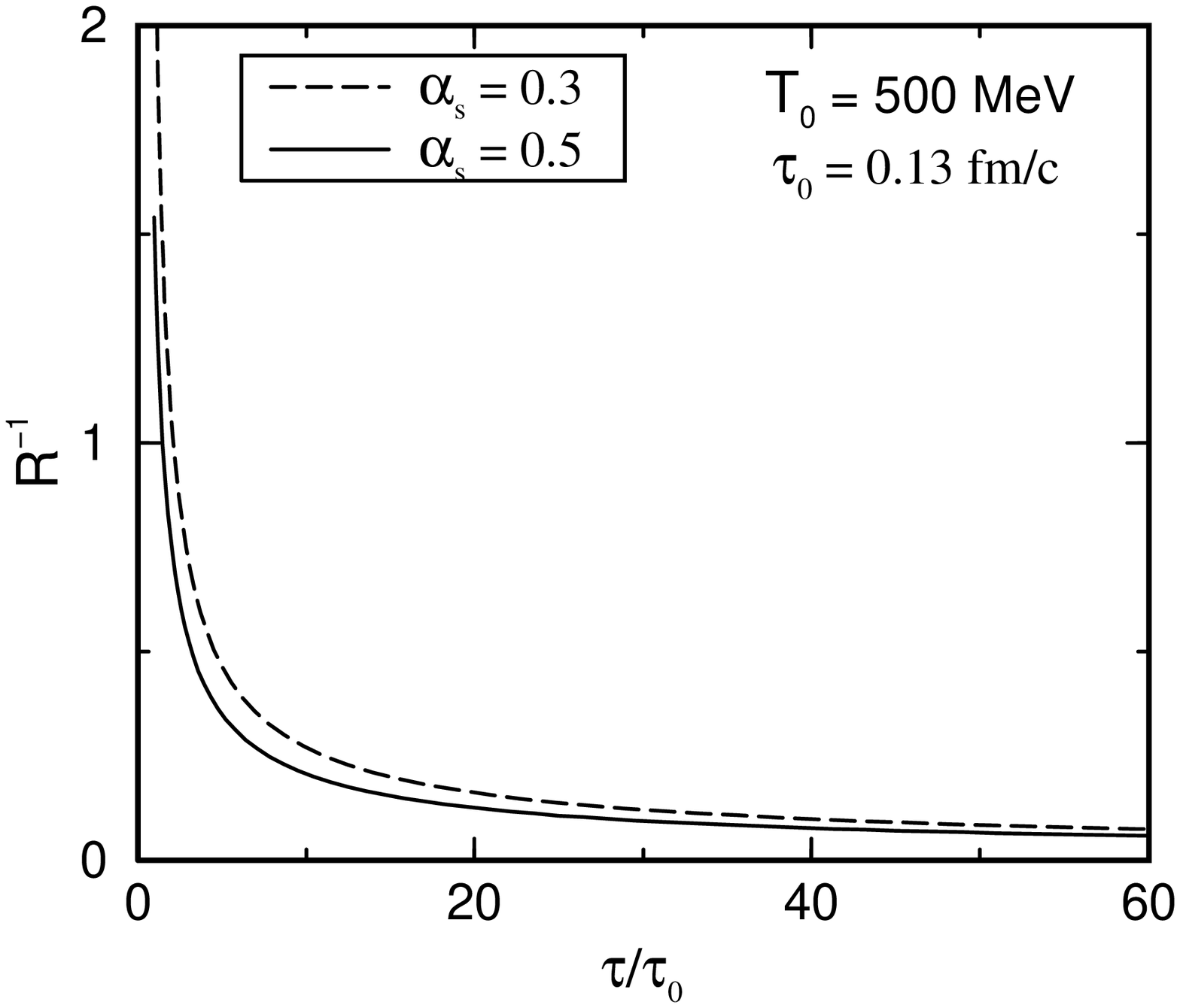,width=9cm}}
\end{minipage}
\caption{The $\tau$ dependence of temperature $T$, entropy density $s$ and
 the inverse of Reynolds number $R^{-1}$ in the scaling solution 
 with $s$ and $\eta$ are proportional to $T^3$. }
\label{fig:TsR1}
\end{figure}

In Fig. \r{fig:TsR1} the $\tau$ dependence of $T,s,$ and $R^{-1}$ with 
two typical values of $\alpha_s$ is compared with the perfect fluid case.
For both values of $\alpha_s$,  $T,s$ and $R^{-1}$ have peaks, while for the
perfect fluid they monotonically decrease and $R^{-1}$ monotonically
decreases. In both cases the initial conditions are those determined from the
uncertainty principle. This way of fixing initial conditions has a strong
restriction on the value of the initial Reynolds number. It is seen in Fig. 
\r{fig:T0t0R0} that $R_0^{-1}$ will always be greater than one. This is the
condition that gives rise to the peaks in temperature, energy density and entropy
density. In the same Fig. \r{fig:T0t0R0} we also show the region for the
condition $R_0 > 1$ in the $T_0\tau_0$--$\alpha_s$ plane for arbitrary 
initial conditions $(T_0,\,\tau_0)$ at a given $\alpha_s$.

The condition for decrease in the thermodynamical variables is
\be
R_0 \geq 1~~~~~ \mbox{or}~~~~~T_0\tau_0 \geq {1\over 3}{b\over a} \K
\ee
where the initial Reynolds number can be calculated from
\be
R = 3 {a\over b} T\tau \p
\ee
The above condition and the initial conditions $\tau_0 \approx 1/(3 T_0)$
implies $b\leq a$. It also determines
the region where one might apply the Navier-Stokes-Fourier laws.

In (1+1) dimensions the (1+1)--dimensional scaling solution implies that the
effective viscosity coefficient is only the bulk viscosity $\zeta$. This is also
the case with the (3+1)--dimensional solution in (3+1) dimensions. This is due to
the relation $\btd^\mu u^\nu = \btu^{\mu\nu}\PD_\ro u^\ro$ resulting from
scaling solution. 
The (1+1)--dimensional scaling solution in (3+1) dimensions implies that the 
effective viscosity coefficient is ${4\eta/3}+\zeta$. 
\begin{figure}[htb]
\begin{minipage}[t]{7.0cm}
\centerline{\psfig{figure=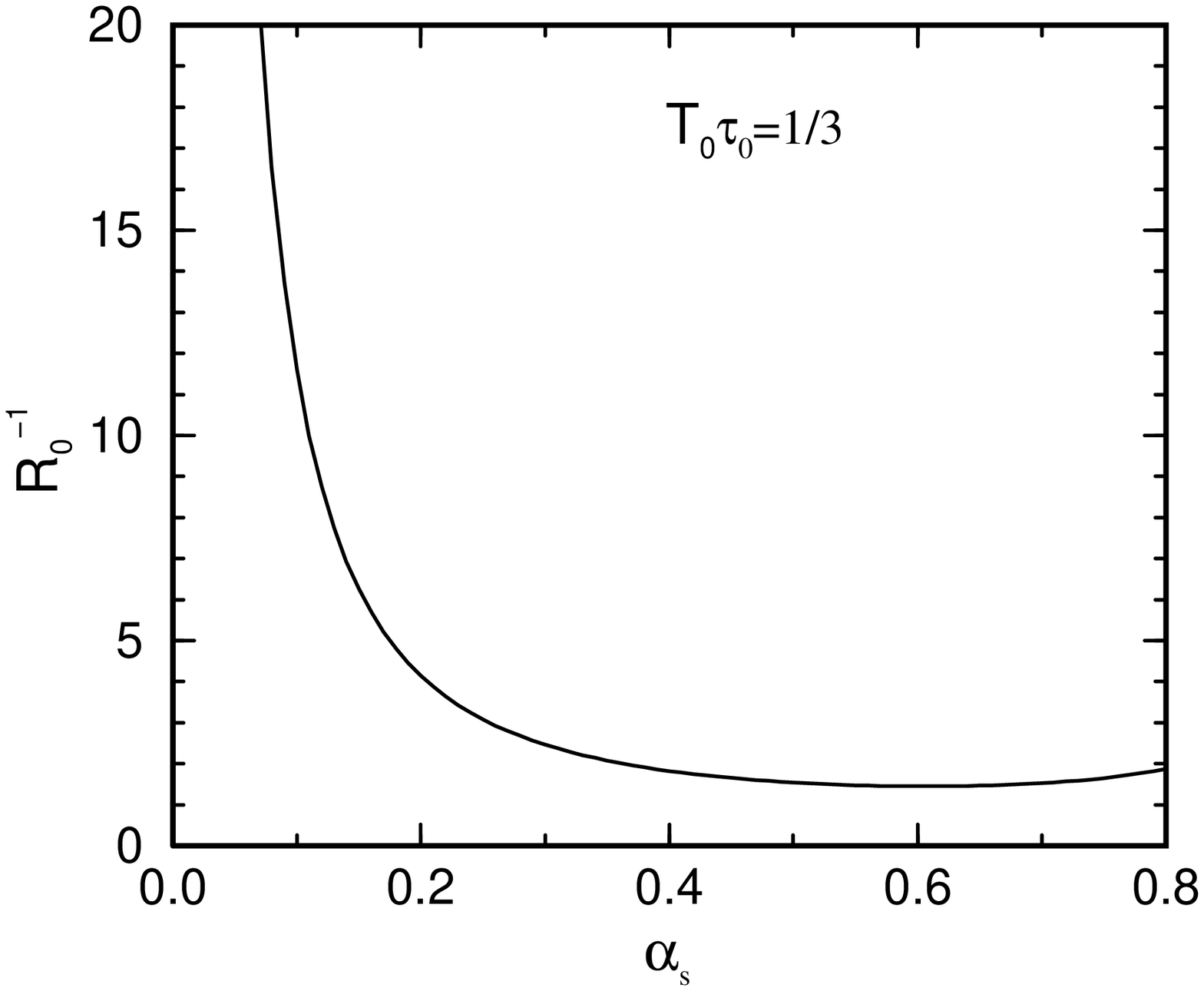,width=7.5cm}}
\end{minipage}
\hfill
\begin{minipage}[t]{7.0cm}
\centerline{\psfig{figure=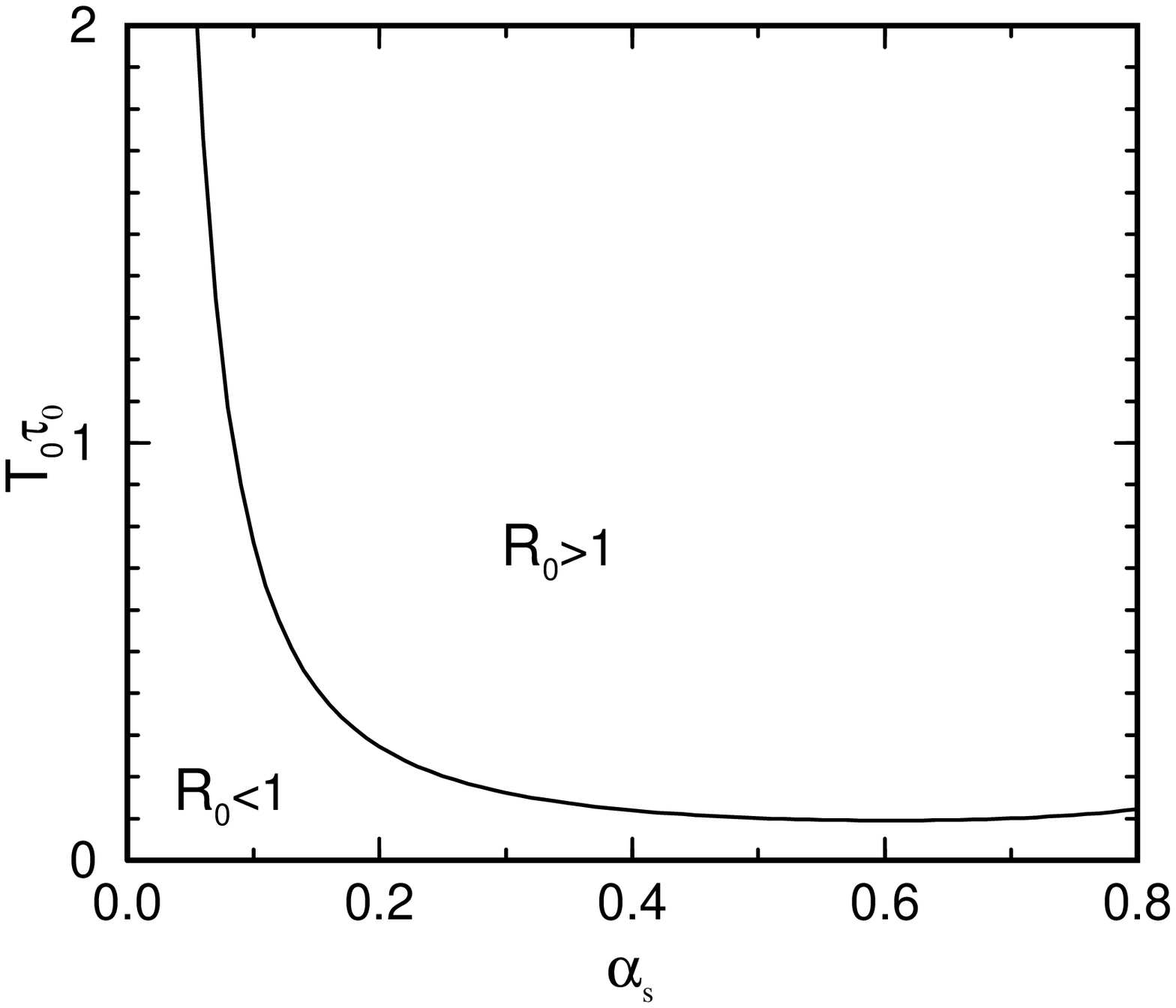,width=7.5cm}}
\end{minipage}
\caption{The $\alpha_s$ dependence of $R_0$ and the region which
satisfies $R_0\geq1$ is shown in the $\alpha_s$--$T_0\tau_0$ plane; 
where the curve is the condition $R_0=1$. }
\label{fig:T0t0R0}
\end{figure}
\begin{figure}[hp]
\begin{minipage}[t]{7.0cm}
\centerline{\psfig{figure=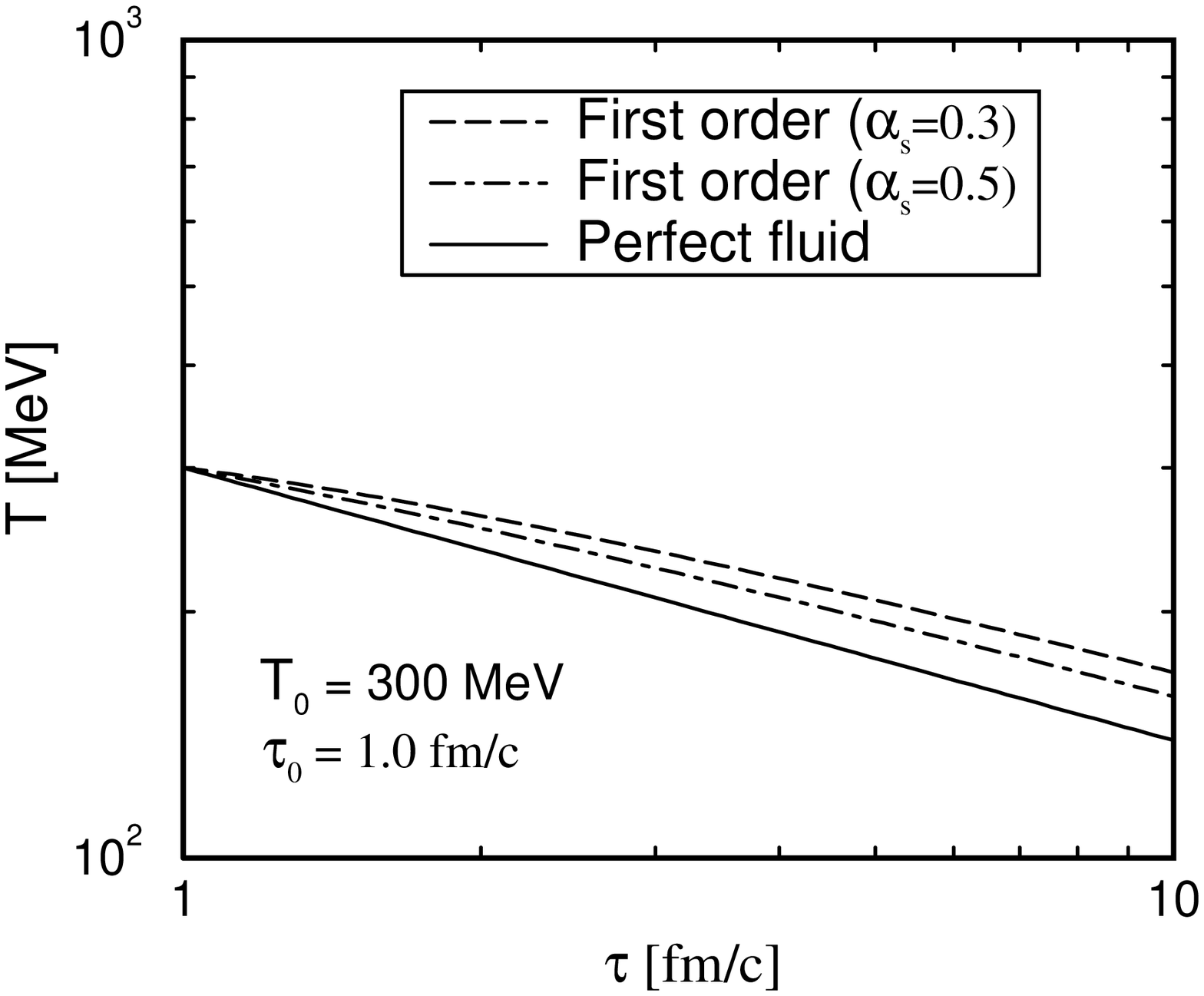,width=7.5cm}}
\end{minipage}
\hfill
\begin{minipage}[t]{7.0cm}
\centerline{\psfig{figure=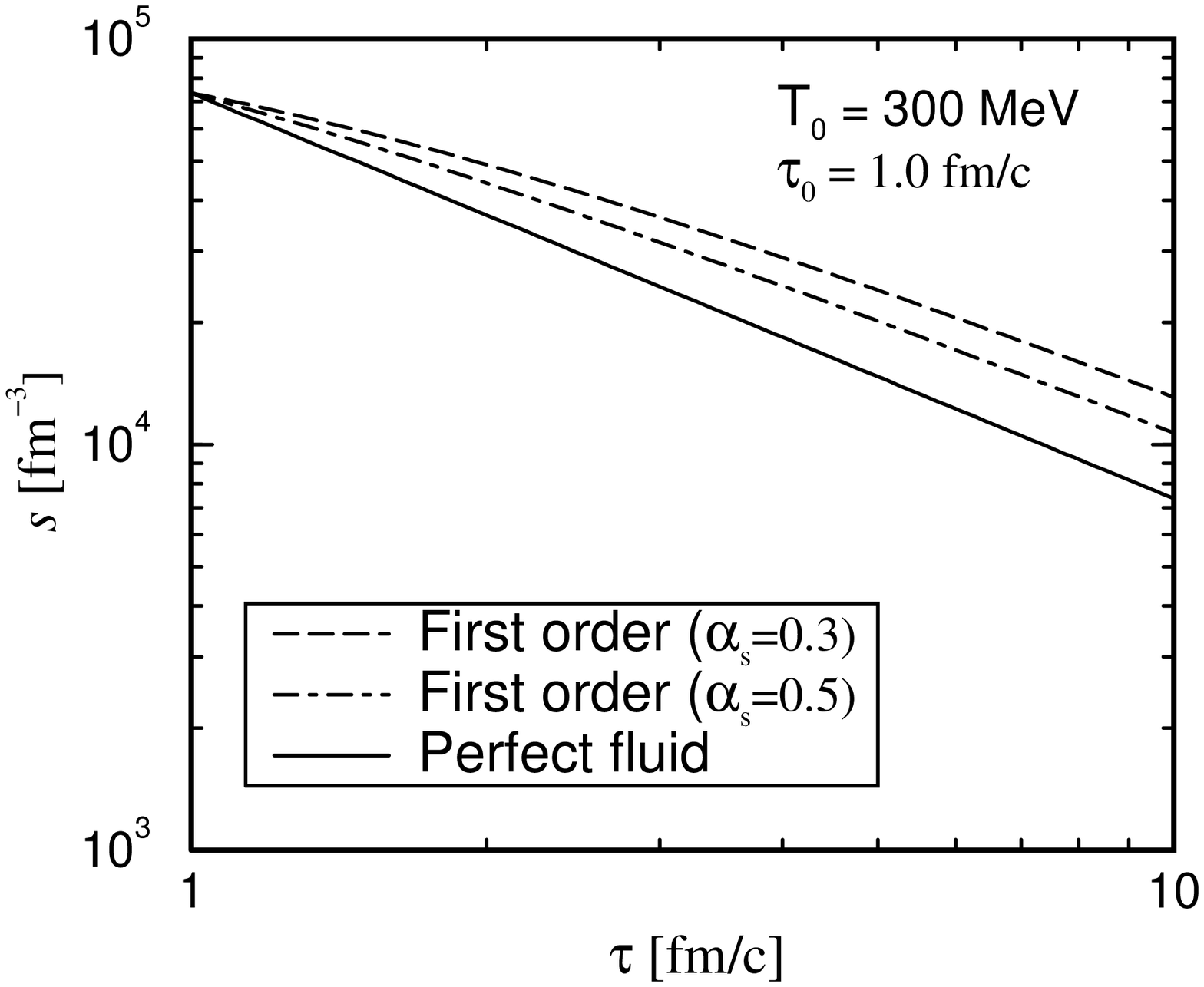,width=7.5cm}}
\end{minipage}

\vspace{0.3cm}

\begin{minipage}[b]{7.0cm}
\centerline{\psfig{figure=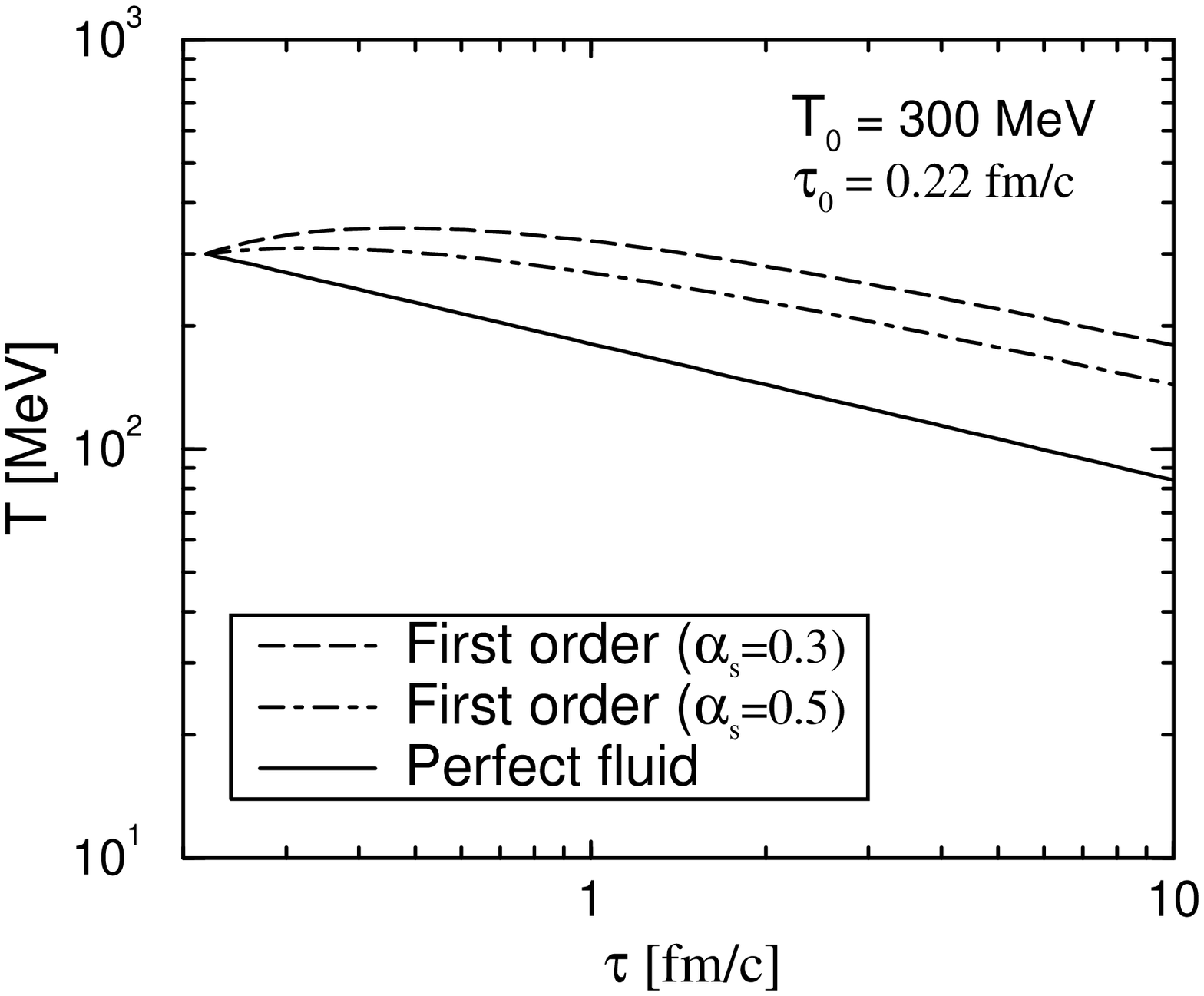,width=7.5cm}}
\end{minipage}
\hfill
\begin{minipage}[b]{7.0cm}
\centerline{\psfig{figure=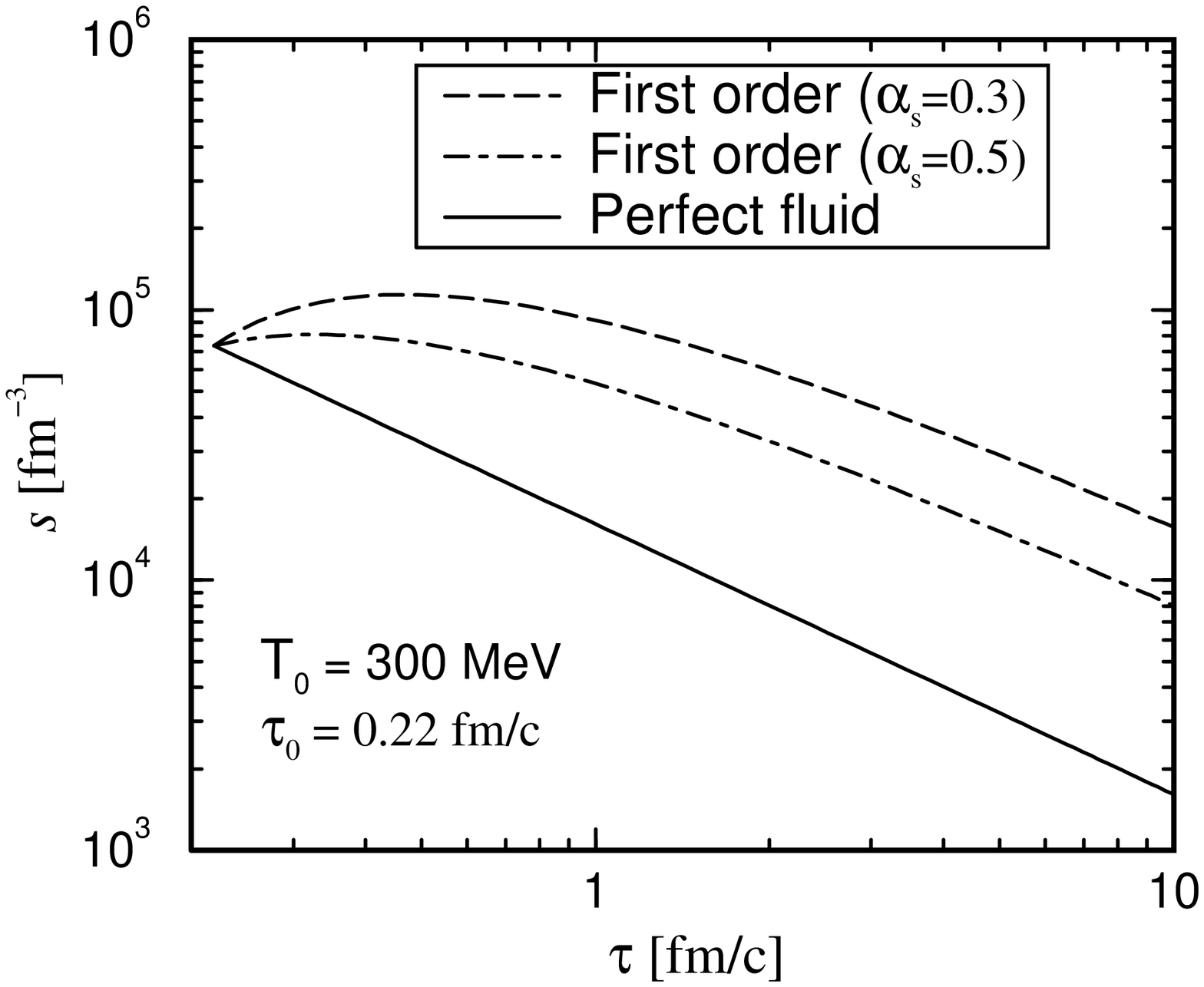,width=7.5cm}}
\end{minipage}
\caption{The $\tau$ dependence of temperature $T$ and entropy density $s$ for
arbitrary initial conditions and for initial conditions from uncertainty
principle.}
\label{fig:Tstau1}
\end{figure}

In Fig. \r{fig:Tstau1} we show how initial conditions affect the overall
space--time evolution. In the top panel we show the results for arbitrarily
chosen initial conditions. For this set of initial conditions dissipative
effects are small. On the other hand the bottom panel with the initial
conditions from uncertainty principle indicates that dissipative effects are
significant.

\subsection{Causal Fluid Dynamics}

We have seen in the previous section the importance of Reynolds' number in
determining the effects of dissipation. One of the mathematical advantages 
of the previous section is the direct connection between the Reynolds number 
and the initial conditions $(T_0,\,\tau_0)$. This is because the 
first order theory does not 
have well--defined initial conditions for the dissipative fluxes, and the 
latter are related to the thermodynamic forces by linear algebraic expressions.

In this section we determine the dissipative fluxes from the transport
equations. In the Landau--Lifshitz frame the transport equations are still
given by (\r{eq:Pieq})--(\r{eq:pieq}) but with slightly different heat coupling
coefficients in the bulk and shear viscous pressure equations. Under the scaling
solution assumption those coupling terms do not contribute to the dynamics of
the system.

The $(1+1)$--scaling solution in $(3+1)$--dimensions yields the following
structure of transport equations 
\bea
{\PD \Pi \over \PD \tau} &=& -{\Pi\over \tau_\Pi} -\half {1\over \beta_0} \Pi
\left[\beta_0 {1\over \tau} + T {\PD \over \PD \tau}\left({\beta_0\over T}\right)\right]
\nonumber \\
&&-{1\over \beta_0} {1\over\tau} \K \label{eq:bulkeqnc}\\
{\PD q^\mu\over \PD \tau} &=& -{q^\mu\over \tau_q} +\half {1\over \beta_1} q^\mu
\left[\beta_1 {1\over \tau} + T {\PD\over \PD \tau} \left({\beta_1\over T}
\right)\right] \K\label{eq:heateqnc}\\
{\PD \pi^{\mu\nu}\over \PD \tau} &=& -{\pi^{\mu\nu} \over \tau_\pi} - \half
{1\over \beta_2} \pi^{\mu\nu} \left[\beta_2 {1\over \tau} +T {\PD\over \PD
\tau}\left({\beta_2\over T}\right)\right] \nonumber\\
&&+{1\over \beta_2} \left[\tilde{\btu}^{\mu\nu} - {1\over 3} \btu^{\mu\nu}
\right]{1\over \tau} \p\label{eq:sheareqnc}
\eea
In the last of the above equations $\tilde{\btu}^{\mu\nu}
=\btu^{\mu\nu}$ for $0\leq\mu,\nu\leq 1$ and zero otherwise 
(because of only one non-vanishing spatial component of the
4--velocity).

For the (1+1)-- dimensional Bjorken similarity fluid flow in
(3+1) dimensions the energy equation (\ref{eq:enercons}) becomes
\begin{equation}
\frac{d\varepsilon}{d\tau}+\frac{\varepsilon+p}{\tau}
-\frac{1}{\tau} \Phi +\Pi \frac{1}{\tau} = 0 \enspace,
\label{eq:scalingenergy}
\end{equation} 
where $\Phi \equiv \pi^{00}-\pi^{zz}$ is determined from the shear viscous 
tensor evolution equation (\ref{eq:sheareqnc})
\begin{eqnarray}
\frac{d}{d\tau}\Phi &=&
-\frac{1}{\tau_\pi}\Phi - \frac{1}{2}\Phi\left(\frac{1}{\tau} +\frac{1}{\beta_2}
T \frac{d}{d\tau}\left(\frac{\beta_2}{T}\right)\right)
+\frac{2}{3}\frac{1}{\beta_2}\frac{1}{\tau} \enspace. \label{eq:scalingshear}
\end{eqnarray}
The equation for $\Pi$ will not be needed because $\zeta=0$.

In this subsection we will distinguish the perfect fluid, first order, and
second order theories by the quantity $\Phi$:
\bea
\Phi &\equiv& 0 ~~~~~~~~~~~~~~~ \mbox{perfect fluid} \enspace,\\
\Phi &=& \frac{4}{3} \eta/\tau ~~~~~~~~~~~~~~~ \mbox{first order theory}
\enspace,\\
\frac{d\,\Phi}{d\,\tau} &=& -{\Phi \over \tau_\pi} -{1\over 2} \Phi
\left({1\over \tau}  +{1\over \beta_2} T {d \over d \tau}\left({\beta_2\over
T}\right)\right) +{2\over 3} {1\over \beta_2}{1\over \tau} \\
&&~~~~~~~~~~\mbox{second order theory} \nonumber\enspace,
\eea
The Reynolds number is generally given by
\be
R = {(\eps+p)\over \Phi} ~~~~~\mbox{or}~~~~~{T s\over \Phi} \p
\ee
R goes to infinity in the ideal fluid approximation.

Let us first discuss the initial condition for $\Phi$. For an ideal fluid $\Phi$
vanishes since there are no dissipative fluxes. For the first order theories the
initial condition for $\Phi$ is not well--defined and is given by the initial
conditions $(T_0,\,\tau_0)$. For the second order theories we have
well--defined initial condition for $\Phi$ since the dissipative fluxes are
found from their evolution equations. 

In deriving the transport equations it is assumed that
the dissipative fluxes are small compared to the primary variables
$(\eps,\,n,\,p)$. For shear flux we require that
\be
\left[\pi^{\mu\nu}\pi_{\mu\nu}\right]^{1/2} =\sqrt{{3\over 2}} \Phi \ll p \p
\ee
In terms of $\Phi$ this condition can be written as
\be
\Phi \ll \sqrt{{2\over 3}} p \p
\ee
Another important quantity in determining the importance of viscous effects is
the ratio of the macroscopic time scale to the microscopic time scale
\be
R_\tau = {\tau \over \tau_\pi(\tau)} \K\l{eq:rt1}
\ee
which in our case is given by
\be
R_\tau = {2\over 3} {a\over b} T \tau \p\l{eq:rt2}
\ee

In first order theories the question of how much a particular dissipative flux
is generated as a response to corresponding thermodynamic/kinematic forces in
nuclear collisions is  governed by the primary initial conditions
$(T_0,\,\tau_0)$. That is, one just reads off the value of $\Phi_0$ from the
the linear algebraic expression for $\Phi$. We have seen that for values of the
Reynolds number less than one, the thermodynamic quantities increase with time.
This might be signalling the instability of the solution. Alternatively this
might imply that we are using the first order theories beyond their domain of
validity .  The primary initial conditions can in principle be extracted from
experiments. These in turn will give us the value of $\Phi_0$.  This value of
$\Phi_0$ will eventually determine how the thermodynamic variables evolve with
time. This is clearly understood by looking at the ratio of the pressure term
to viscous term, namely, $R$ as already discussed in the previous section.

In the second order theories the question of how much a particular
dissipative flux is generated as a response to corresponding
thermodynamic/kinematic forces in the early stages of nuclear reactions is not
trivial but interesting. In order to find the viscous contribution to the time
evolution of thermodynamic quantities we need to solve the differential
equation for $\Phi$. Therefore one has to determine the initial conditions for
$\Phi$. Although we don't know the exact form of the initial value for $\Phi$
we will discuss the limiting cases. The first and most important limiting case
is based on the assumption made when deriving the second order theory transport
equations, namely, that the dissipative fluxes be small compared to the primary
variables. For the shear viscous flux this means that the shear viscous
stress--tensor must be small compared to the pressure. The value of $\Phi$ will
always be less than $p$, hence the initial value $\Phi_0$ will always be less
than $p_0$. This has an interesting consequence: the initial Reynolds number is
always greater than one. Thus in second order theory under these conditions
there will be no increase (and hence no peaks) of thermodynamic variables with increasing time.
However in general the thermodynamic quantities will decrease with time for as
long as the condition 
\be
\Phi \leq \eps+p \K
\ee 
is satisfied, which in the present case implies that $\Phi_0 \leq 4 p_0$.  
However, values of $\Phi$  greater than the pressure $p$ leads to unphysical
negative effective enthalpy. Unlike in the first order theories where it is not
always possible to address this problem of negative effective enthalpy, in the
second order theories we are guided by the limitations which are embedded in the
valid application regimes of the theories, namely, the condition
$(\pi^{\mu\nu}\pi_{\mu\nu})^{1/2} \ll p$. This condition guarantees that there
effective enthalpy is always positive.

There are other two ways of determining the initial conditions for $\Phi$. The
first one is by using the existing microscopic models such as VNI
\cite{Geiger95},  (HIJING) \cite{Wang97} and  (UrQMD) \cite{Bass98} to extract 
the various components of $\pi^{\mu\nu}$ from $T^{\mu\nu}$.  Since we are dealing
here with a partonic gas VNI seems to be a good choice for the present work. We
will use the results from the improved version of VNI  \cite{BMS02} to fit our
calculation in order to extract the initial value for  $\Phi$.  Another way of
determining the initial value for $\Phi$ is to extract the initial value of the
Reynolds number experimentally.  Two of the most experimentally accessible
quantities are the multiplicity per unit rapidity $\d N/\d y$ and the
transverse energy per unit rapidity $\d E_T/\d y$.   A detailed study for the
initial and boundary conditions for dissipative fluxes is needed to fully
incorporate these fluxes into the dynamical equations for the thermodynamic
quantities.

In this section we focus mainly on the results of second order theories and
compare them to the first order theories and the perfect fluid 
results. As it is the main focus of this work we will look at the dynamics 
of the quark--gluon plasma or partonic gas.

The energy equation
(\ref{eq:scalingenergy}) and the viscous stress
equation (\ref{eq:scalingshear}) can be written as 
\begin{eqnarray}
\frac{d}{d\tau}T &=& -\frac{T}{3\,\tau} + \frac{T^{-3}\Phi}{12\,a\,\tau}
\enspace,\label{eq:temp}\\
\frac{d}{d\tau} \Phi &=& - \frac{2\,a\,T\Phi}{3\,b} -
\frac{1}{2}\Phi\left(\frac{1}{\tau}-5 \frac{1}{T}\frac{d}{d\tau}T\right) +\frac{8\,a\,T^4}{9\,\tau}
\enspace. \label{eq:shear}
\end{eqnarray}
For a perfect fluid and a first order theory the energy equation
(\ref{eq:temp})
can be solved analytically to give (\r{eq:0Tsol}) and (\r{eq:1Tsol})
%\begin{eqnarray}
%T(\tau) &=& T_0\left[\frac{\tau_0}{\tau}\right]^{1/3}
%\label{eq:perfect} ~~~~~\mbox{perfect fluid}\enspace,\\
%T(\tau) &=& T_0\left[\frac{\tau_0}{\tau}\right]^{1/3} \left\{1+\frac{b}{6\, a}
%\frac{1}{T_0\,\tau_0}
%\left(1-\left[\frac{\tau_0}{\tau}\right]^{2/3}\right)\right\}
%\label{eq:1st}\\
%&&\mbox{first-order theory} \enspace. \nonumber
%\end{eqnarray}
In the first order theory we do not have any relaxation coefficients. Then
equation (\ref{eq:scalingshear}) gives $\Phi = (4 \eta/3)/\tau$. Eqs.
(\r{eq:0Tsol}), (\r{eq:1Tsol}) and the numerical solution to the second order
equations (\ref{eq:temp}) and (\ref{eq:shear}) will be used to study the proper
time evolution of temperature. The other thermodynamic quantities, namely,
energy density and entropy density, are  related to the  temperature by the
equation of state. It is important to show the entropy results due to the
importance of entropy in the theory of irreversible extended thermodynamics and
due to the fact that entropy is related to multiplicity.

We start by showing the dependence of the temperature evolution on the initial value
of $\Phi$. In Fig. \r{fig:Ttauphi} we show this dependence for two values of
$\alpha_s$. The dependence of first order results on $\alpha_s$ is clear from
the results presented before.  
The dependence of the second order theory on $\alpha_s$ is different. A large value of
$\alpha_s$ modifies the power dependence for $\tau\gsim 1$ fm/c to be
more or less the same as the power dependence for $\tau \lsim1$ fm/c. Thus the
effect of $\alpha_s$ are important for $\tau > 1$ fm/c as can be seen 
from the top right panel of Fig.
\r{fig:Ttauphi} where a large value of $\alpha_s$ is shown.

In studying the dependence of the results on the initial conditions for $\Phi$
we have also included some unphysical choices for illustrative purposes. For
$0.1 <\Phi_0/p_0 <1 $ the choice of $\Phi_0$ is important, but below
$\Phi_0=0.1 p_0$ the equation for $\Phi$ gives the same contribution to the
evolution of thermodynamic quantities.
\begin{figure}[hp]
\begin{minipage}[h]{7.0cm}
\centerline{\psfig{figure=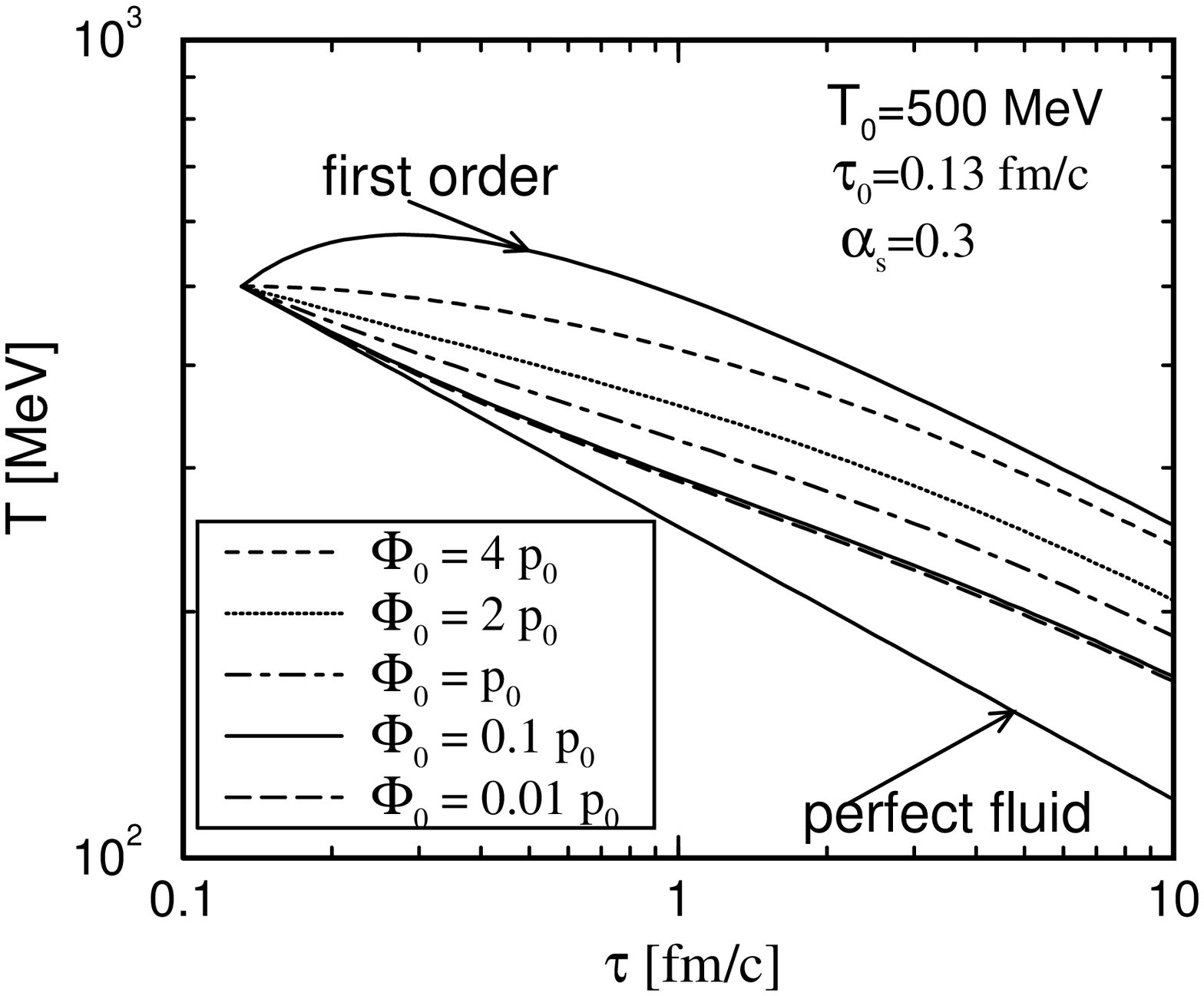,width=7.5cm}}
\end{minipage}
\hfill
\begin{minipage}[h]{7.0cm}
\centerline{\psfig{figure=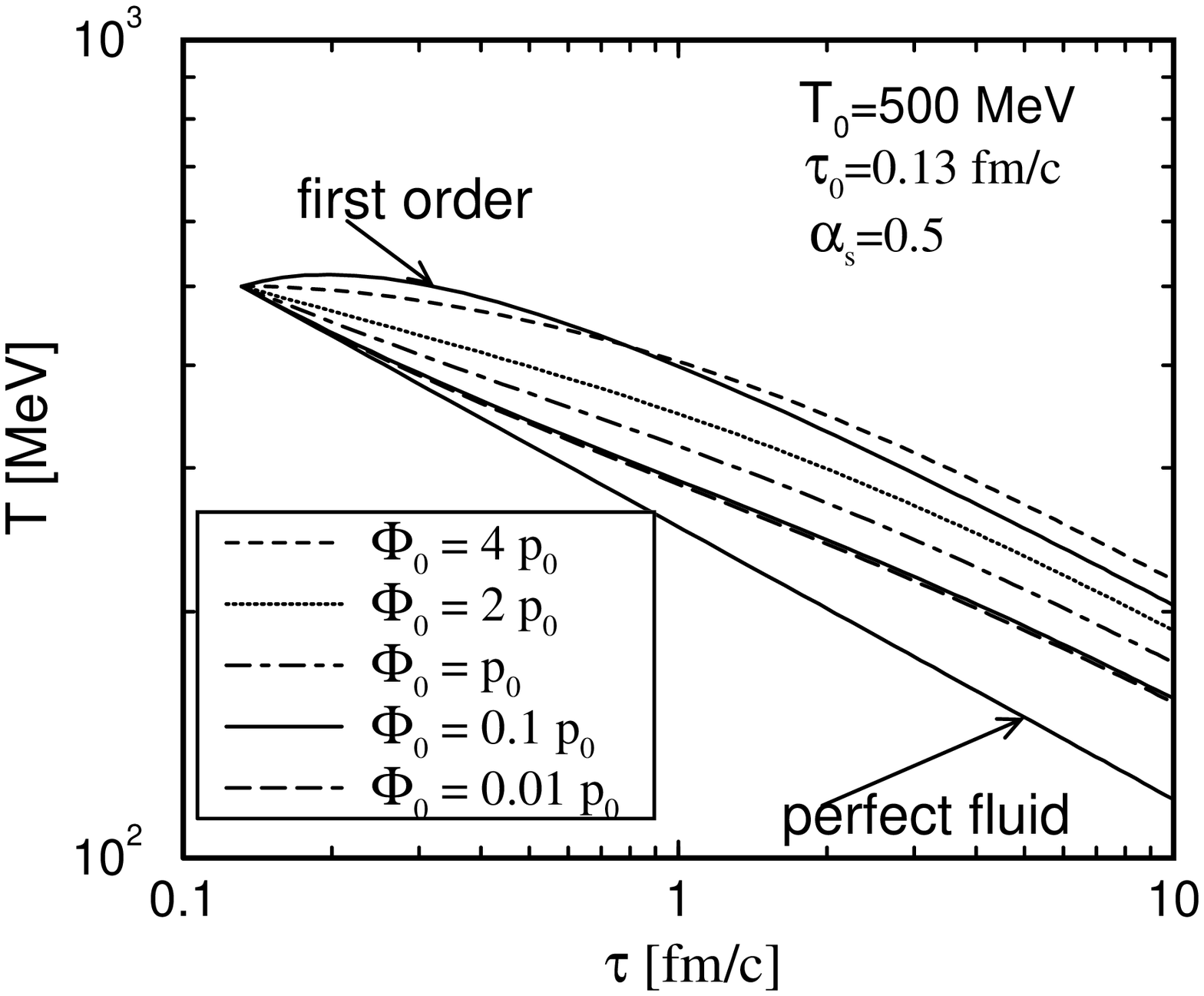,width=7.5cm}}
\end{minipage}

\vspace{0.3cm}
\centering
\begin{minipage}[b]{7.0cm}
\centerline{\psfig{figure=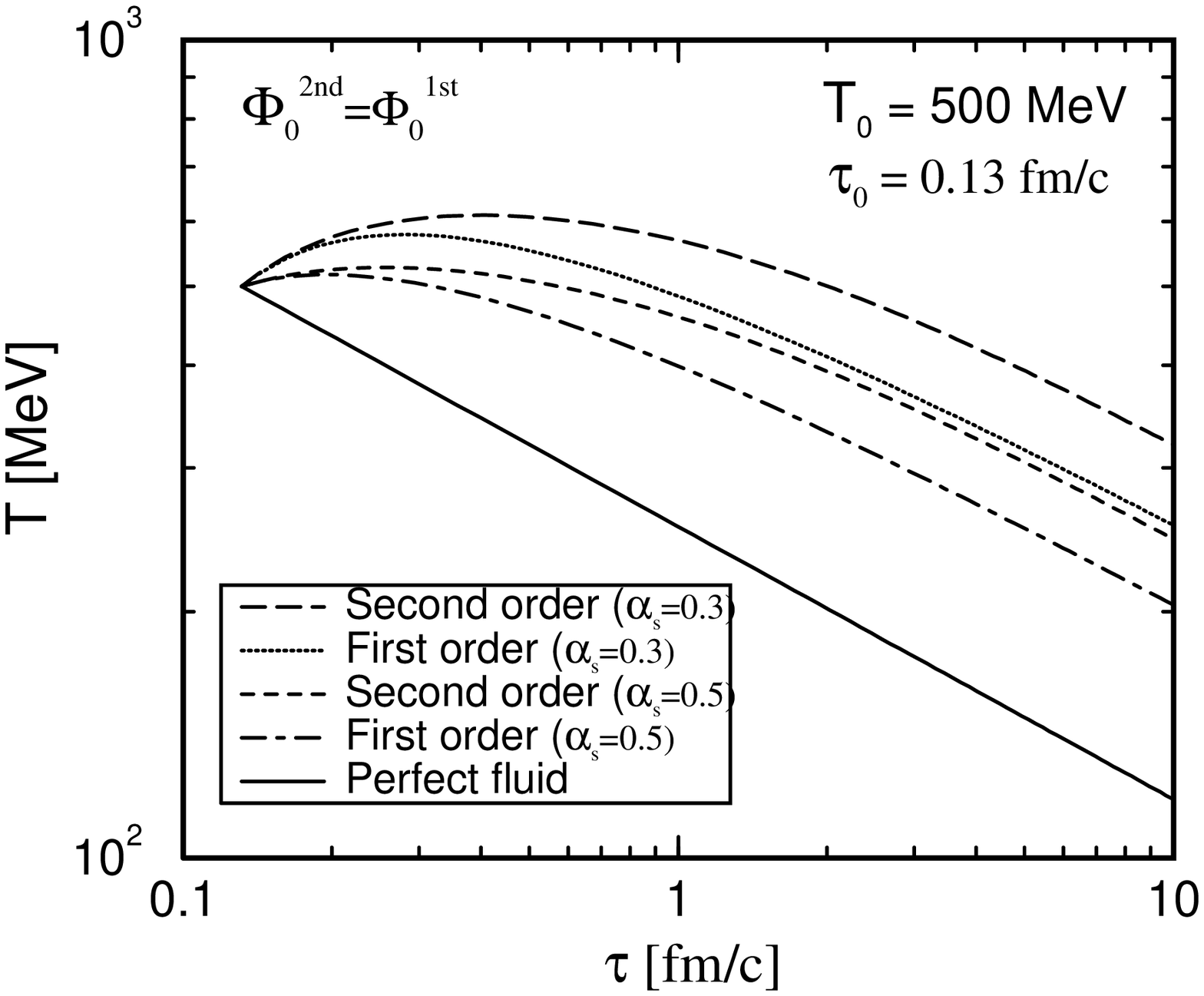,width=7.5cm}}
\end{minipage}
\caption{The time evolution of temperature for different choices of $\Phi_0$ 
for given primary initial conditions $(T_0,\,\tau_0)$, and two different values 
of $\alpha_s$. The curves for different values of $\Phi_0$ are from the second 
order theory. The bottom panel is for $\Phi_0^{2nd}=\Phi_0^{1st}$.}
\label{fig:Ttauphi}
\end{figure}
It is also tempting to choose the initial conditions for the second order
theory to match what the first order theory predicts to be the initial value of
$\Phi$. Note the order of the curves on the bottom panel of Fig.
\r{fig:Ttauphi} . The second  order theory predicts larger deviations than the
first order theory. This should be exactly the same picture  if both theories
are synchronized in a regime where both are valid. Unfortunately, it is not
trivial to make the reverse match of initial conditions. This situation will
arise in natural way when both first order theory and second order theories are
applied in the situation where they are both valid, as we will see later.

Under physical initial conditions the second order theory gives a Reynolds
number that is always greater than one. This can be seen from Fig. 
\r{fig:Re2nd} where for illustrative purposes we also include curves for
unphysical initial conditions for $\Phi$. Note that $\Phi_0 = 4 p_0$ is the
maximum value before the solutions becomes
unstable. This is a critical value that gives a 
Reynolds number $R_0 =1$. As expected the first order theory gives $R<1$ at the same
time. Throughout this work, unless otherwise stated so, we use the primary
initial conditions based on the uncertainty principle as already discussed and as
we will briefly describe below. Under this prescription of primary initial
conditions, which might be relevant for RHIC and LHC, the first order theories 
are not suitable in describing the dynamics of thermodynamic quantities. 
On the other hand, the
second order theories are suitable in describing the physical process happening
at earlier times. An advantage of the extended
irreversible thermodynamics, or second order theories, are their ability to be
applicable over a wide range of regimes. However, for a different choice of
initial conditions both theories might yield similar results, as we shall see. 

\begin{figure}[htb]
\centering
\psfig{figure=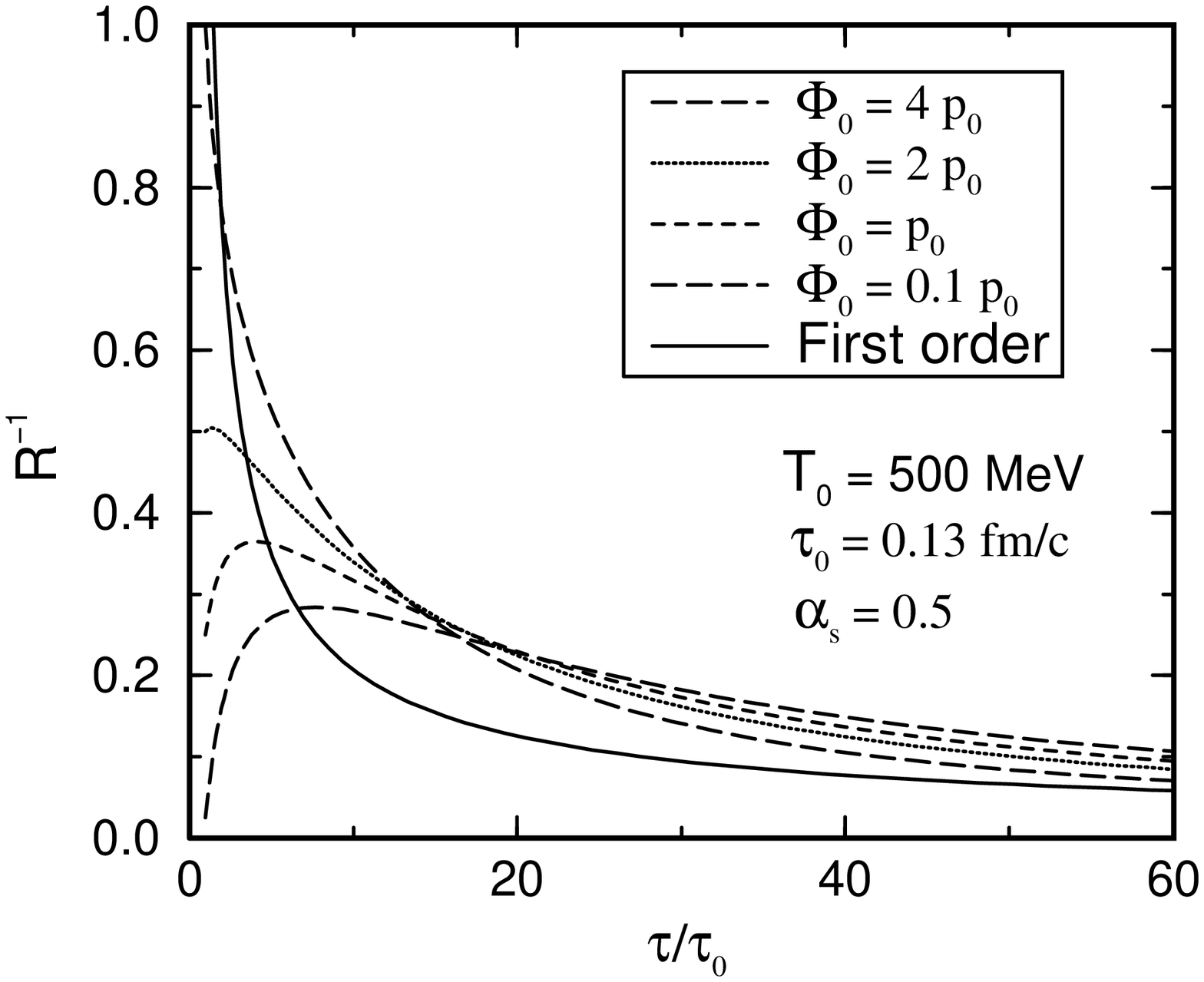,width=3.5in,height=3.5in}
\caption{Proper time evolution of the inverse Reynolds number for different
values of $\Phi_0$ for the given primary initial conditions.}
\label{fig:Re2nd}
\end{figure}

In what follows we will try to get close to the conditions that are realized in
the laboratory. We will consider scenarios close to those at RHIC 
and LHC. We will use the most common primary initial conditions as discussed below.
But first, we have to estimate the initial value of $\Phi$ for these two
scenarios. We will use the recent results from VNI calculations for the proper 
time evolution \cite{BMS02}. We will make a fit to the data points and extract 
the initial value of $\Phi$.

\begin{figure}[htb]
\begin{minipage}[h]{7.0cm}
\centerline{\psfig{figure=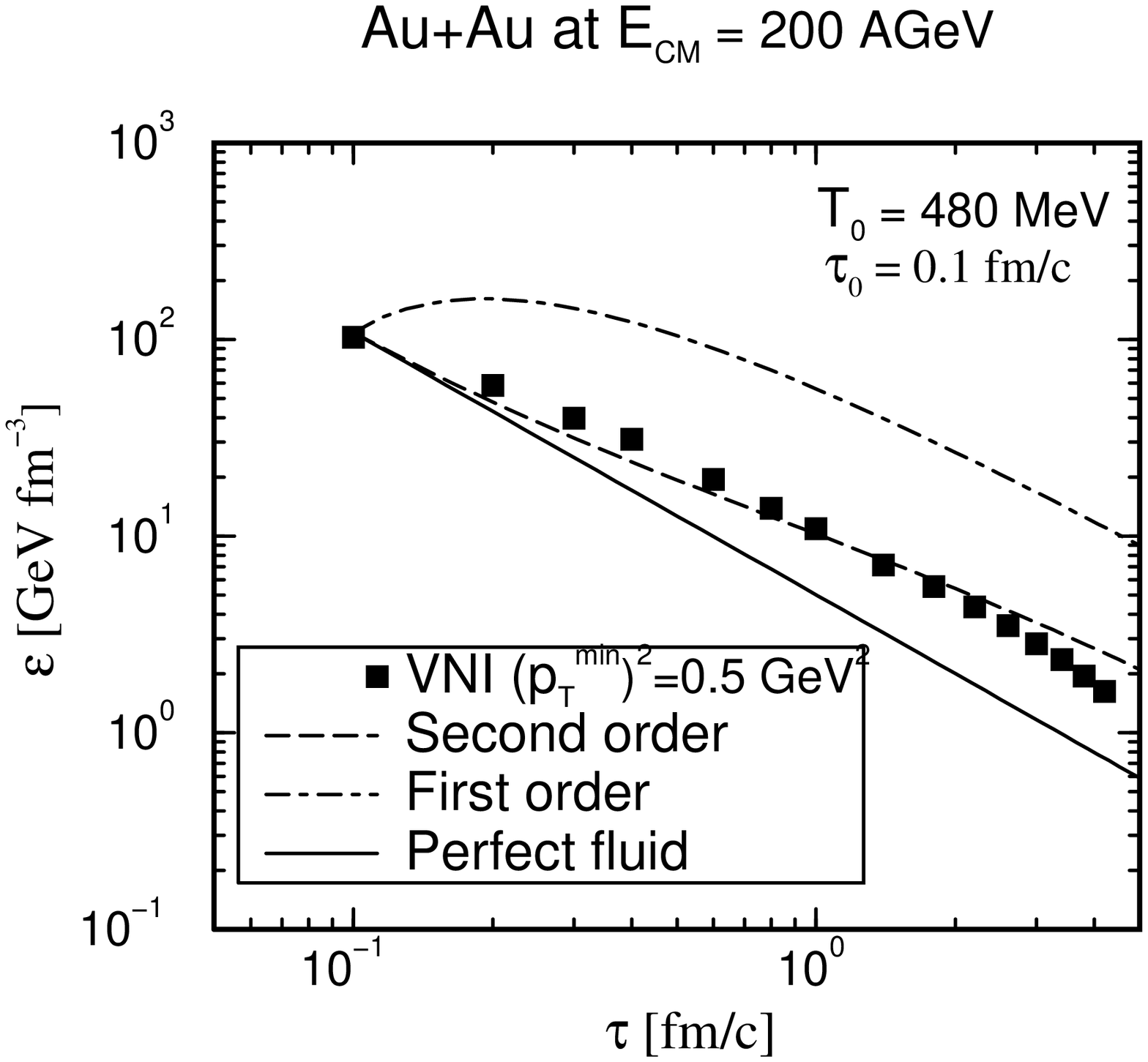,width=7.5cm}}
\end{minipage}
\hfill
\begin{minipage}[h]{7.0cm}
\centerline{\psfig{figure=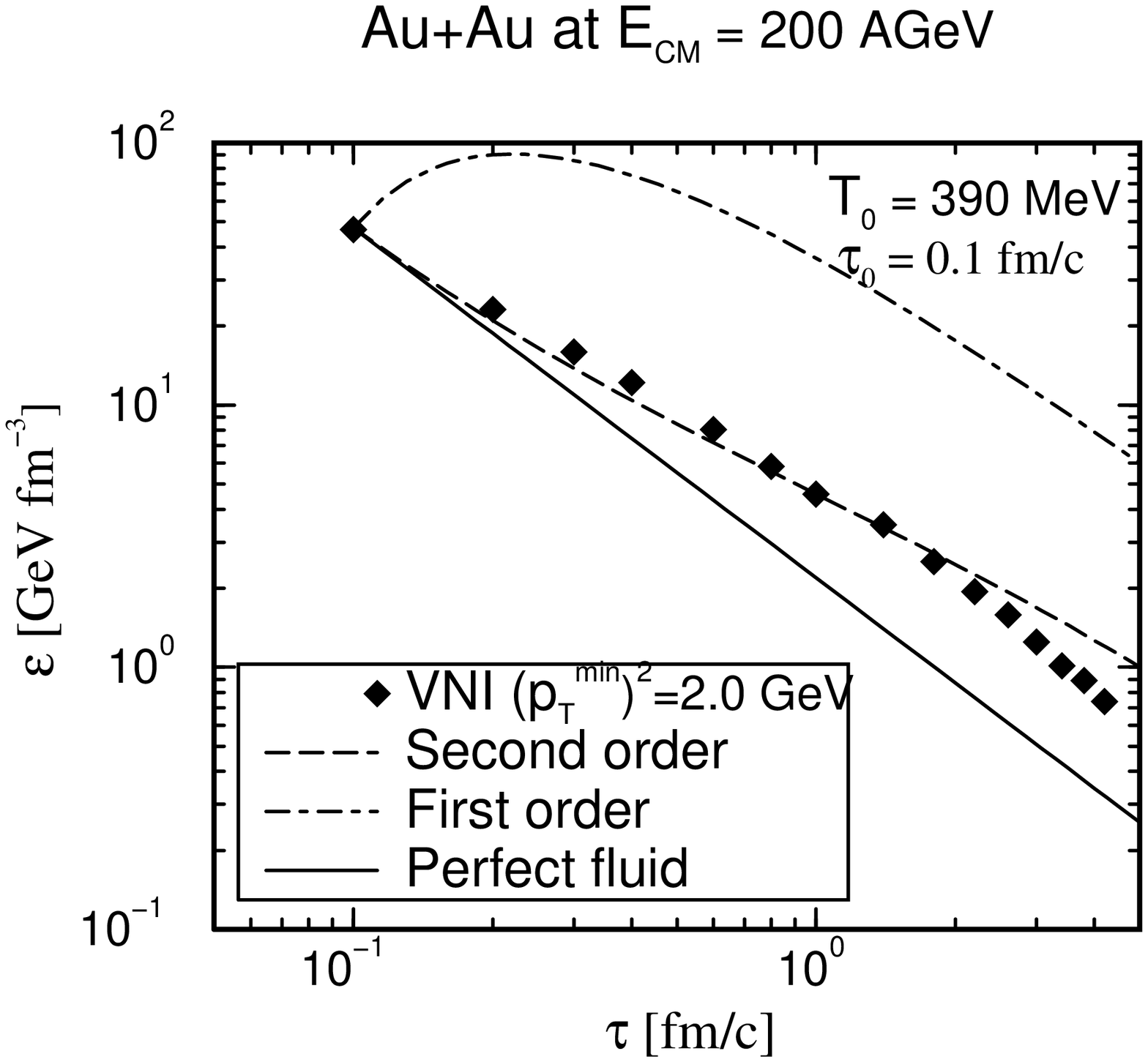,width=7.5cm}}
\end{minipage}
\caption{The proper time evolution of energy density. The data points are
results from VNI simulations and the curves are fits.}
\label{fig:VNIeps}
\end{figure}

Even though the motive is to extract an initial condition for $\Phi$, there is
something interesting in Fig. \r{fig:VNIeps}. In this figure a  comparison
between the perfect fluid approximation, the first order theory, and the second
order theory is clear. The kinetic theory result, of course,  differs
significantly from the perfect fluid dynamics result. The first order  theory
obviously fails terribly. The essential point, however, is that the  second
order theory is in good agreement with the VNI results. Due to the preliminary
nature of VNI results we cannot yet claim perfect agreement between the two
approaches. However, the fact that both have similar power laws is striking. In
the beginning it looks like $\tau^{-1}$ and then   later on $\tau^{-4/3}$ for
the VNI results. One expects that when the full three dimensional problem is
studied within the fluid dynamical approach we  might have even better
agreement.   The fitted value of $\Phi_0$ is found to be about $0.2 p_0$ which
is, of course, a physical value. The value of $\alpha_s$ used is about $0.5$.
For all RHIC results presented here we will use the expected primary initial
conditions with  $\Phi=0.2 p_0$ and $\alpha_s=0.5$. For the LHC scenario we
will use  the expected primary initial conditions with $\Phi_0=0.3 p_0$ and 
$\alpha_s=0.4$. 

As a benchmark both the fluid dynamical and cascade models have been solved
numerically for same primary initial conditions and equation of state. This is
done for consistency.  It is apparent that hyperbolic models performs better
than the parabolic ones, in agreement with VNIBMS simulations. Also for energy
density there is a peak in the parabolic model which is absent in hyperbolic
model. This spurious unphysical result highlights the difference between the
parabolic and hyperbolic model in region of large gradients. We remark that the
initial state under consideration presents very steep velocity gradients.
Therefore this is an ideal benchmark for testing fluid dynamical models against
transport models. Comparisons of Navier--Stokes--Fourier results with transport
models were made in \cite{Gyulassy} with NSF failing terribly for smaller cross
sections. In that particular study the NSF also brought in the problem of
negative effective pressure. The transport results however gave a much better
description. What is important however is that the second order theory seems to
do a better job even in this case. The latest results on this latter point to
be published elsewhere are still under investigation and comparison to previous
work on the effective pressure of a saturated Gluon plasma \cite{Dumitru00} is
done.

\begin{figure}[htb]
\begin{minipage}[t]{7.0cm}
\centerline{\psfig{figure=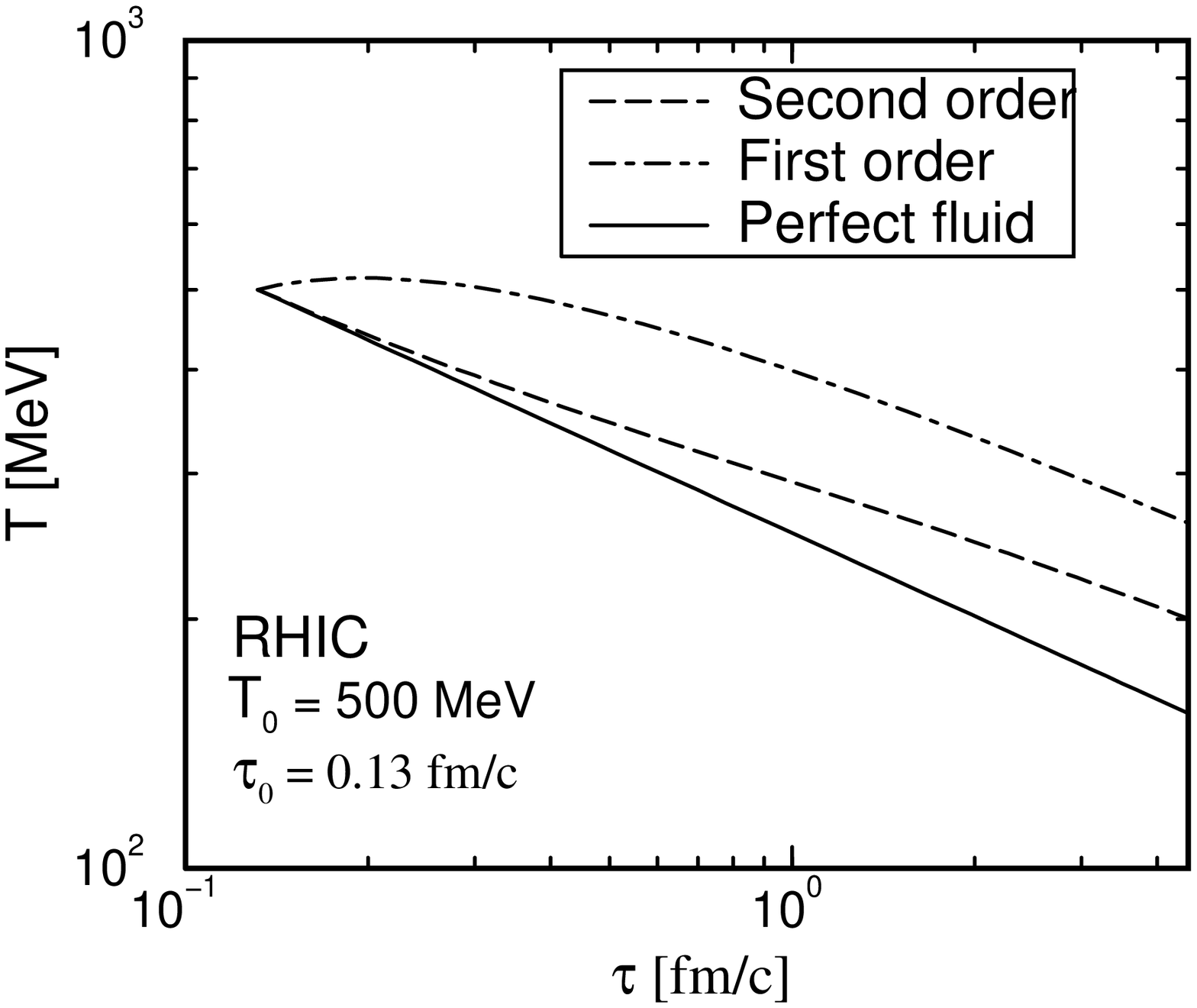,width=7.5cm}}
\end{minipage}
\hfill
\begin{minipage}[t]{7.0cm}
\centerline{\psfig{figure=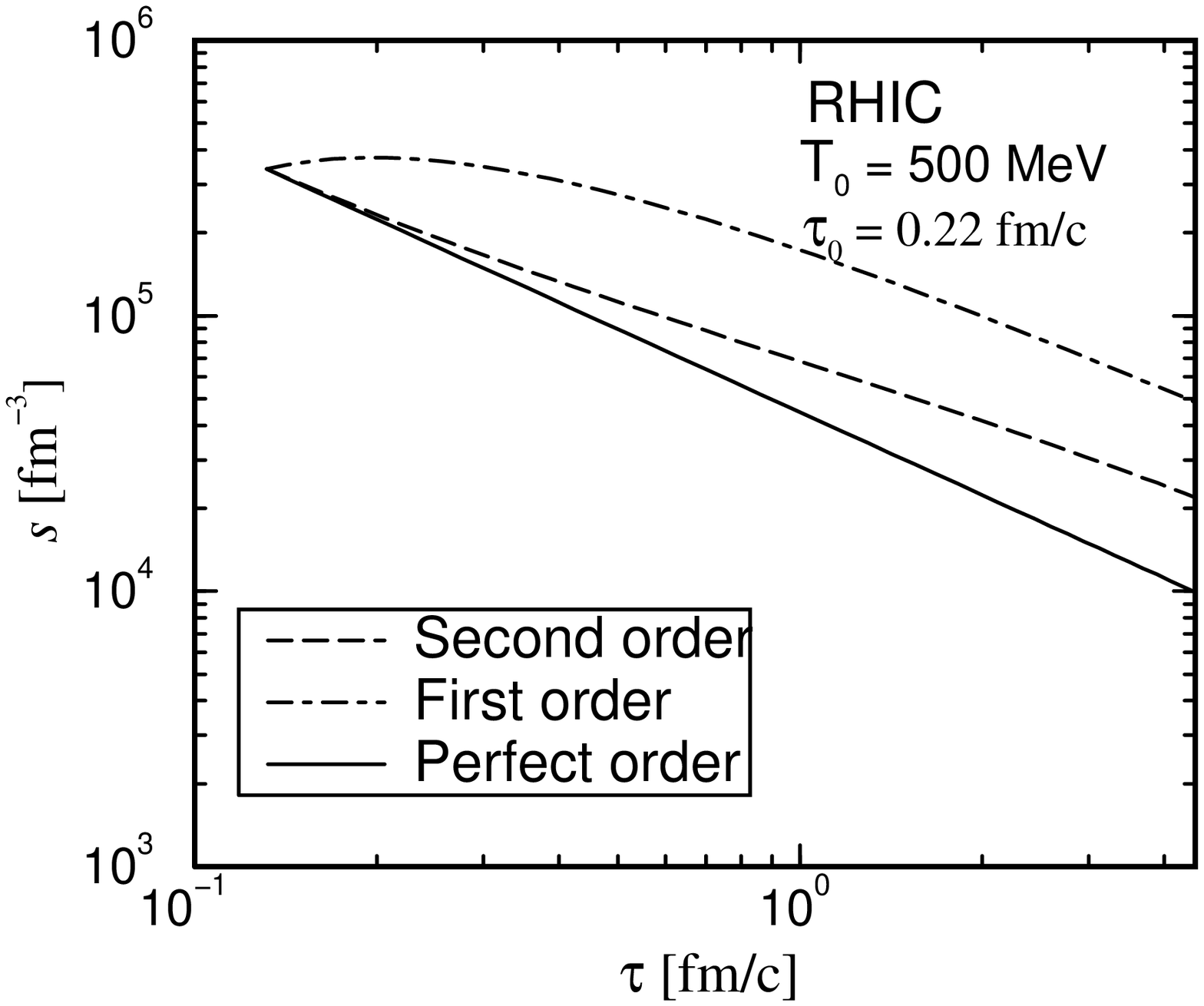,width=7.5cm}}
\end{minipage}
\caption{The proper time evolution of temperature and entropy for given
primary initial conditions $(T_0,\,\tau_0)$ for a RHIC scenario.}
\label{fig:TstauRHIC}
\end{figure}

\begin{figure}[htb]
\begin{minipage}[b]{7.0cm}
\centerline{\psfig{figure=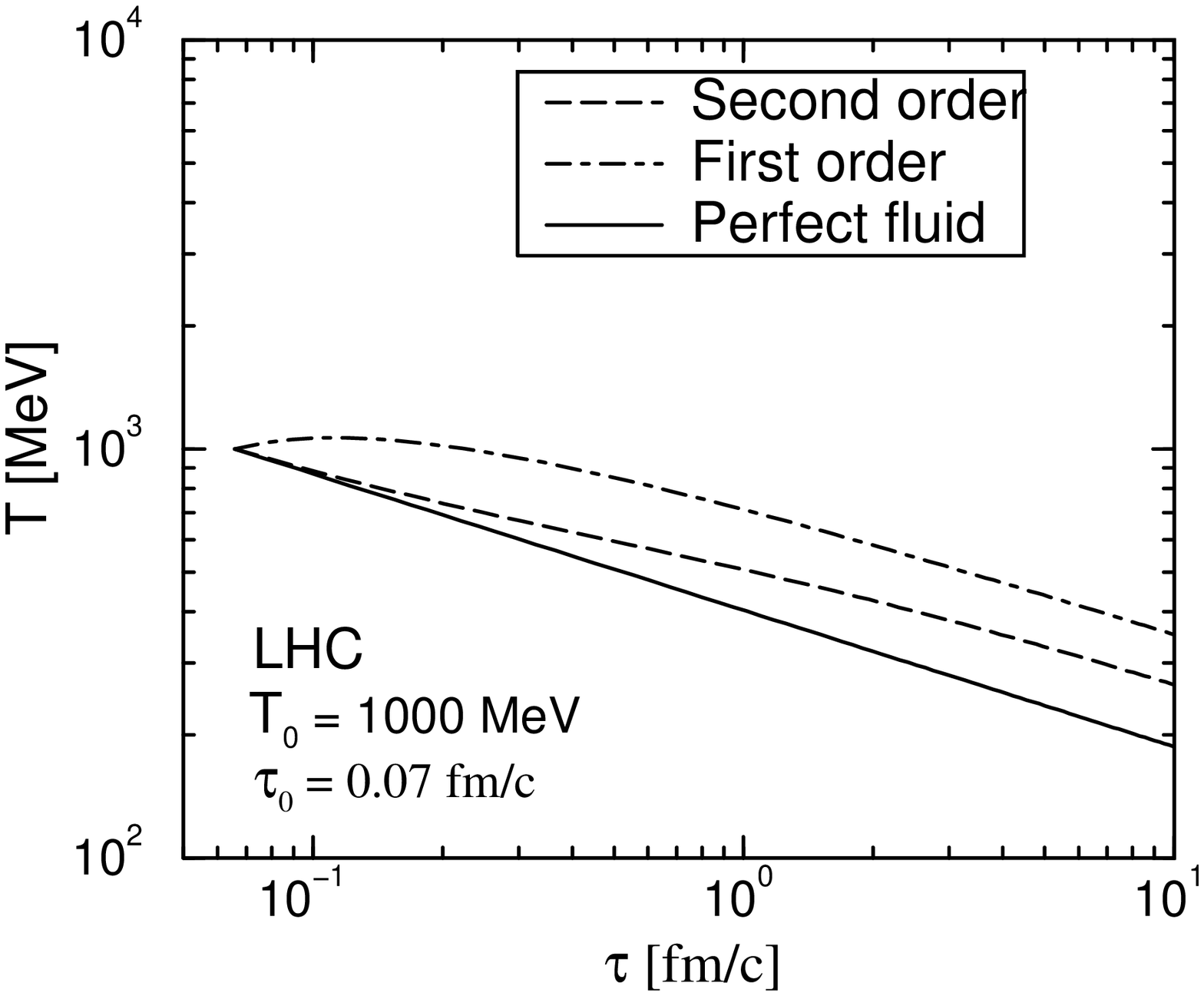,width=7.5cm}}
\end{minipage}
\hfill
\begin{minipage}[b]{7.0cm}
\centerline{\psfig{figure=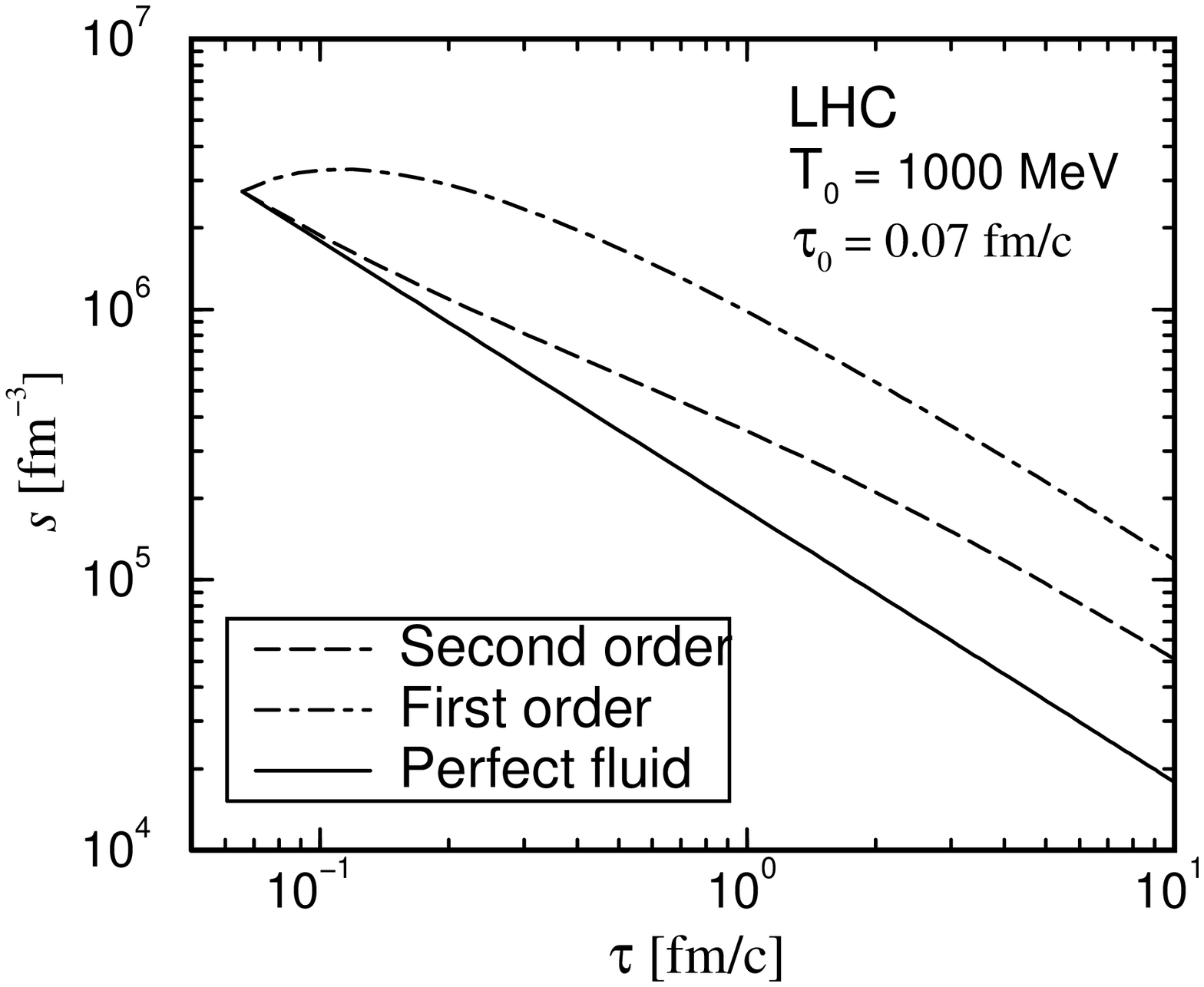,width=7.5cm}}
\end{minipage}
\caption{The proper time evolution of temperature and entropy for given
primary initial conditions $(T_0,\,\tau_0)$  for LHC scenario.}
\label{fig:TstauLHC}
\end{figure}

\begin{figure}[htb]
\begin{minipage}[t]{7.0cm}
\centerline{\psfig{figure=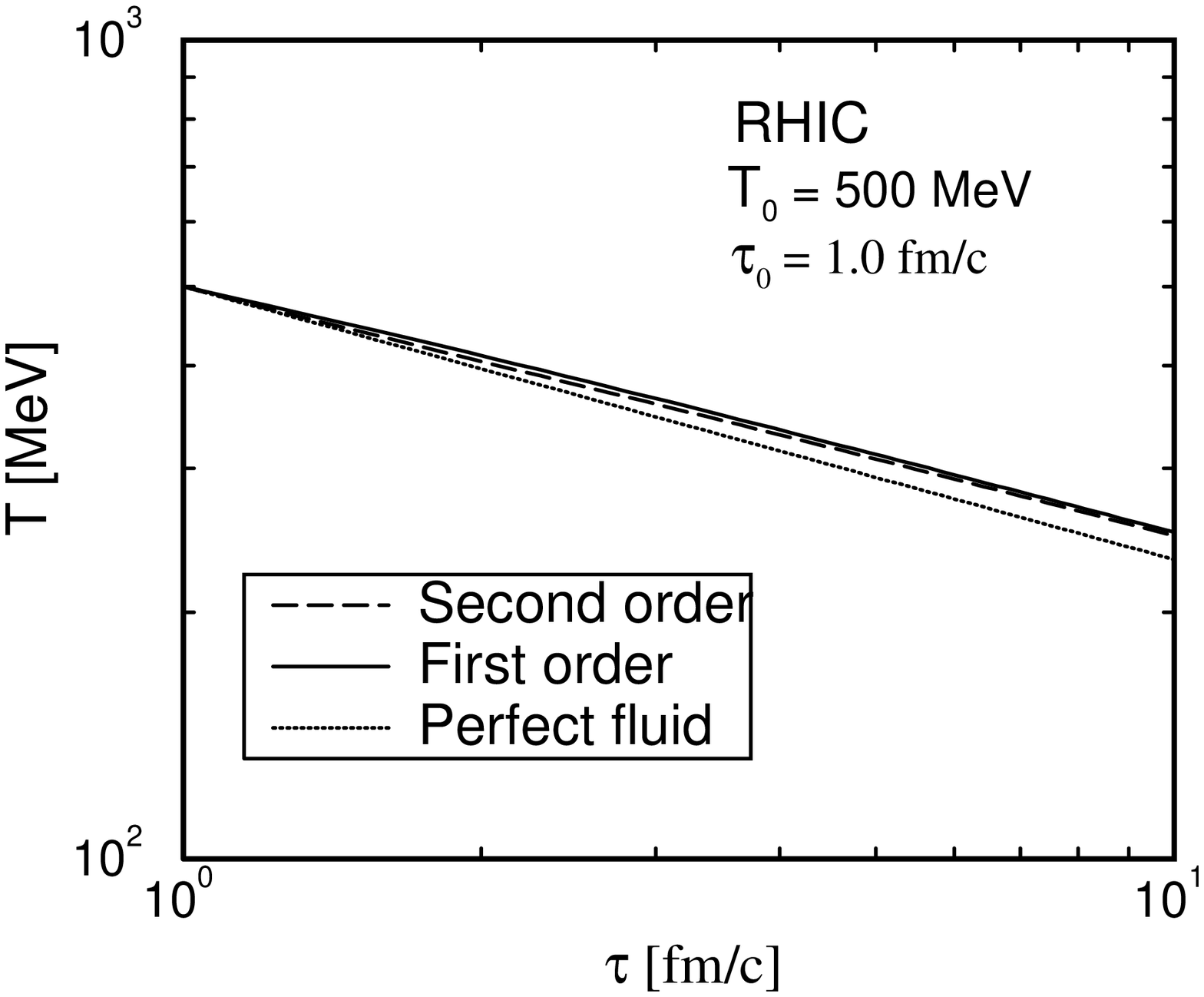,width=7.5cm}}
\end{minipage}
\hfill
\begin{minipage}[t]{7.0cm}
\centerline{\psfig{figure=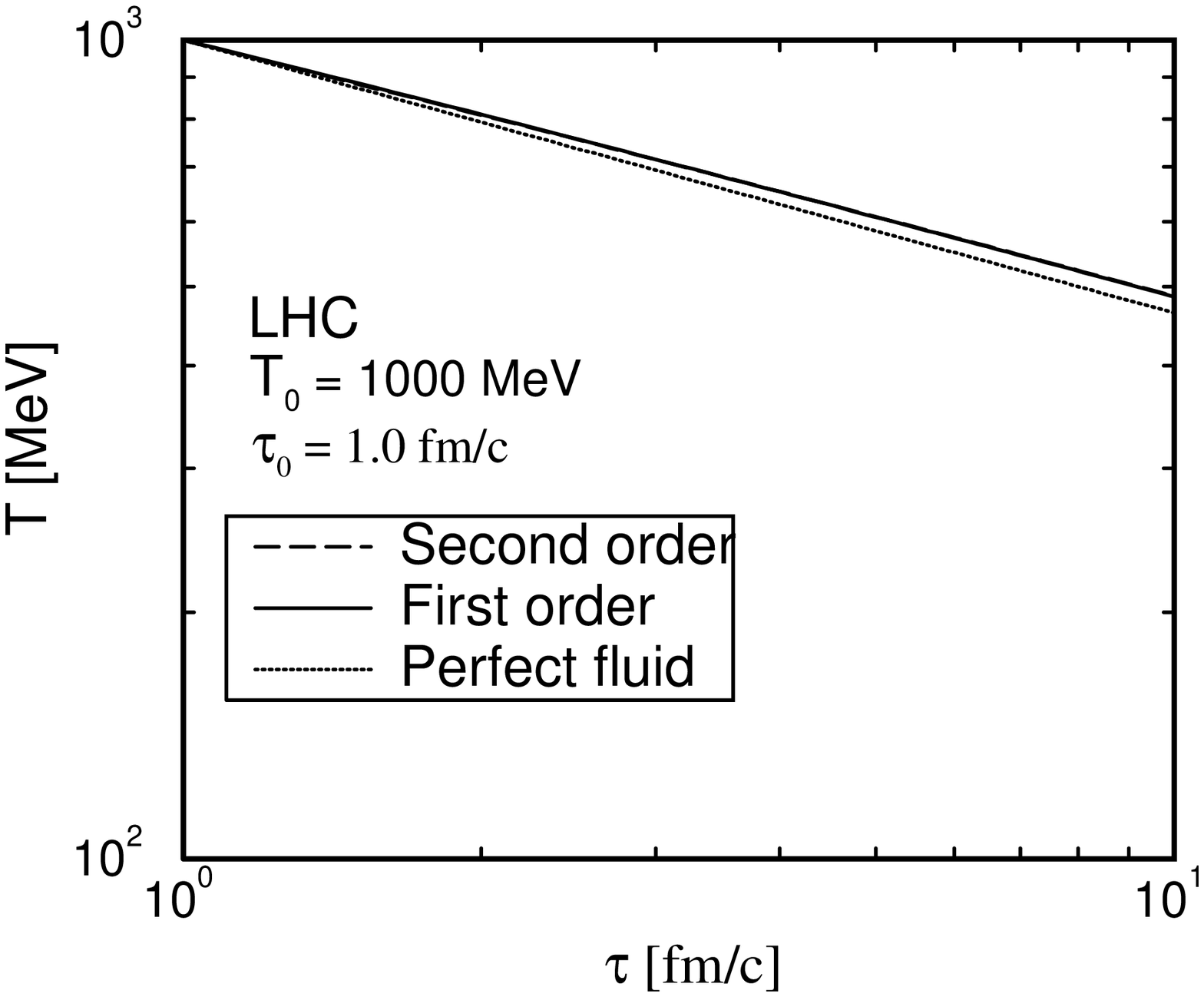,width=7.5cm}}
\end{minipage}
\caption{The proper time evolution of temperature for RHIC scenario and LHC
scenario. The initial time is arbitrarily chosen. The first order and second
order theories overlaps.}
\label{fig:TtauRLpick}
\end{figure}

The initial temperatures expected are $T_0$=500 MeV at RHIC and $T_0$=1000 MeV
at LHC. In  Figs. \ref{fig:TstauRHIC} and \ref{fig:TstauLHC} the initial time
$\tau_0$ is estimated by using the uncertainty principle \cite{Kapusta92}: 
$\tau_0 \cdot {\langle E \rangle}_0 \sim 1$ where ${\langle E \rangle}_0 \sim 3
T_0$ for massless particles. This results in  $\tau_0$=0.13 fm/c at RHIC and
$\tau_0$=0.07 fm/c at LHC.  The initial value used for $\Phi$, must be
specified independently for the second order theory. These values are
legitimate since the second order theory is based on the assumption that the
dissipative fluxes are small compared to the primary thermodynamic variables,
namely $p,n, $ and $\varepsilon$.  The effect of dissipation is more pronounced
at the very early stages of heavy ion collisions when gradients of
temperature,  velocity, etc., are large. This can be seen by comparing Figs.
\r{fig:TstauRHIC}, \r{fig:TstauLHC} with Fig. \r{fig:TtauRLpick}.  At late
times the effect of dissipation diminishes greatly. For comparison,  in Fig.
\ref{fig:TtauRLpick}  we take a constant initial time  $\tau_0$=1.0 fm/c, which
is the characteristic hadronic time scale. From Figs. \r{fig:TstauRHIC} and
\r{fig:TstauLHC} Euler hydrodynamics predicts the fastest cooling. At
hadronization times of order of 4 fm/c the perfect fluid dynamics solution
decreases by a factor of about three. The first order theory fails badly even
for this case where we have a very high initial parton density.  The first
order theory significantly under-predicts the work done during the expansion
relative to the  M\"uller-Israel-Stewart and Euler predictions. Thus the
temperature decreases more slowly with the inclusion of dissipative effects.
This would lead to greater yields of photons and dileptons. Also, the
transverse energy and momentum would be reduced as the collective velocities
are dissipated into heat. The system takes longer to cool down. This will delay
freeze--out. The entropy, $s=4\,a\,T^3$, is enhanced. This is important because
entropy production can be related to the final multiplicity.

A legitimate question to ask is: Do we really want to synchronize the initial
conditions for both ideal fluid, first order and second order theories?
Given some initial conditions we want to investigate the importance of second
order theories as compared to first order theories and perfect fluids.
That is, if one is given a set of well known initial conditions from experiment
we want to see which of the theories best describe the dynamics of the 
system. Given an observable and a set of primary initial conditions we would
like to see whether the microscopic cascade models, the ideal fluid, the first
order theory or second order theory best describes the evolution of the system. 

Let us now analyze the differences between the second order and first order
theories.
The first thing we notice is that the Eckart--Landau--Lifshitz theory predicts
that at early times the temperature will rise before
falling off. This is more pronounced when we have small initial times. 
 Naively one would
expect that the system would cool monotonically as it expands, even in the case of dissipation 
where energy-momentum is conserved. On the other hand, it is
seen that for large initial times and high temperatures the three theories
have a similar time evolution. As can be seen from Fig. 
\ref{fig:TtauRLpick}, all three cases start at the same point and then fall off 
with time.
The difference stems from the fact that in the second order theory the
transport equations of the dissipative fluxes describe the evolution of these
fluxes from an arbitrary initial state to an equilibrium state. The
first order theory, though, is just related to the thermodynamic forces which, if
switched off, do not demonstrate relaxation. Hence they are sometimes
referred to as {\em quasi--stationary} theories.
As can be seen from Fig. \ref{fig:TtauRLpick}, it is before the establishment of an
 equilibrium state that the two theories differ significantly. 


\subsection{Hadronic Fluid Dynamics}

So far our focus has been on the quark--gluon plasma where the
composition of the parton fluid enters the description through the form of the
conservation laws and the equation of state. 
Now we study the dynamics of a pion fluid. Pions are the lightest
hadrons. They are produced in abundance in ultra--relativistic collisions compared
to heavier hadrons, particularly in the central region. It is therefore important
to study their influence on the expansion. If pions are produced by
hadronization of quark--gluon plasma, then dissipation encountered during their
subsequent expansion may change the observables. The expansion in the central
region conserves pion energy and momentum. Since pions carry baryon number zero,
their total number is not conserved. Therefore, we expect the equilibrium
number density of pions in a given volume to vary with temperature.

The equation of state is approximated by that of a massless pion gas. 
Thus the pressure is given by $p=a\,T^4$ with $a = g_h \pi^2/90$ 
where $g_h = 3$ is the number of degrees of freedom. The energy density 
and entropy density 
are given by $\eps= 3\,a\,T^4$ and $s=4\,a\,T^3$ respectively.
The bulk viscous pressure equation does not contribute for
massless particles, since $\zeta\longrightarrow 0$ \cite{Weinberg72}. For the 
(1+1)--dimensional Bjorken-type hydrodynamics the heat term in the energy equation will
not contribute. Thus we need only the shear viscous pressure for this
presentation. 
The energy density evolution equation is determined by 
\begin{equation}
\frac{d \,\eps}{d\,\tau} = -\frac{(\eps + p)}{\tau}
+\frac{\Phi}{\tau}\K
\end{equation}
where
\begin{eqnarray}
\Phi &\equiv& 0~~~~~~~~~~~~~~~ \mbox{perfect fluid} \enspace,\\
\Phi &=& \displaystyle \frac{4}{3} \eta/\tau~~~~~~~~~~ \mbox{standard theory}
\enspace,\\
\tau_\pi \frac{d\,\Phi}{d\,\tau} &=&  \displaystyle -\Phi -{1\over 2} \Phi \tau_\pi
\left({1\over \tau}  +{1\over \beta_2} T {d \over d \tau}\left({\beta_2\over
T}\right)\right) + \frac{4}{3}\eta/\tau \\
&&~~~~~~~~~~~~~~~\mbox{extended theory} \nonumber \enspace,
\end{eqnarray}
where $\eta = \tau_\pi/(2\beta_2)$. 
For massless particles $\beta_2=3/(4\,p)$, and this is used
in the expression for $\tau_\pi$. 
The primary transport coefficients of a massless pion gas are not that
well-known. For chiral pions the expressions for shear viscosity and thermal
conductivity are given in \cite{Prakash93}. 
We will estimate the shear viscosity from the mean collision time of the
pions. The mean time between collisions of pions moving at $\lla v \gra \approx
1$ is given by
\be
\tau_\pi = {1\over \sigma n} \K
\ee
where $n(T) = b T^3$ with $b = 3 \zeta(3)/\pi^2$ is the pion density and $\sigma
\approx 1$ fm$^2$ is an effective cross section. The quantity 
$\tau_\pi T^3 = 1/(\sigma b)$ fm$^2$ is roughly constant for temperatures 
$T >100$ MeV. 
The shear viscosity can therefore be represented by 
\be
\eta = f_\eta T~~~~~ \mbox{with}~~~~~f_\eta 
= {2\over 3} {a\over b} {1\over \sigma} \p
\ee

Using the transport and thermodynamic properties outlined here the energy 
and transport equations can be written as
\bea
{\d T \over \d \tau} &=& -{1\over 3} {T\over \tau} +{1\over 12} {1\over a T^3
\tau}\Phi \K\\
{\d \Phi \over \d \tau} &=& -\sigma b T^3 \Phi -{1\over 2} \left({1\over \tau} -
5{1\over T} {\d T \over \d \tau}\right) + {8\over 9} {a T^4\over \tau} \p
\eea 

The energy equation can be solved analytically for the
perfect fluid and the first order (provided $\eta$ is constant) cases. But since we
want $\eta$ to depend on temperature or time one must then solve the equations
numerically, or first find the temperature evolution as done in the previous
section.
In the case of the perfect fluid and the first order theories the energy equation 
is readily solved.
\bea
T(\tau) &=& T_0\left[\tau_0\over \tau\right]^{1/3} ~~~~~~~~~~\mbox{perfect
fluid} \K\\
T(\tau) &=& T_0\left[\tau_0\over \tau\right]^{1/3} \left\{1 + {f_\eta\over 3 a T_0^3\tau_0}
\ln\left[\tau_0\over \tau\right] \right\}^{1/3} ~~~~~\mbox{first order} \p  
\eea

Hydrodynamics applies to the intermediate stages of the evolution until the
system eventually freezes out of local equilibrium into freely streaming
fragments as it expands. At freeze--out hydrodynamics must break down, since the
pion collision rate drops below the local expansion rate. 
The system will be purely longitudinal as long as its transverse motion can
be neglected. Once the transverse rarefaction wave reaches the center, however,
the flow becomes effectively three dimensional, so that the density diminishes
as $\tau^{-3}$. Freeze out then occurs because the
mean free path, which then increases as $\tau^3$, grows faster than the expansion
time scale $\mid n/\dot n\mid =\tau/3$.

\begin{figure}[htb]
\begin{minipage}[h]{7.0cm}
\centerline{\psfig{figure=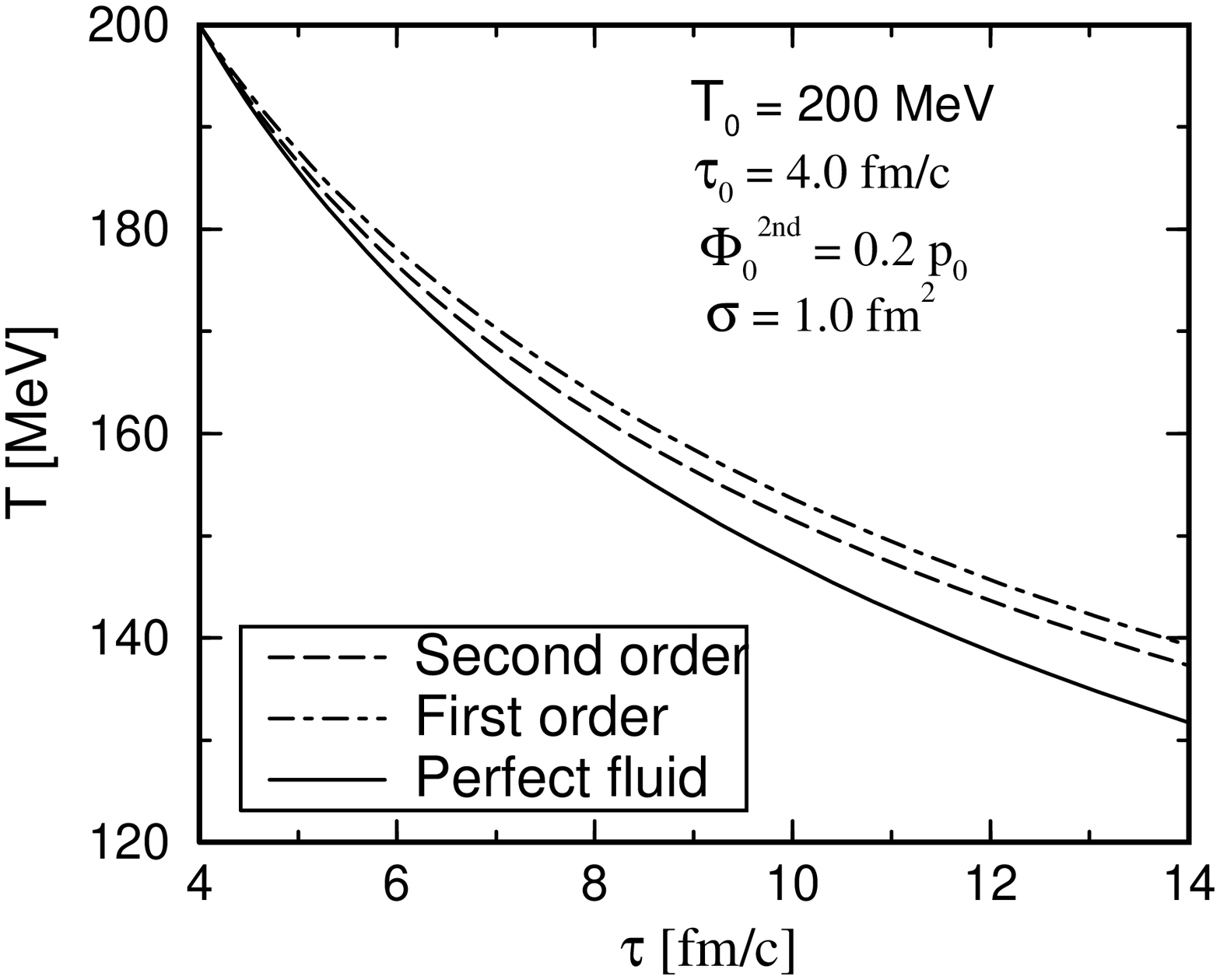,width=7.5cm}}
\end{minipage}
\hfill
\begin{minipage}[h]{7.0cm}
\centerline{\psfig{figure=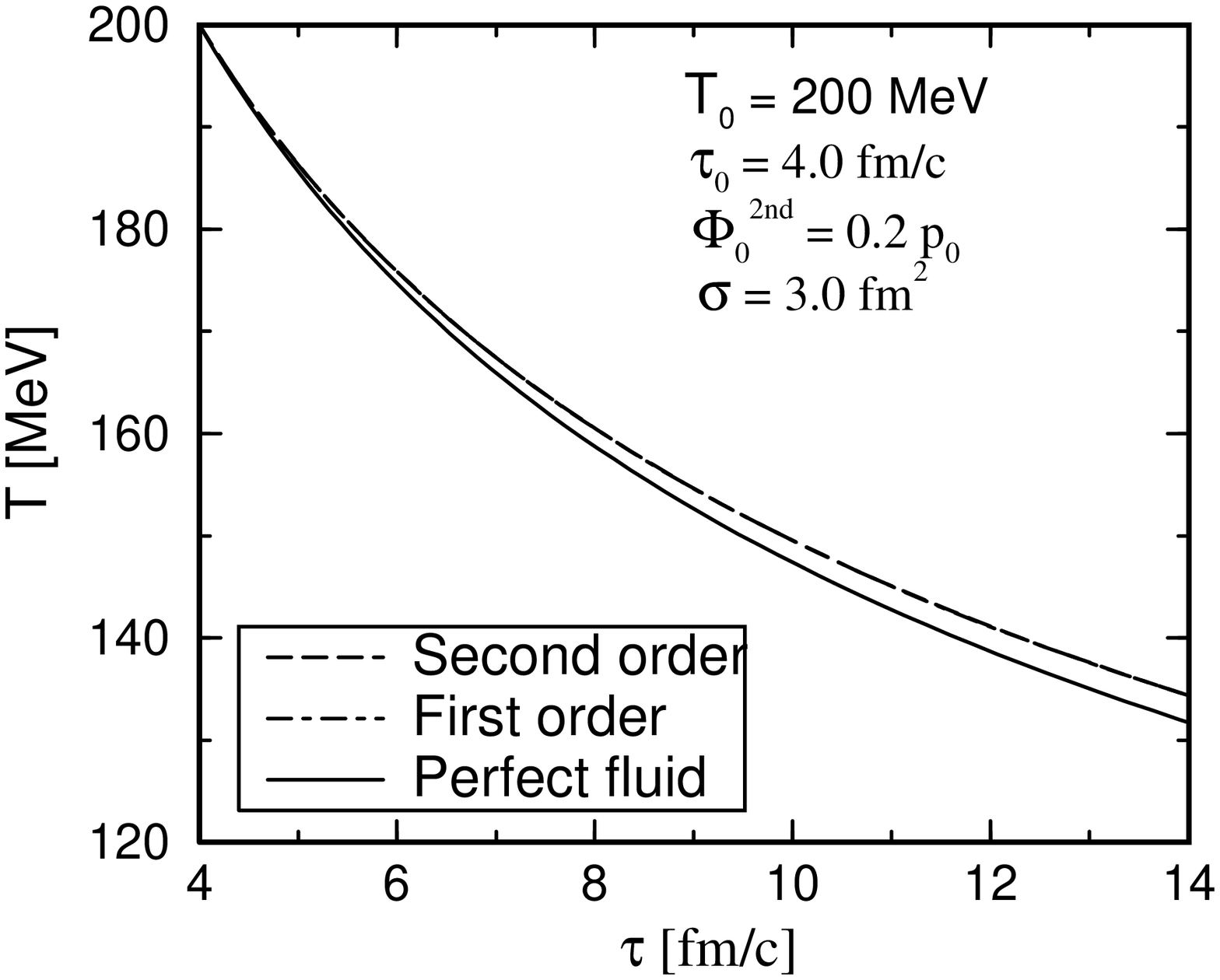,width=7.5cm}}
\end{minipage}
\caption{The proper time evolution of temperature for or $\s=1.0$ fm$^2$ and
$\s=3.0$ fm$^2$. The initial conditions ($T_0,\,\tau_0$) are arbitrarily chosen.
The first order and second order theories overlaps.}
\label{fig:Ttaupion4}
\end{figure}

In Fig. \ref{fig:Ttaupion4}  we show the $\tau$ dependence of temperature
for the three different cases: a perfect fluid, a first order theory
of dissipative fluids and a second order theory of dissipative fluids. 
Here we assume that the pion gas is produced at hadronization of quark--gluon
plasma at $\tau=4$ fm/c. As expected, in this regime, with the given initial
conditions, the first order and second order theories converge. This convergence 
is faster with increasing cross sections. 
The effects of
viscosity are small but non--negligible. 

\begin{figure}[htb]
\begin{minipage}[h]{7.0cm}
\centerline{\psfig{figure=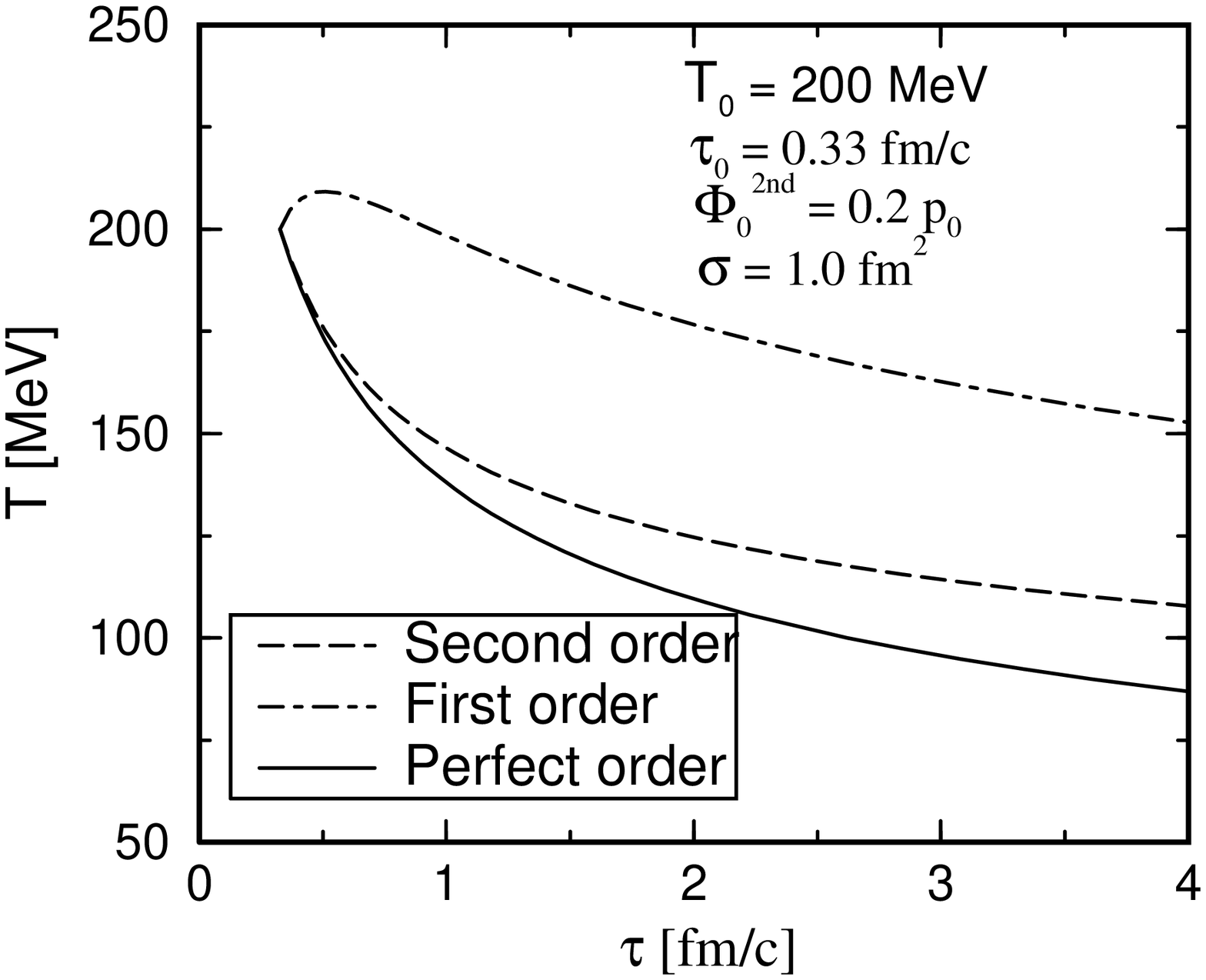,width=7.5cm}}
\end{minipage}
\hfill
\begin{minipage}[h]{7.0cm}
\centerline{\psfig{figure=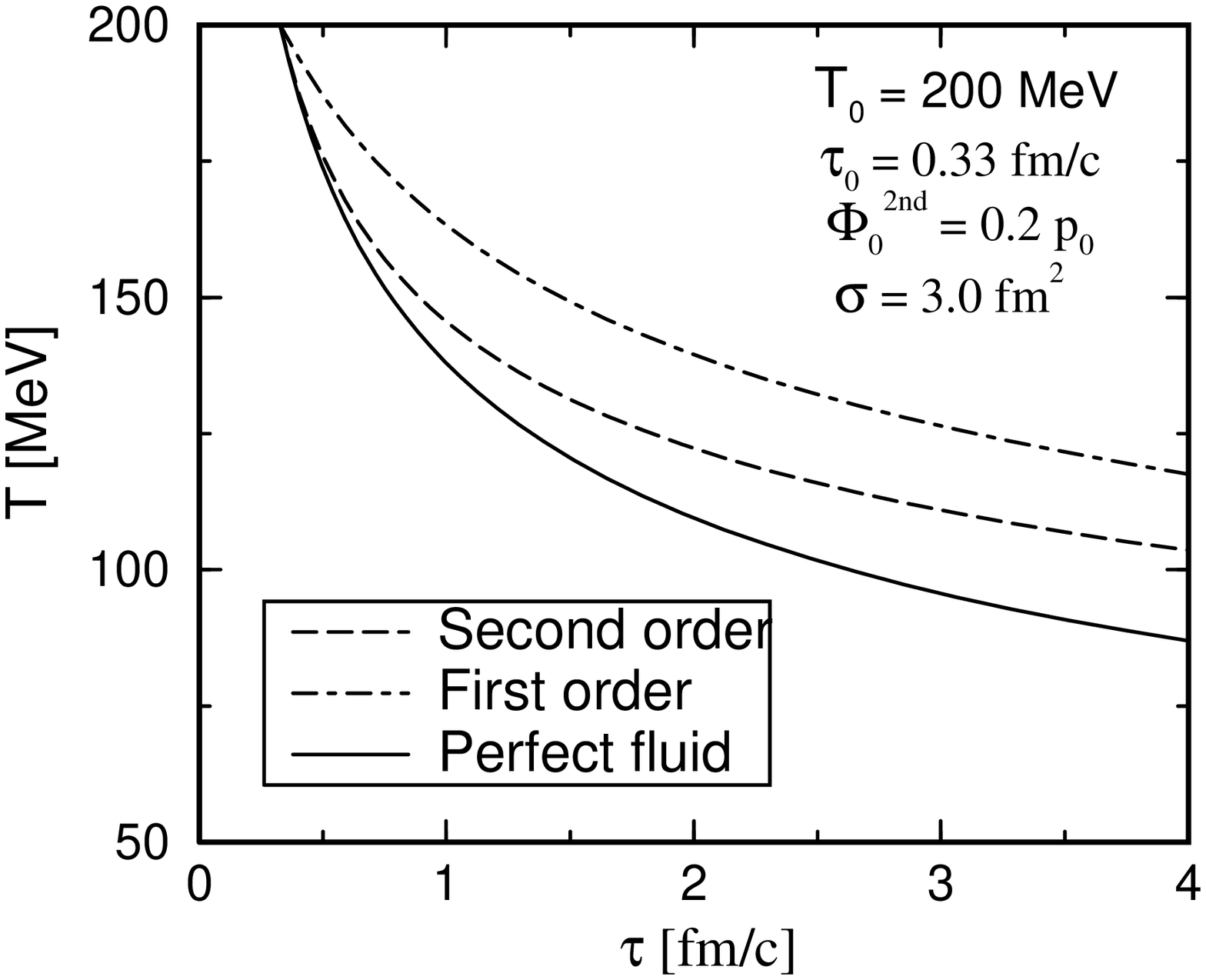,width=7.5cm}}
\end{minipage}
\caption{The proper time evolution of temperature for $\s=1.0$ fm$^2$ and
$\s=3.0$ fm$^2$. The primary initial conditions are those from uncertainty
principle.}
\label{fig:Ttaupionthird}
\end{figure}

In Fig. \ref{fig:Ttaupionthird}  we assume that the pion gas is formed at
$\tau_0 \sim 1/(3 T_0)$. As we know by now, the difference between the
three theories is noticeable and first order theories are not suitable. We see
here also that the convergence of first order theory results and second order theory
results will occur for large cross sections.

The presence of dissipation in heavy ion reactions will have profound effects on
the space--time evolution of the system. 
The freeze--out will be delayed. Temperature and energy density decrease more 
slowly. 
Enhancement of entropy production will increase the
production of observed particles since the two can be related. Since the system
takes longer to cool this will lead to an enhancement in the production of
thermal signals (dileptons and photons).

\section{Conclusions}
\label{sect:summary}

In this work I have given a comprehensive exposition of the non--equilibrium
properties of a new state of matter produced in heavy ion reactions. In doing
so I presented some basic features of non--equilibrium fluid dynamics. I 
studied the space--time description of high energy nuclear collisions.  I chose
to follow the phenomenological approach and left kinetic theory approach for
future work. 
 
The ultimate aim is to bridge the phenomenological theory  with the kinetic
theory of the matter produced in heavy ion coliisions. In doing so I made use 
of the dissipative fluid dynamics.
The connection between the macroscopic theory and microscopic theory enters
through the transport coefficients of the matter.  The equation of state
provided closure to the system of conservation equations.

I demonstrated that extended irreversible thermodynamics provides a consistent
framework to simulate and study the space--time evolution of ultra-relativistic
nuclear collisions, from some initial time  to the final particle yield.
Although this approach relies on a number of fundamental assumptions and is far
from providing an accurate quantitative description, it has the advantage of
wide applicability.

The advent of accelerators such as RHIC and  LHC provide  an opportunity for
studying the dynamics and properties of the matter at very high energy
density.  In the description of the evolution of such a system, it is mandatory
to evaluate, as accurately as possible, the order of magnitude of different
characteristic time scales, since their relationship with the time scale of
observation will determine, along with the relevant  equations, the evolution
pattern. This is rather general when dealing with dissipative systems. It has
been my purpose here, by means of simple model with simple equation of state 
and arguments related to a wide range of time scales, to emphasize the
convenience of resorting to hyperbolic theories when dissipative processes,
either outside the steady--state regime or when the observation time is of the
order of or shorter than some characteristic time of the system, are under
consideration. Furthermore, dissipative processes may affect the way in which
the system tends to equilibrium, thereby affecting  the future of the system 
even for time scales much larger than the relaxation time.

In the early stages of heavy ion collisions, non--equilibrium effects play a 
dominant role. A complete description of the dynamics of heavy ion reactions
needs to include the effects of dissipation through dissipative or
non--equilibrium fluid dynamics. As is well--known, hyperbolic theories of
fluid dissipation were formulated to get rid of some of the undesirable
features of parabolic theories, such as acausality. It seems appropriate
therefore to resort to hyperbolic theories instead of parabolic theories in
describing the dynamics of heavy ion collisions.  Thus in ultra--relativistic
heavy ion collisions, where the fluid evolution   occurs very rapidly, the
second order theories, due to their universality, should be used to analyze  
collision dynamics. 

Unlike in first order theories, where the transport equations are just the
algebraic relations between the dissipative fluxes and the thermodynamic
forces,  second order theories describe the evolution of the dissipative fluxes
from an arbitrary initial state to a final steady-state using the transport
equations. The  presence of relaxation terms in second order theories makes the
structure of  the resulting transport equations hyperbolic and they thus
represent a  well-posed initial value problem.  

The consequences of non-ideal fluid dynamics, both first order (if applicable) 
and second order were demonstrated here in a simple situation, that of scaling
solution assumption and simple equation of state.  A more careful study of  the
effects  of the non-ideal fluid dynamics on the observables is therefore
important.  Conversely, measurements of the observables related to
thermodynamic quantities  would allow us to determine the importance and
strength of dissipative processes in heavy ion collisions.

Here we have used only a simple equation of state in a simplified model of high
energy nuclear  collisions.  A more realistic situation (including transverse
expansion) will  require careful analysis of both the transport coefficients
and the equation of state which are employed in the full set of the equations. 
It is then that one may have a better understanding of when to use either of
these theories in the context of relativistic heavy ion collisions. 

\subsection{The Need for Hyperbolic Theories of Relativistic Dissipative Fluid
Dynamics for Nuclear Collisions}

The need for hyperbolic theories of dissipative fluid dynamics has been
emphasized before \cite{hyperbolic}. But it is in ultra--relativistic nuclear
collisions where it needs to be stressed.

The first order theories of dissipative fluid dynamics have a long history and
proved very useful, in some special cases in understanding observables.  
However, they exhibit some undesirable features, such as acausality,  that
prompted the formulation  of hyperbolic theories of dissipative fluid dynamics
to get rid of them.  This was  achieved at the price of extending the set of
field variables by including the dissipative fluxes, such as the heat current
and viscous stresses,   on the same footing as the primary ones, such as energy
densities, equilibrium pressures, and number densities, thereby giving rise to
a set of more physically  satisfactory but involved theories from the
mathematical point of view. These theories have the additional advantage of
being backed by a kinetic theory of gases (Grad's 14--moment approximation). 
Second order theories, being hyperbolic in structure, lead to well-posed
initial-value (Cauchy) problems. They also lead  to causal propagation. Unlike
the first order theories, second order theories have  relaxation terms which
permit us to study the evolution of the dissipative fluxes.

In addition to causality violation, the Eckart theory leads to unstable
equilibrium states under small perturbations. A dissipative hot and dense
nuclear matter may very rapidly tend towards a quasi--stationary state that is
adequately described by the Eckart theory. However, this needs to be checked a
priori.  The Eckart theory can describe steady--state regimes. However it is 
clearly unable to deal with the evolution towards these regimes,
or with the overall dynamics of the fluid, in a satisfactory way.

It can be argued that the infinite speeds of propagation predicted by parabolic
theories and the unphysical behaviour of the space--time evolution of
thermodynamic quantities is just a consequence of using them outside their
range of applicability \cite{Danielewicz85}. However hyperbolic theories also
share this constraint but they naturally predict speeds which are of the order
of the characteristic speeds of the problem at hand. Therefore hyperbolic
theories have a much wider range of applicability than parabolic theories.

Aside from the relaxation time $\tau$, hyperbolic theories introduce a certain 
number of new coefficients that couple the different dissipative fluxes. This is 
rather natural, since these theories are designed to explain more
complex phenomena than parabolic theories do. These quantities are not free 
parameters that one could choose at will. 
These new coefficients are restricted by the convexity of the
entropy function, and can be calculated explicitly with the help of kinetic
theory or fluctuation theory.  It turns out that they are functions of the
equation of state.

In summary, although parabolic theories have proved very useful for many
practical purposes, they  appear to fail hopelessly in describing the dynamics
of heavy ion collisions. In contrast, hyperbolic theories successfully  give a
better description in agreement with transport models and hopefully they will be
able to predict the experimental results.  Thus  hyperbolic theories are more
reliable. In the steady--state, under the conditions mentioned before, and for
times exceeding $\tau$ both theories converge. Based on the results presented
here one can only stress the convenience of using hyperbolic transport
equations when parabolic theories either fail or the problem under
consideration happens to lie outside the range of applicability of parabolic
theories.

I will conclude by pointing out some of the advantages and challenges of the
second order theories.  The challenge we face is the increase in the space of thermodynamic
variables. We now have, in addition to the transport coefficients,  new
coefficients in the problem. These are the relaxation coefficients $\beta_i$
and the coupling coefficients $\alpha_i$. These new coefficients depend on the
primary thermodynamic variables, such as  $n,\varepsilon$ and $p$ and therefore
are determined by the equation of state. Like viscosity and thermal
conductivity, which are required to be positive by the second law of
thermodynamics, these new coefficients are  constrained by hyperbolicity
requirements. In principle , in order to solve the second order relativistic
dissipative fluid dynamic problem, one still needs the equation of state,
initial conditions and the transport coefficients.
Another challenge involves solving
the equations numerically. Finally one would like  to compare the results of
non--equilibrium fluid dynamics to observables. This will require an effort to
solve the full system of equations numerically. Efforts in this direction are
under way \cite{Muronga03a}

The study of non-ideal or non--equilibrium fluid dynamics will be important for
constructing hydro-molecular dynamic schemes \cite{Bass00}.   Hyperbolic
theories might prove to be convenient in such scheme. A resulting
hydro-molecular dynamic scheme can then be compared to the studies of
multi-fluid dynamics \cite{Brachmann97}.

\section{Outlook}

There are important questions that need to be investigated in order to tackle
the challenges faced by hyperbolic theories.  An important question is the
measurability of the dissipative fluxes. The heat flux through a system may be
simply evaluated by measuring the amount of energy transported per unit area
and time. The viscous pressure can be measured from the tangential shear force
exerted per unit area. In practice, it may be difficult to evaluate these
quantities at each instant and at every point. From kinetic theory these fluxes
can be simulated from microscopic transport models such as HIJING, VNI and
UrQMD.

A thorough study of transport coefficients is needed. The results from lattice
QCD should give clear predictions for these coefficients. So far there are no
reliable phenomenological expressions for the coefficients. The shear viscosity
results presented in \cite{Muronga03b} using the microscopic model UrQMD serve as a starting
point for future calculations of transport coefficients.

A thorough study of the equation of state is required in order to be consistent
with the non--equilibrium description of matter. In addition to the
initial/boundary conditions for the primary/equilibrium variables a thorough
study of initial/boundary conditions for the dissipative fluxes is necessary.
This is required for the evolution equations of the fluxes. For example, one
would like to know how much of a particular dissipative flux is generated as  a
response to an associated thermodynamics force in the early stages of the heavy
ion collisions. This in turn gives us information about the initial entropy
generated as a result of dissipation. 

For the complete description of the dynamics of viscous, heat conducting
matter  we need to consider more realistic situations: a system that expands in
both the longitudinal and transverse directions \cite{Muronga03a}  and we
need a  full (3+1)--dimensional solutions to the conservation and evolution
equations. This will require extensive numerical computation. This is a
challenging but interesting  problem. In order to understand the
observables we need a full formulation of hyperbolic theory that should be
tested against other models/theories.

To solve imperfect hydrodynamics amounts to knowledge of   (i) a realistic
equation of state, (ii) reliable   transport coefficients (ii) realistic
initial/boundary conditions and  (iv) numerical computational algorithm. In
order to understand the  observables from RHIC and LHC knowledge of these
requirements is needed  and that is what I have started to do.  I hope that
this will gather momentum in due course.

\acknowledgments
I am grateful to Adrian Dumitru for reading the manuscript and for valuable
discussions. I also thank Joe Kapusta , Pasi Huovinen, Dirk Rischke and  Tomoi
Koide for valuable comments.
This work was supported by the US Department of Energy grant
DE-FG02-87ER40382.

\end{document}